\documentclass[aps,prd,twocolumn,floatfix,showpacs,superscriptaddress,nofootinbib]{revtex4-1}  
                
\usepackage{graphicx}  
\usepackage{dcolumn}   
\usepackage{bm}        
\usepackage{amssymb}   
\usepackage{epsfig}  
\usepackage{latexsym}
\usepackage{amssymb} 
\usepackage[english]{babel}
\usepackage{amsmath}
\usepackage{enumerate}
\usepackage{color}
\usepackage{amsmath}
\usepackage{mathrsfs,dsfont}
\usepackage{bm}

\newcommand{\changetext}[1]{#1}

\newcommand{\coefalpha}{0.75}
\newcommand{\threej}[6]{{\left(\begin{array}{ccc} #1 & #2 & #3 \\ #4 & #5 & #6 \end{array}\right)}}

\newcommand{\sovast}{Sov. Astron.}
\newcommand{\procspie}{Proc. SPIE}
\newcommand{\apjl}{Astrophys. J. Lett.}
\newcommand{\aap}{Astron. Astrophys.}

\newcommand{\aaps}{Astron. Astrophys. Supp. Ser.}
\newcommand{\aj}{Astron. J.}

\newcommand{\mnras}{Mon. Not. R. Astron. Soc.}

\newcommand{\physrep}{Phys. Rep.}

\newcommand{\jcap}{Journal of Cosmology and Astroparticle Physics}

\newcommand{\beq}{\begin{equation}}
\newcommand{\eeq}{\end{equation}}
\newcommand{\bga}{\begin{gathered}}
\newcommand{\ega}{\end{gathered}}

\usepackage{natbib}
\usepackage{graphicx}
\def\be{\begin{equation}}
\def\ee{\end{equation}}
\def\({$}
\def\){$}
\def\bea{\begin{eqnarray}}
\def\eea{\end{eqnarray}}

\def\lf{\left (}
\def\rt{\right )}
\def\f{\frac}

\begin{document}
\widetext    
\title{Detecting primordial gravitational waves with circular polarization of the 
redshifted 21 cm line: II. Forecasts}
\author{Abhilash Mishra}
\email{abhilash@astro.caltech.edu}
\affiliation{Theoretical Astrophysics Including Relativity (TAPIR),\\
Caltech, M/C 350-17, Pasadena, California 91125, USA}
\author{Christopher M. Hirata}
\email{hirata.10@osu.edu}
\affiliation{Center for Cosmology and Astro Particle Physics (CCAPP),\\
The Ohio State University,  191 West  Woodruff Lane, Columbus, Ohio 43210, USA}    
 
\begin{abstract} 
In the first paper of this series, we showed that the CMB quadrupole at high redshifts results in a small circular polarization of the emitted 21 cm radiation. In this paper we forecast the sensitivity of future radio experiments to measure the CMB quadrupole during the era of first cosmic light ($z\sim 20$). The tomographic measurement of 21 cm circular polarization allows us to construct a 3D remote quadrupole field. Measuring the $B$-mode component of this remote quadrupole field can be used to put bounds on the tensor-to-scalar ratio $r$. We make Fisher forecasts for a future Fast Fourier Transform Telescope (FFTT), consisting of an array of dipole antennas in a compact grid configuration, as a function of array size and observation time. We find that a FFTT with a side length of 100 km can achieve $\sigma(r)\sim 4\times 10^{-3}$ after ten years of observation and with a sky coverage $f_{\mathrm{sky}}\sim 0.7$.  The forecasts are dependent on the evolution of the Lyman-$\alpha$ flux in the pre-reionization era, that remains observationally unconstrained. Finally, we calculate the typical order of magnitudes for circular polarization foregrounds and comment on their mitigation strategies. We conclude that detection of primordial gravitational waves with 21 cm observations is in principle possible, so long as the primordial magnetic field amplitude is small, but would require a very futuristic experiment with corresponding advances in calibration and foreground suppression techniques.
\end{abstract} 
        
\pacs{98.70.Vc, 98.80.Bp, 98.80.Es} 
\maketitle

\section{Introduction and Motivation}
The idea that the early universe underwent a period
of inflationary expansion, is one of the cornerstones
of modern cosmology. Inflation was originally invoked
as a solution to the flatness and horizon problems~\cite{1981PhRvD..23..347G} but proved to be a powerful explanation for the 
generation of initial perturbations in the early universe, that
eventually evolved to the large scale structure we see today~\cite{1982PhRvL..49.1110G, 1983PhRvD..28..679B,
1982PhLB..115..295H, 1982PhLB..108..389L, 1981ZhPmR..33..549M}.
Increasingly precise cosmological tests have verified the predictions
of the simplest single-field-slow-roll inflationary models; that the
primordial density perturbations are adiabatic, nearly Gaussian,
and nearly (but not exactly) scale-invariant~\cite{2016A&A...594A..20P, 2016A&A...594A..17P, 2016ASSP...45...41M}.

Beyond the predictions for primordial density (scalar)
perturbations, inflation also predicts the existence of 
a stochastic gravitational wave background,
with a nearly scale-invariant power spectrum~\cite{1984NuPhB.244..541A, 1982PhLB..115..189R,
1983PhLB..125..445F, 1979JETPL..30..682S}. Detection of these inflationary gravitational waves would be a smoking
gun for inflation, and their detection would open up a completely new
window into both the physics of the very early universe and physics at otherwise inaccessible energy scales, $V_{\rm inf}^{1/4}\sim 10^{-3}(r/0.01)^{1/4}M_{\rm Pl}$, where $r$ is the tensor-to-scalar ratio and $M_{\rm Pl}$ is the Planck mass. 

The principal near-term strategy to detect inflationary gravitational waves relies on the fact that waves with wavelengths comparable to the horizon size would induce a gradient free ``B-mode'' pattern in the polarization of the CMB via Thomson scattering \cite{1997PhRvD..55.7368K, 1997PhRvL..78.2058K, 1997ApJ...482....6S, 1997PhRvL..78.2054S, 1997PhRvD..55.1830Z}. There are several experimental efforts underway to detect the B-mode pattern in the CMB polarization, including ABS (Atacama B-mode Search)~\cite{2010arXiv1008.3915E}, ACTPol~\cite{2014JCAP...10..007N}, BICEP2/Keck Array~\cite{2014ApJ...792...62B, 2015ApJ...811..126B} and POLARBEAR/Simons Array~\cite{2014SPIE.9153E..1FA}. The search for inflationary gravitational waves remains the top scientific priorities for future CMB experiments (see the CMB S4 Science Book~\cite{2016arXiv161002743A}). 

The strength of the inflationary gravitational waves
is encoded in the tensor-to-scalar ratio $r$,
which is related to the Hubble rate during inflation 
and in turn depends on the energy scale 
at which inflation takes place. It is defined as $r =\Delta_{h}^{2}/\Delta_{\zeta}^2$
where,
\be
\Delta_{\zeta}^2 (k)\equiv \f{k^3}{2\pi^2} \langle |\zeta|^2 \rangle 
\ee
is the power spectrum of the curvature perturbations and 
\be
\Delta_{h}^{2} (k)\equiv 2 \f{k^3}{2\pi^2} \langle |h_{k} |^2 \rangle = \f{2}{\pi^2}\f{H^2}{M_{\mathrm{pl}}^2}
\ee
is the gravitational-wave power spectrum (summed over two polarizations), where $H$ is the Hubble rate during inflation.  The value of $r$ depends on the model of inflation considered.
Current constraints on $r$ from the combination of the CMB $B$-mode and other (more model-dependent) observables are $r<0.07$ (95\%\ CL) \cite{2016PhRvL.116c1302B}.

Galactic foregrounds, primarily due to dust emission, make the detection of tensor modes using the CMB particularly challenging. Gravitational lensing due to scalar perturbations also produce a B-mode pattern and might fundamentally limit the values of $r$ that can be probed using the CMB. In the event that future CMB experiments do detect B-modes due to inflationary GWs, it is important to devise methods, with different systematic errors, that will conclusively prove that the GW signal is indeed primordial. Furthermore, in the event that the value of $r\lesssim 0.001$, planned CMB experiments are unlikely to be able to detect $B$-modes. It is thus appropriate to investigate alternative methods to detect inflationary gravitational waves. 

In Paper I of this series (Hirata et al.\ 2017) we calculated the effect of the CMB quadrupole during the Dark Ages of the universe on  the splitting of the $F = 1$ hyperfine excited level of neutral hydrogen at high redshifts. We showed that unlike the Zeeman effect, where $M_{F}=\pm 1$ have opposite energy shifts, the remote CMB quadrupole shifts $M_{F}=\pm 1$ together relative to $M_{F}= 0$. This leads to a small circular polarization of the emitted 21cm photon, which is in principle observable.

Measurement of the circular polarization of the 21cm signal using future radio interferometers can allow us to construct a 3D {\it remote CMB quadrupole} field (i.e. the quadrupole component of the CMB skies observed by hydrogen atoms at high redshifts) during the cosmic Dark Ages. Just like the CMB polarization field, this field can be decomposed into $E$ and $B$ modes. The measurement of $B$ modes of this new remote quadrupole field, can then be used to put bounds on $r$.

In this paper (Paper II) we forecast the ability of future radio experiments to measure the remote quadrupole of the CMB  using  the circular polarization of the 21 cm line. We show that a very large Fast Fourier Transform Telescope (FFTT)~\cite{2009PhRvD..79h3530T} can in principle construct a remote quadrupole field at high redshifts ($z\sim 20$), and we make forecasts for the measurement of $r$ as a function of array size and survey duration.

This paper is organized as follows: we summarize the main
results of Paper I and outline our method in Sec.~\ref{s:outline}. 
In Sec.~\ref{s:remotequad} we make forecasts for the measurement of the remote
quadrupole of the CMB using Fast Fourier Transform Telescopes. In
Sec.~\ref{sec:sensitivity} we compute the power spectrum of the remote CMB
quadrupole and sensitivity to $r$. In Sec.~\ref{sec:foregrounds} we discuss
various foregrounds that are relevant to our measurement, and 
in Sec.~\ref{sec:discussion} we summarize and discuss the implications of our results.

\section{Outline of the Method}
\label{s:outline}

\begin{figure}[t]
  \centering
  \includegraphics[height= 3cm, width=8.5cm]{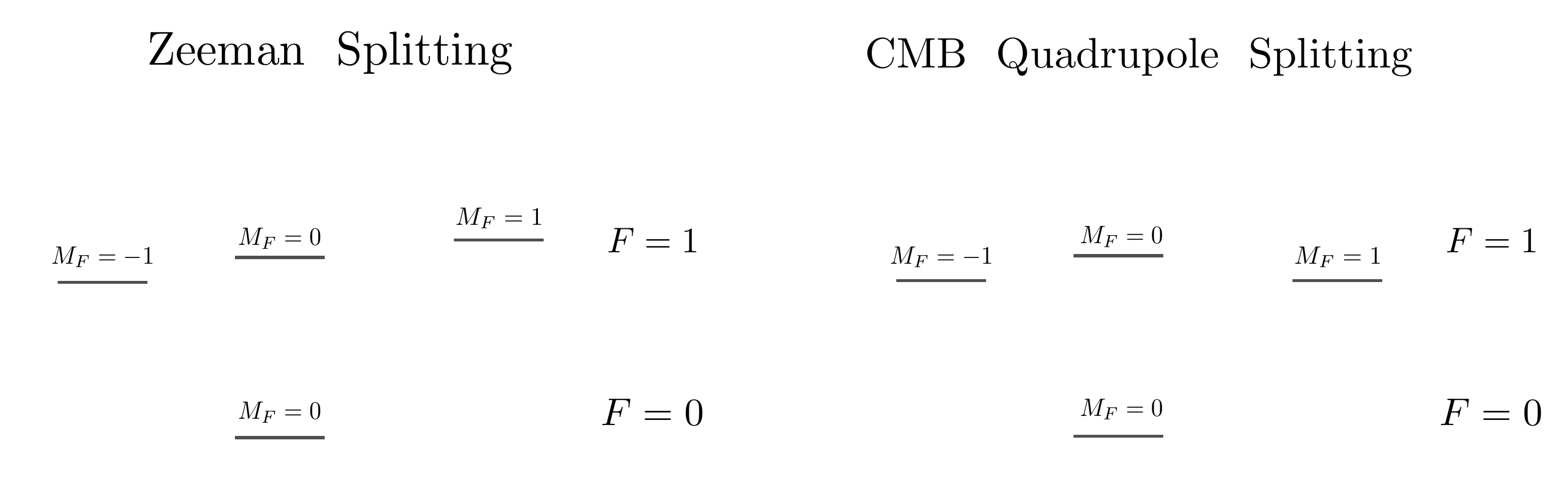}
  \caption{Energy Level Splitting}
  \label{fig:splitting}
\end{figure}

Scattering processes between photons and neutral
hydrogen atoms in the early universe can affect 
21cm observables, and lead to novel probes of physics 
at high redshifts. An extensive
review of the physics of the 21 cm transition can be found in 
Furlanetto et al.~\cite{2006PhR...433..181F}. Recently, Venumadhav et al.~\cite{2014arXiv1410.2250V} and 
Gluscevic et al.~\cite{2017PhRvD..95h3011G} considered
the effect of magnetic fields in the early universe on 
the splitting of the $F = 1$ hyperfine excited level of hydrogen.
At high redshifts, a neutral hydrogen atom is bathed in 
an anisotropic 21 cm radiation bath 
due to density fluctuations in the gas. Such
an anisotropic radiation field leads to spin polarization of the neutral
hydrogen atoms in the $F=1$ state, and hydrogen atoms
in the excited $F=1$ state align with the quadrupole of the incident
21 cm radiation. The presence of an external magnetic field leads
to the precession of atoms in the $F=1$ state, and the emitted
21 cm radiation is misaligned with the incident 21 cm quadrupole. 
Gluscevic et al.~\cite{2017PhRvD..95h3011G} showed that
this effect can in principle be used to probe large scale magnetic fields of 
the order of $10^{-21}$ Gauss comoving in the early universe. 

Beyond the Zeeman splitting due to an external magnetic field,
the CMB anisotropy at high redshifts also leads to a splitting
of the $F=1$ level but the symmetry properties is different from the magnetic
field case. In the case of an external magnetic field 
(Zeeman effect), the energy levels of the $M_{F}=\pm1$ levels shift in opposite directions,
while the CMB anisotropy leads to a shift in the same direction (see Fig.~\ref{fig:splitting}).
The emitted 21 cm photon in the latter case has small circular polarization. 

In Paper I we showed that the degree of circular polarization 
emitted by the neutral hydrogen atom as it transitions from $F=1$
to the $F=0$ state is related to the CMB quadrupole by
\begin{eqnarray}
V_{\rm obs}\changetext{({\bm k})} &=& -\frac{\sqrt{2\pi}}{25\sqrt{3}}\,\frac{T_s T_\star K_{\rm mag}  f \tau^2\delta\changetext{({\bm k})}}{T_{\gamma 0}A (1+\coefalpha\tilde x_{\alpha})(1+\tilde x_{c} +\tilde x_{\alpha})} 
\nonumber \\
&& \times \left( 1 - \frac{T_\gamma}{T_s} \right)
\Im [
a_{21} Y_{21}(\hat{\bm k})
+2a_{22} Y_{22}(\hat{\bm k})
],
\label{eq:Vobs}
\end{eqnarray}
where ${\bm k}$ is the \changetext{Fourier wave vector and $\hat{\bm k}$ is its direction}; 
$\tilde x_{\alpha}$ and $\tilde x_{c}$ parametrize the 
rates of depolarization of the ground state by optical pumping
and atomic collisions respectively; $T_{s}$ and $T_{\gamma}$ are the 
spin temperature and the CMB temperature at redshift $z$;
and $a_{2m}$ is the CMB quadrupole at the redshift $z$.
The spin-zero spherical harmonics $Y_{2m}$ are defined in the usual way (see Paper I
for details).

\changetext{Note that the derivation of Eq.~(\ref{eq:Vobs}) treated the CMB quadrupole moments $a_{2m}$ as constant. Equation~(\ref{eq:Vobs}) is thus applicable in the limit of a separation of scales: the scale on which the CMB quadrupole varies (the horizon scale during the pre-reionization epoch) is much larger than the wavelength $2\pi/k$ of the density perturbations probed in 21 cm radiation.}

\begin{figure}[t]
  \centering
  \includegraphics[width=8cm]{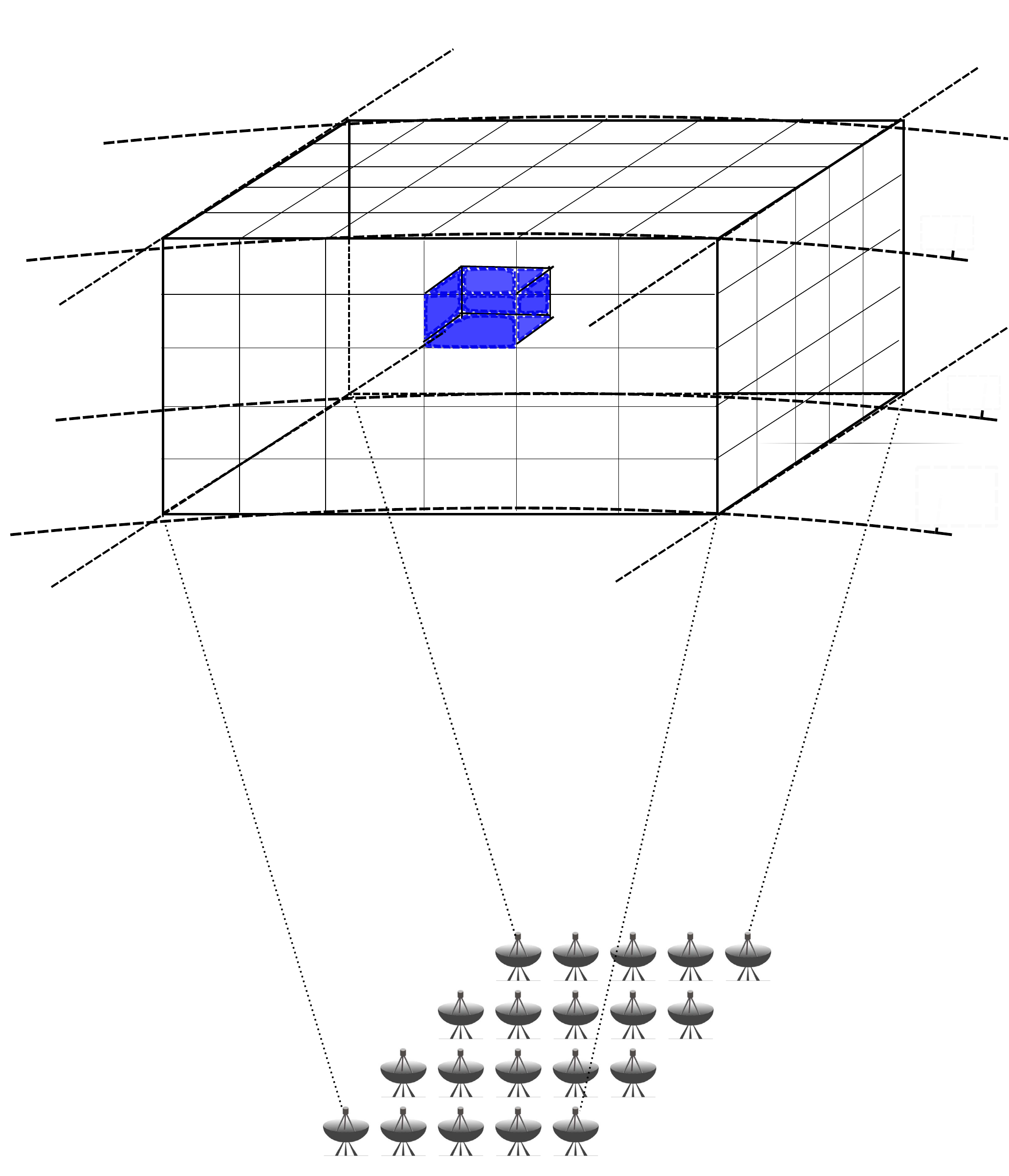}
  \caption{Tomographic measurements by Fast Fourier Transform Telescopes (FFTTs) would allow us to measure
  the remote quadrupole of the CMB $a_{2m}(z)$ ($m=1,2$)  in volume pixels (``voxels") of volume ${\cal V}_{\rm c}$
  in narrow slice of redshift space. Creating a map of remote quadrupole moments across many voxels allows us to construct a
   a spin-weight $m$ field, which can be decomposed into $E$ and $B$ modes. Measurement of the $B$-modes of this field
   allows us to put bounds on the tensor-to-scalar ratio $r$.}
  \label{fig:cartoon}
\end{figure}

The measurement of the new circular polarization power-spectra
can allow us to measure the {\it remote quadrupole}
of the CMB, in a given volume-pixel \changetext{(``voxel'')} in the sky, at a high redshift ($z>10$). 
For a wide-angle, tomographic 21 cm survey, we can measure
the remote quadrupole of the CMB in many voxels in the sky,
allowing us to construct a 3D remote quadrupole field at high redshifts.
The 3D remote CMB quadrupole field in turn can be decomposed into
$E$ and $B$ modes in analogy with the decomposition of the CMB
polarization field~\cite{1997PhRvD..55.7368K, 1997PhRvL..78.2058K, 1997PhRvL..78.2054S, 1997PhRvD..55.1830Z, 1997ApJ...482....6S}. We show that the power-spectra of the ``B-modes"
of the remote quadrupole field can be used to measure the
tensor-to-scalar ratio $r$. A schematic of our method is shown 
in Fig.~\ref{fig:cartoon}. 

One way of thinking about at our method is to imagine  neutral hydrogen in all the voxels in Fig.~\ref{fig:cartoon} to be independent CMB-quadrupole detectors. The construction of the new  remote quadrupole field during the dark ages allows for the statistical measurement of the $E$ and  $B$ modes which in turn contains information about primordial tensor modes (i.e. gravitational waves). Our method is similar to the one proposed in Ref.~\cite{1997PhRvD..56.4511K}, but the authors suggest the use  of discrete clusters to reconstruct the CMB quadrupole moments at their locations. Our method, in principle, allows for construction of a continuous field of remote quadrupole moments, and probes higher redshifts than those accessible to the cluster method (see also Refs.~\cite{2003PhRvD..67f3505C,
2004ApJ...612...81D,2004astro.ph..2474S,2004PhRvD..70f3504P,2006PhRvD..73l3517B,2012PhRvD..85l3540A}).

\changetext{Finally, we note that there are in fact two stages of statistical inference in our proposed method. In the first stage (Sec.~\ref{s:remotequad}), one uses the 21 cm fluctuations in a given voxel to estimate the CMB quadrupole $a_{2m}$ at the position of that voxel. In this stage, the short-wavelength density perturbations are random variables, and the CMB quadrupole is an unknown constant in each voxel whose value we are trying to determine. In the second stage of statistical inference (Sec.~\ref{sec:sensitivity}), the $a_{2m}$ are themselves random variables, and from their measured values we are trying to infer $r$. Such two-stage chains of inference are common in cosmology; for example, in a weak lensing experiment, we would have a first stage where we take galaxy images and infer the lensing shear (assumed constant over the size of a galaxy image), and then a second stage where the shear is itself a random variable whose power spectrum carries cosmological information.}

\changetext{The tensor-to-scalar ratio $r$ appears in the power spectrum of the remote quadrupole moments. That is, the tensor power spectrum $\Delta_h^2(k)$ or the tensor-to-scalar ratio $r$ is quadratic in the estimators for $a_{2m}$ (we will see this in Eq.~\ref{eq:Bpower}), which themselves are constructed from the local $TV$ power spectrum. Thus, one could in principle think of the estimator for $r$ as being constructed from the $TVTV$ trispectrum (just as one treats CMB lensing estimators as being constructed from the trispectrum \cite{2001PhRvD..64h3005H}). However, given the separation of scales, the two-stage ``power spectrum of a local power spectrum'' approach in this paper seems more intuitive and closer to the physics. We also expect that an eventual analysis of 21 cm data would use the two-stage approach (at least in one branch of the analysis), since the computational techniques and understanding of systematics for power spectra are so much more advanced than for trispectra.}

\section{Measuring the Remote Quadrupole of the CMB}
\label{s:remotequad}

In this section we compute the sensitivity of future tomographic 21 cm surveys to measure the remote quadrupole of the CMB at high redshifts. We begin by reviewing some basic notation relevant to remote CMB quadrupole measurements. The experimental setup ideal for this measurement is the Fast Fourier Transform Telescope (FFTT) setup, due to its excellent surface brightness sensitivity compared to sparsely sampled arrays \cite{2009PhRvD..79h3530T}. We review the FFTT setup and make Fisher matrix forecasts for the measurement of the remote quadrupole for different FFTT configurations.
\changetext{In this section, the CMB quadrupole $a_{2m}$ is simply assumed to be constant over each voxel; our objective is to determine the uncertainty on $a_{2m}$ for a given FFTT configuration and observing time. In Sec.~\ref{sec:sensitivity}, we will promote $a_{2m}$ to a random variable and use our measurements of $a_{2m}$ to constrain $r$.}

\subsection{Relation of the 21 cm power spectrum to the remote quadrupole of the CMB}
\label{ss:remote_quadrupole}

The central idea of our technique is that the circular polarization of the emitted 21cm radiation from the high-redshift hydrogen cloud depends on the remote quadrupole of the CMB at that redshift depends on through Eq.~(\ref{eq:Vobs}). Specifically, the existence of a CMB quadrupole at some position results in the creation of new power spectra involving the circular polarization that would otherwise be zero. We focus on the temperature-circular polarization cross-power spectrum $TV$, since its signal-to-noise ratio scales with the amplitude of the CMB quadrupole effect (SNR$\,\propto a_{2m}$) instead of the case of the circular polarization auto-power spectrum $VV$ (SNR$\,\propto a_{2m}^2$). \changetext{In a given voxel, the $a_{2m}$ are treated as constant, and give rise to a local cross-power spectrum $P_{TV}({\bm k})$:}
\begin{equation}
P_{TV}(k) = \frac{\partial T_{\rm obs}}{\partial \delta}(k)\, \frac{\partial V_{\rm obs}}{\partial \delta}(k)\, P_\delta(k).
\label{eq:PTV}
\end{equation}
\changetext{Note that this is the {\em local} power spectrum in this voxel, averaged over the ensemble of short-wavelength density perturbations, but with $a_{2m}$ fixed. The CMB quadrupole moments $a_{2m}$ vary on much larger scales than those directly observed with 21 cm arrays (the scale on which $a_{2m}$ varies is of order the horizon scale at that redshift), and so in Eq.~(\ref{eq:PTV}) we have {\em not} averaged over them. If we did average over realizations of $a_{2m}$, then $\langle a_{2m}\rangle=0$ and hence we would have no contribution to $P_{TV}({\bm k})$.}

\changetext{We now turn to the transfer functions in Eq.~(\ref{eq:PTV}).} The temperature perturbation is given by the usual relation,
\begin{equation}
\frac{\partial T_{\rm obs}}{\partial \delta} = 37.3\,{\rm mK} \left(\frac{1+z}{20}\right)^{1/2} \left( 1 - \frac{T_\gamma}{T_s} \right)[1 + (\hat{\bm k}\cdot\hat{\bm n})^2].
\end{equation}
From Eq.~(\ref{eq:Vobs}) we can see that the circular polarization transfer function $\partial V_{\rm obs}/\partial\delta$  is given by
\begin{eqnarray}
\frac{\partial V_{\rm obs}}{\partial \delta} &=& -8.6\,{\rm mK} \left(\frac{1+z}{20}\right)^{2} \frac{T_\gamma}{T_s}\left( 1 - \frac{T_\gamma}{T_s} \right)
\nonumber \\
&& \times \frac{1}{(1+\coefalpha\tilde x_\alpha)(1+\tilde x_c+\tilde x_\alpha)}
\nonumber \\ && \times 
\Im [a_{21} Y_{21}(\hat{\bm k})
+2a_{22} Y_{22}(\hat{\bm k})
].
\end{eqnarray}
The circular polarization transfer function depends on the direction of the wavenumber $\hat{\bm k}$.

The \changetext{local} power spectrum $P_{TV}(k)$ is thus sensitive to 4 of the 5 types of quadrupole moments of the CMB. Each of these 4 quadrupole moments leads to a quadrupole dependence of the $TV$ spectrum:
\begin{list}{$\bullet$}{}
\item An $xz$ CMB quadrupole ($\Re a_{21}<0$) leads to a positive $TV$ spectrum for $k_yk_z<0$ and negative for $k_yk_z>0$.
\item A $yz$ CMB quadrupole ($\Im a_{21}>0$) leads to a positive $TV$ spectrum for $k_xk_z>0$ and negative for $k_xk_z<0$.
\item An $x^2-y^2$ CMB quadrupole ($\Re a_{22}>0$) leads to a positive $TV$ spectrum for $k_xk_y<0$ and negative for $k_xk_y>0$.
\item An $xy$ CMB quadrupole ($\Im a_{22}<0$) leads to a positive $TV$ spectrum for $k_x^2-k_y^2>0$ and negative for $k_x^2-k_y^2<0$.
\item The observable $P_{TV}(k)$ is not sensitive to the $m=0$ CMB quadrupole mode that is symmetric around the line of sight.
\end{list}

\subsection{Local power spectrum and detectability}
\label{ss:subfisher}
In this section we evaluate sensitivity 
of future tomographic 21 cm surveys to the 
remote quadrupole of the CMB.
The ability to measure the remote CMB quadrupole can be determined using the Fisher formalism: 
in a region of comoving volume ${\cal V}_{\rm c}$, we have
\begin{equation}
F_{\mu\nu} = \int \frac{d^3\bm k}{(2\pi)^3} \frac{{\cal V}_{\rm c} [\partial P_{TV}({\bm k})/\partial p_\mu][\partial P_{TV}({\bm k})/\partial p_\nu]}{[P_{TT}({\bm k}) + N_{TT}({\bm k})][P_{VV}({\bm k})+N_{VV}({\bm k})]},
\label{eq:3fisher}
\end{equation}
where ${\cal V}_{\rm c}$ is the comoving volume and $p_\mu$ are the parameters -- in this case the 4 measurable quadrupole components: $\Re a_{21}$, $\Im a_{21}$, $\Re a_{22}$, and $\Im a_{22}$. Here $N_{TT}({\bm k})$ is the temperature noise power spectrum, and $N_{VV}({\bm k})$ is the circular polarization noise power spectrum. For a dual-polarization interferometer with the same noise temperature in both polarizations, $N_{VV}({\bm k}) = N_{TT}({\bm k})$. We discuss the 
noise power spectrum in Sec.~\ref{ss:FFTT}. 
Under the further assumption of noise power spectra that are symmetric around the line of sight (which occurs when the distribution of baselines is nearly circularly symmetric), the Fisher matrix reduces to
\begin{equation}
F_{\mu\nu} = {\cal V}_{\rm c} \left( \begin{array}{cccc}
w_1 & 0 & 0 & 0 \\ 0 & w_1 & 0 & 0 \\ 0 & 0 & w_2 & 0 \\ 0 & 0 & 0 & w_2
\end{array} \right).
\end{equation}
That is, there is an inverse variance per component of $w_m$ (units: Mpc$^{-3}$) per unit comoving volume, which may be different for the $m=1$ and $m=2$ quadrupole components. The Fisher estimate of the variance in $\Re a_{2m}$ or $\Im a_{2m}$ is $1/(w_m{\cal V}_{\rm c})$. Two diagonal elements of Eq.~(\ref{eq:3fisher}) suffice to determine $w_1$ and $w_2$.

\subsection{Fast Fourier Transform Telescopes}
\label{ss:FFTT}
The ideal experimental setup for measuring the remote quadrupole of the CMB 
using the circular polarization of 21 cm is the proposed 
Fast Fourier Transform Telescope (FFTT) as described in~\cite{2009PhRvD..79h3530T}.
The FFTT consists of a tightly packed 
array of simple dipole antennas in a regular rectangular grid. 
The electric field is digitized at the antennae and 
subsequent correlations and Fourier transforms are done 
digitally.  The FFTT is based on the simple idea that if the antennae are 
arranged on a rectangular grid, Fast Fourier Transforms
can be used to scale the cost as $N\mathrm{log}_2N$ instead of $N^2$
(where $N$ is the number of antennae).
The FFTT concept allows for mapping of a very wide field of view
with very high sensitivity, making it ideal for 
21 cm tomography experiments.

A schematic of the FFTT design we consider for the forecasts in this paper 
is shown in Fig.~\ref{fig:cartoon}. We consider a square
array design with a compact grid of dipole antennas with side length $L$, 
effective area $L^2$,  that observes for a time $\tau$ with a bandwidth $\delta\nu$  around
some frequency $\nu$.  In principle the array can observe the entire visible sky at any given time.
The figure shows how we split the 3D volume of the universe at high redshifts
observed by the array into smaller volume pixels (``voxels"). Our goal
is to estimate the detectability of the remote CMB quadrupole in each of these voxels.

The experiment is characterized by three key parameters: the length of the array $L$,
the time of observation $\tau$ and the system temperature $T_{\mathrm{sys}}$. The noise power
spectrum per mode ${\bm k}$ (in intensity units) is given by
\be
N_{TT}({\bm k})= \f{\lambda^4 c (1+z)^2 D_{M}^2(z)}{\Omega_{\mathrm{beam}}\tau H(z)\nu_{21}}  \f{T_{\mathrm{sys}}^2}{A_{e}^2 n_{\mathrm{base}}({\bm k})},
\label{eq:noise}
\ee
where $D_{M}(z)$ is the comoving distance to the redshift $z$, $A_{e}$ is  the collecting area, and $n_{\mathrm{base}}({\bm k})$ is the number density of baselines 
that observe a
given mode ${\bm k}$ at a given time. Here noise is reported in temperature units, $T$ in K and $N_{TT}({\bm k})$ in K$^2$ Mpc$^3$ \cite{2017PhRvD..95h3011G}.

A given mode in the sky ${\bm k}$, will be observed by many baselines of the 
FFTT during an observation campaign. Hence the noise power spectrum needs to be
weighed by the number of baselines observing a given mode ${\bm k}$. The number of baselines 
observing a mode $\vec{k}$ is given by
\begin{eqnarray}
\langle n_{\mathrm{base}}({\bm k})\rangle = \lf\f{L}{\lambda}\rt^2 -\f{4}{\pi}\f{L}{\lambda}\f{D_{M}(z)}{2\pi}k \sin\theta_{k} 
\nonumber \\
+\f{1}{\pi} \left[ \f{D_{M}(z)}{2\pi}k \sin\theta_{k} \right]^2,
\end{eqnarray}
where $\theta_{k}$ is the polar angle and $\phi_k$ the longitude in a coordinate system where the line of sight is along the $z$ axis. The number of baselines is averaged over $\phi_k$, which is appropriate if at least $\sim 90^\circ$ of Earth rotation occurs over the course of an observing window.

\begin{figure}
\includegraphics[trim=0cm 0cm 0cm 0cm, clip=true, width=0.5\textwidth]{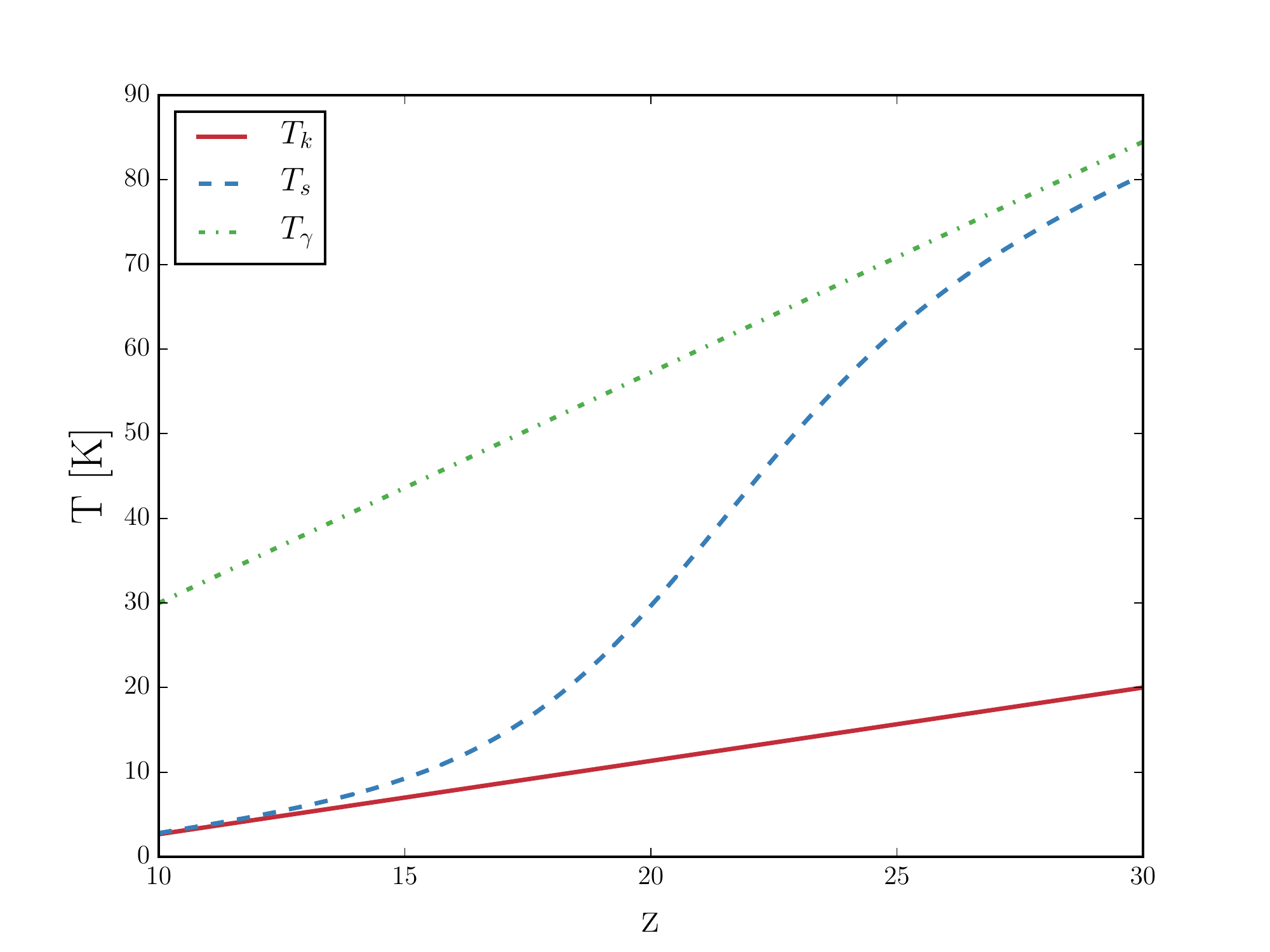}
\caption{Inputs used for the sensitivity calculation, computed
for standard cosmology using the 21CMFAST code. The plot shows the fiducial models for spin, kinetic,
and CMB temperatures.}
\label{fig:temp}
\end{figure}

\begin{figure}
\includegraphics[trim=0cm 0cm 0cm 0cm, clip=true, width=0.5\textwidth]{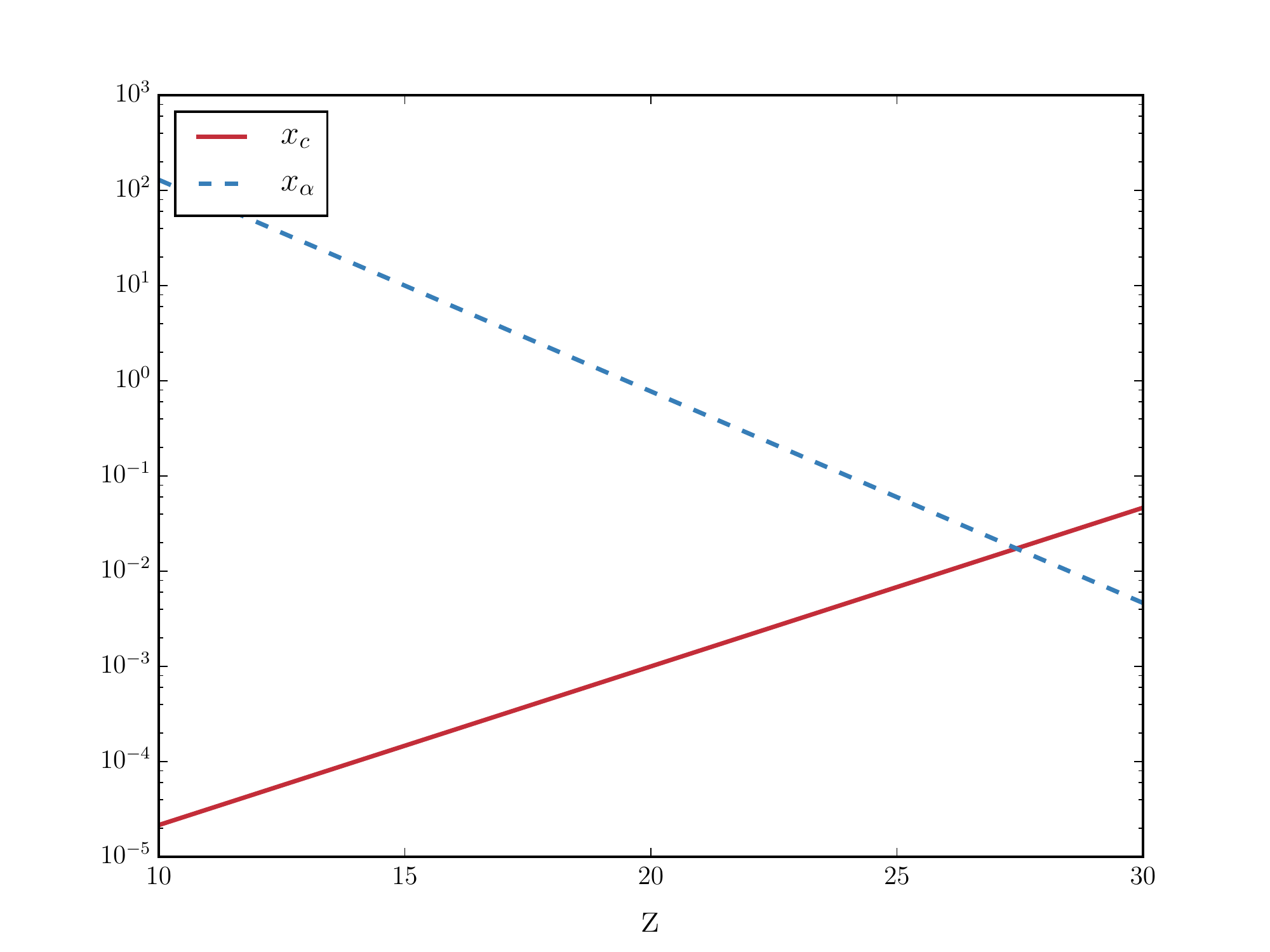}
\caption{Inputs used for the sensitivity calculation, computed
for standard cosmology using the 21CMFAST code. The plot shows the fiducial models for
quantities that parametrize the rate of depolarization of the
ground state by optical pumping and atomic collisions as discussed in the text and in~\cite{2017PhRvD..95h3011G}}.
\label{fig:reion}
\end{figure}

\subsection{Results for reference experiments}
\label{ss:refexperiments}

\begin{figure}
\includegraphics[trim=0cm 0cm 0cm 0cm, clip=true, width=0.5\textwidth]{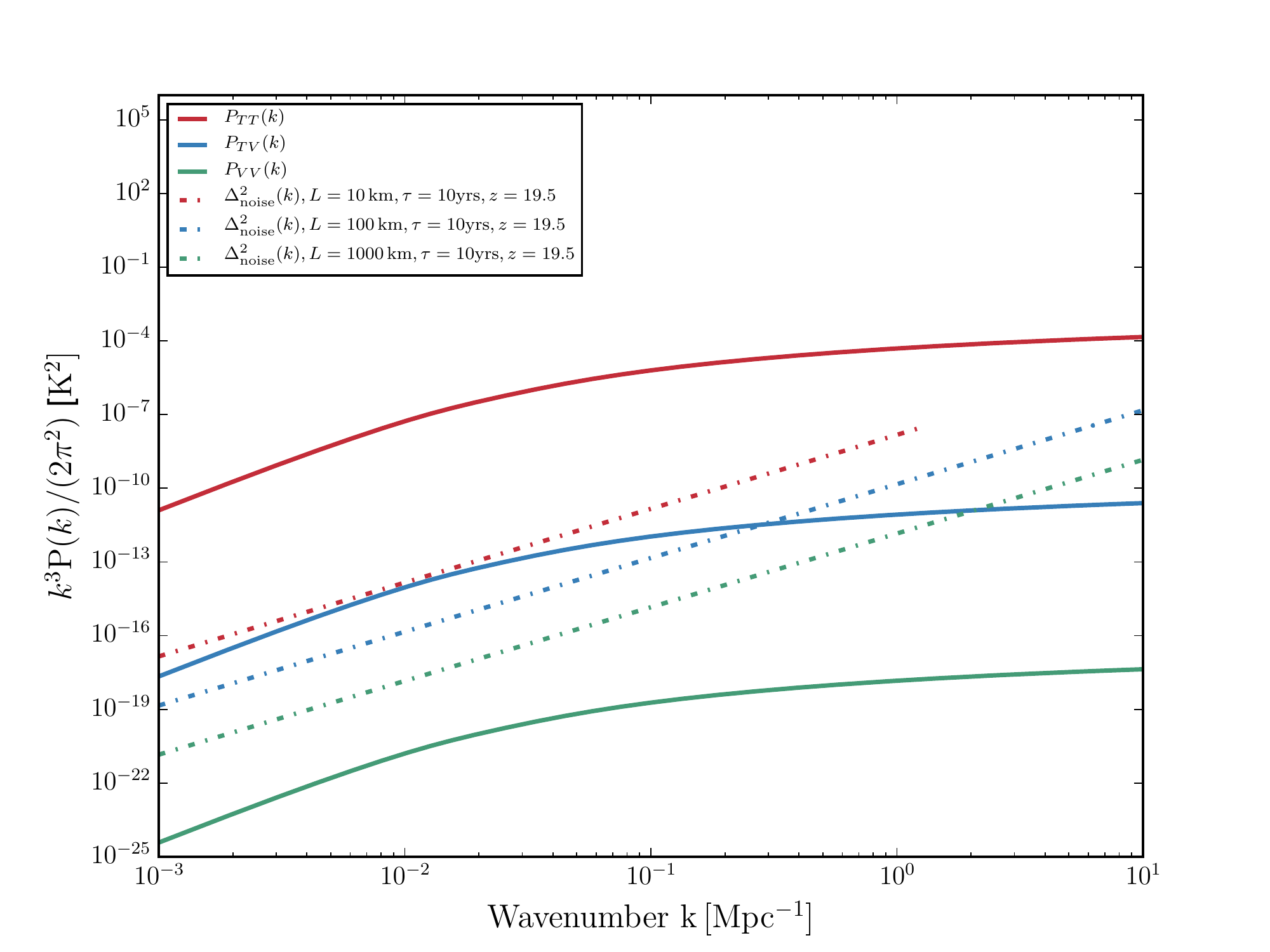}
\caption{Temperature, circular polarization, and noise power spectra relevant to the Fisher calculation. We compute
the power-spectra for observations corresponding to $z=19.5$. Noise power-spectra for two different configurations of
FFTTs are shown.}
\label{fig:powerspectra}
\end{figure}

We now proceed to evaluate the sensitivity of a 
tomographic 21 cm survey to measure the remote
quadrupole of the CMB during the pre-reionization epoch, at a given redshift $z$ and for a ``voxel" of volume
${\cal V}_{\rm c}$. Specifically, we compute the elements of 
the Fisher matrix (Eq.~\ref{eq:3fisher}), for different FFTT configurations and observation times. 

We consider a square-grid configuration for the FFTT with a length
$L$ and collecting area $A_{e} = L^2$. The time $\tau$ for computing the
noise spectra in Eq.~(\ref{eq:noise}) is the {\em observing time}, which is smaller than the wall-clock time since a given portion of the sky is visible
for only part of the day. 
We assume a system temperature of $T_{\mathrm{sys}}=1000$K.

Other inputs to the Fisher matrix computation includes the spin temperature $T_s$,
the kinetic temperature $T_k$ of the IGM, and the CMB temperature $T_{\gamma}$
as a function of redshift (Fig.~\ref{fig:temp}). We also compute quantities that
parametrize the rate of depolarization of the ground state
by optical pumping $x_\alpha$ and atomic collisions $x_c$ (Fig.~\ref{fig:reion}). The quantities are calculated 
using the {\sc 21cmFast} code~\cite{2011MNRAS.411..955M}. 
For the {\sc 21cmFast} runs, we set the sources responsible for early heating to Population III stars and use a  
star formation efficiency \verb|F_STAR|=$0.0075$. 
For more details about the parameters used 
for the {\sc 21cmFast} outputs see Ref.~\cite{2017PhRvD..95h3011G}. 
We use standard cosmological parameters ($H_0 = 67 \, \mathrm{km}\,\mathrm{s}^{-1}\,\mathrm{Mpc}^{-1}, 
\Omega_{\mathrm{m}} =0.32, \Omega_{\mathrm{K}} = 0, n_s = 0.96, \sigma_{8} = 0.83, w = -1$)
consistent with Planck measurements~\cite{2016A&A...594A..13P}.

A FFTT can in principle observe the entire sky
above the horizon. However, the image degrades
rapidly near the horizon and the useful field of view
is about half $\Omega \sim \pi$. The angular resolution 
of a FFTT is $\theta_{\mathrm{res}} \sim \lambda/ \sqrt{A}$. 
The angular scale of the 
``voxel" in which the CMB quadrupole is measured
to be approximately ten times the angular resolution of the telescope.
The maximum comoving wavenumber probed by the FFTT ($k_{\rm max}$) is given by
\be
 k_{\rm max}= \frac{2\pi}{d_{A}(z)\,\theta_{\mathrm{res}}}
 .
\ee
Note that every super-pixel can be observed simultaneously and
so $\tau$ for a super-pixel is the total time that the FFTT observes
a given patch of the sky. The Fisher integral takes place over a super-pixel and
we take $k_{\rm max}$ corresponding to the angular resolution of
the survey. The minimum wavenumber probed is taken to be several orders
of magnitude smaller than $k_{\rm max}$ (the Fisher integral is not 
sensitive to the choice of $k_{\rm min}$). 
\begin{figure}
\includegraphics[trim=0cm 0cm 0cm 0cm, clip=true, width=0.5\textwidth]{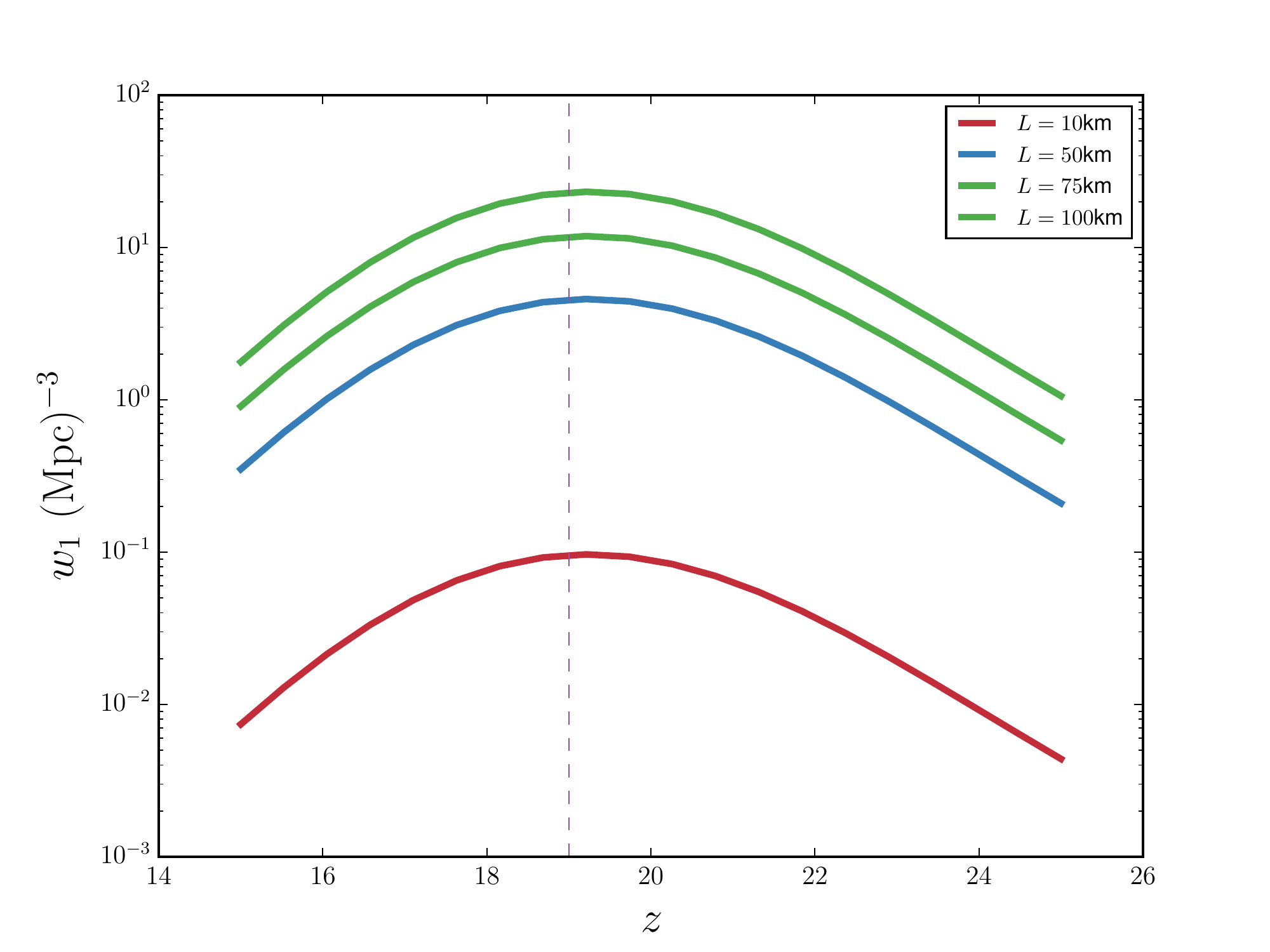}
\includegraphics[trim=0cm 0cm 0cm 0cm, clip=true, width=0.5\textwidth]{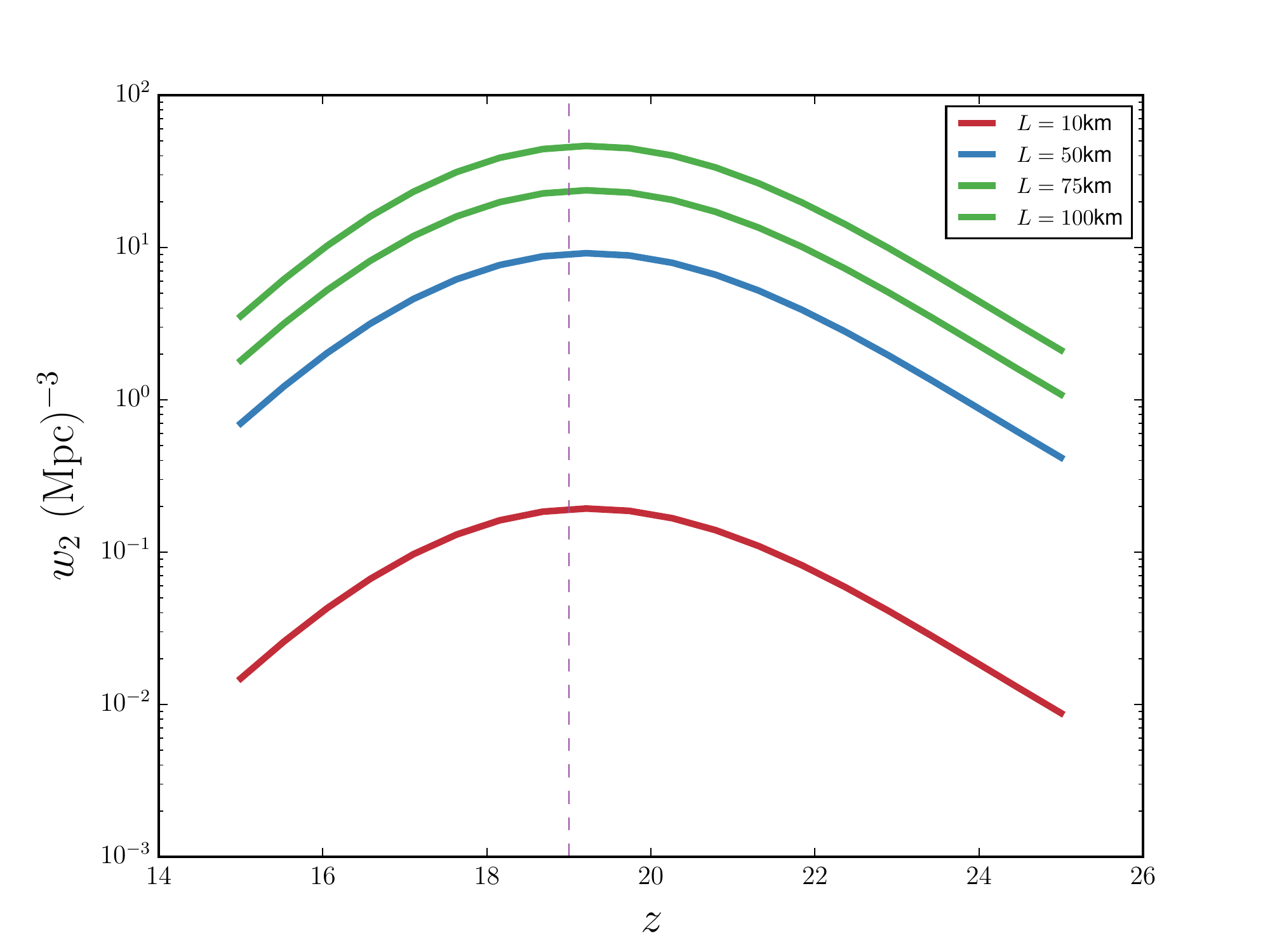}
\caption{Elements of the Fisher matrix $w_1$ and $w_2$  as a function of redshift $z$, computed for
a model of reionization described in the text. For our fiducial model, both $w_1$ and $w_2$ peak at $z=19.5$, i.e.\ the redshift where the Lyman-alpha coupling becomes efficient ($\tilde x_\alpha\sim 1$).}
\label{fig:w_z}
\end{figure}

To estimate the Fisher integral we plot the relevant power spectra
in Eq.~(\ref{eq:3fisher}), including the noise power spectrum for different
configurations of the FFTT in Fig.~\ref{fig:powerspectra}. From the figure
we note that $P_{TT}(k)\gg N_{TT}(k)$
and $N_{VV}(k)\gg P_{VV}(k)$.  The Fisher integral in Eq.~(\ref{eq:3fisher}) can then be approximated to give
\begin{eqnarray}
w_1 &=& \f{1}{(2\pi)^3} \f{ \left(8.6\,{\rm mK} \left(\frac{1+z}{20}\right)^2 \left( 1 - \frac{T_\gamma}{T_s} \right) \left(\frac{T_\gamma}{T_s} \right)\right)^2}{N_{VV}(1+\coefalpha\tilde x_\alpha)^2(1+\tilde x_c+\tilde x_\alpha)^2} \nonumber \\ && \times
 \int^{k_{\rm max}}_{k_{\rm min}} d^{3}k  (\Im (Y_{21}(\theta,\phi)))^2 P_{\delta}(k)
\end{eqnarray} 
and 
\begin{eqnarray}
w_2 &=& \f{2}{(2\pi)^3} \f{ \left(8.6\,{\rm mK} \left(\frac{1+z}{20}\right)^2 \left( 1 - \frac{T_\gamma}{T_s} \right) \left(\frac{T_\gamma}{T_s} \right)\right)^2}{N_{VV}(1+\coefalpha\tilde x_\alpha)^2(1+\tilde x_c+\tilde x_\alpha)^2} \nonumber \\ && \times 
 \int^{k_{\rm max}}_{\rm k_{min}} d^{3}k  (\Im (Y_{22}(\theta,\phi)))^2 P_{\delta}(k).
\end{eqnarray} 

The value of $w_1$ and $w_2$ is a function of redshift and depends on the reionization and spin-excitation history of the universe during the pre-reionization era. In particular it is sensitive to the Lyman-$\alpha$ flux during this epoch which is unconstrained by observations. We use the fiducial model shown in Fig.~\ref{fig:reion} and described in Ref.~\cite{2017PhRvD..95h3011G}. As seen from the figure, for our fiducial model, $w_1$ and $w_2$ peak around $z=19.5$ and our technique is most sensitive in this redshift range. Note that this is likely to change when the Lyman-$\alpha$ flux in the pre-reionization era becomes better constrained.

\section{Power spectrum of the remote CMB quadrupole and sensitivity to the tensor-to-scalar ratio}
\label{sec:sensitivity}

We now consider how well we can measure the tensor-to-scalar ratio using remote quadrupole measurements. This requires us to consider the remote quadrupole moments \changetext{$a_{2m}$} as a statistical field, compute their power spectrum, compare this to the noise computed in \changetext{Sec.}~\ref{ss:subfisher}, and finally perform the Fisher matrix sum over modes.

\subsection{$E$- and $B$-mode decomposition of remote CMB quadrupoles}

The ``derived data product'' from the analysis of \S\ref{ss:subfisher} is a map of the CMB quadrupole moments $a_{2q}$ ($q\neq 0$: we use $q$ here instead of $m$ to avoid confusion below) in each super-pixel of comoving volume ${\cal V}_{\rm c}$. These moments are measured with respect to the local coordinate basis vectors $\{\hat{\bm e}_\theta, \hat{\bm e}_\phi, \hat{\bm n}=\hat{\bm e}_r\}$. This quadrupole is derived from the local power spectrum $P_{TV}({\bm k})$ in each super-pixel. Viewed from the perspective of the observer, $a_{2q}^\ast$ is a spin-weight $q$ field. In analogy to the decomposition of the CMB polarization field \cite{1997PhRvD..55.1830Z}, we may perform a spin-weighted spherical harmonic transformation:
\begin{equation}
a_{2q}^\ast(\chi,\hat{\bm n}) = \sum_{\ell=|q|}^\infty \sum_{m=-\ell}^\ell 
b_{q\ell m}(\chi) \,{_q}Y_{\ell m}(\hat{\bm n}).
\end{equation}
The dependence on comoving distance $\chi$ is retained since we do not decompose the radial direction into eigenmodes. The symmetry property $a_{2,-q}(\chi,\hat{\bm n}) = (-1)^q a_{2q}^\ast(\chi,\hat{\bm n})$ implies that
\begin{equation}
b_{q\ell m}^\ast(\chi) = (-1)^m b_{-q,\ell,-m}(\chi).
\end{equation}
Furthermore, parity inversion results in the transformation $a_{2q}(\chi,\hat{\bm n}) \rightarrow a_{2,-q}(\chi,-\hat{\bm n})$, or equivalently $b_{qlm}(\chi)\rightarrow (-1)^{l} b_{-q,lm}(\chi)$. We now define the electric and magnetic-parity versions of these quadrupole moments: for $q>0$,
\begin{eqnarray}
b^{E,q}_{\ell m}(\chi) &=& \frac12[ b_{qlm}(\chi) +  b_{-q,lm}(\chi) ]
{\rm ~~~and} \nonumber \\
b^{B,q}_{\ell m}(\chi) &=& \frac1{2i}[ b_{qlm}(\chi) - b_{-q,lm}(\chi) ].
\label{eq:EB}
\end{eqnarray}
These moments obey the same complex conjugation properties as usual electric and magnetic moments, i.e. 
$b^{E,q\ast}_{\ell m}(\chi) = (-1)^m b^{E,q}_{\ell,-m}(\chi)$ and
$b^{B,q\ast}_{\ell m}(\chi) = (-1)^m b^{B,q}_{\ell,-m}(\chi)$. Under parity inversion, $b^{E,q}_{\ell m}(\chi)$ picks up a factor of $(-1)^\ell$, whereas $b^{B,q}_{\ell m}(\chi)$ picks up a factor of $-(-1)^\ell$.

The statistics of the CMB quadrupole moment fields can thus be described in terms of the cross-power spectra of these fields at the various comoving distances, e.g.
\begin{equation}
C_\ell^{B1,B2}(\chi,\chi') = \langle b^{B1\ast}_{\ell m}(\chi) b^{B2}_{\ell m}(\chi') \rangle.
\end{equation}
Parity considerations imply a vanishing cross-spectrum between the $E,q$ and $B,q'$ moments. Furthermore, there is no primordial scalar contribution to the $B1$ or $B2$ moments.

\subsection{$B$-mode power spectrum of the remote CMB quadrupole}

We compute the power spectrum of the remote CMB quadrupole by the standard method -- that is, we consider first a single Fourier mode (a plane primordial gravitational wave) with wave vector along the $z$-axis, then rotate it to an arbitrary angle, and finally perform a stochastic average using the power spectrum in the initial conditions.

Consider a gravitational wave with wave number $K$ and strain $h_{\rm R}$ propagating in the $z$-direction and with right-circular polarization, i.e. with metric
\begin{equation}
g_{\mu\nu} = a^2 \left( \begin{array}{cccc} -1 & 0 & 0 & 0 \\ 0 & 1+\frac1{\sqrt2}h_{\rm R}e^{iKx^3} & \frac1{\sqrt2}ih_{\rm R}e^{iKx^3} & 0 \\
0 & \frac1{\sqrt2}ih_{\rm R}e^{iKx^3} & 1-\frac1{\sqrt2}h_{\rm R} e^{iKx^3} & 0 \\ 0 & 0 & 0 & 1 \end{array} \right).
\end{equation}
The normalization is chosen to coincide with the common normalization of tensor perturbations (e.g.\ \cite{2009arXiv0907.5424B}) with $r = \Delta_h^2(k)/\Delta_\zeta^2(k) = 16\epsilon$ in slow-roll single-field inflation. This plane gravitational wave leads to a tensor $\ell=2$ CMB multipole moment
\begin{equation}
\frac{\Delta T({\bm r},\hat{\bm p},\eta)}{\bar T} = h_{\rm R0} e^{iKx^3} \sum_{\ell=2}^\infty (-i)^\ell
\sqrt{\frac{4\pi}{2\ell+1}} \Theta^{\rm T}_\ell(\eta) Y_{\ell 2}(\hat{\bm p}), 
\end{equation}
at position ${\bm r}$, for photons traveling in direction $\hat{\bm p}$, and at conformal time $\eta$ defined as,
\be
\eta(z) =\int^{t(z)}_{0} \f{dt}{a}
\ee
$\Theta^{\rm T}_\ell$ are the tensor multipole moments generated by a unit-amplitude gravitational wave and $h_{\rm R0}$ is the initial amplitude. Rotational symmetry guarantees that only $m=2$ terms exist in the sum over spherical harmonics. The $\ell = 2$ multipole moments measured at some position on the sky and some comoving distance $\chi(z)$ are then
\begin{eqnarray}
a_{2m}(\chi,\hat{\bm n}) &=& -\sqrt{\frac{4\pi}{5}}
h_{\rm R0} e^{iK\chi\cos\theta} \Theta_2^{\rm T}(\eta_0-\chi)
\nonumber \\ && \times
[{\cal D}_2(\phi,\theta,0)]_{m,2},
\label{eqn:quadrupole}
\end{eqnarray}
where ${\cal D}_2(\phi,\theta,0)$ is the passive rotation matrix associated with rotating the reference frame for the multipoles from $\{\hat{\bm e}_1,\hat{\bm e}_2,\hat{\bm e}_3\}$ to $\{\hat{\bm e}_\theta,\hat{\bm e}_\phi,\hat{\bm n}\}$.

The $\ell = 2$ multipole moments from Eqn.(\ref{eqn:quadrupole}) can be re-written in terms of the spin-weighted spherical harmonics,
\begin{equation}
a_{2q}^\ast(\chi,\hat{\bm n}) = -\frac{4\pi}{5}
h_{\rm R0} e^{-iK\chi\cos\theta} \Theta_2^{{\rm T}\ast}(\eta_0-\chi) \,_qY_{2,-2}(\hat{\bm n}).
\end{equation}
The solution for $b_{q\ell m}(\chi)$ can then be written as
\begin{widetext}
\begin{eqnarray}
b_{q\ell m}(\chi) \!\! &=& \!\! \int a_{2q}^\ast(\chi,{\bm n}) \,_qY_{\ell m}^\ast(\hat{\bm n})\,d^2\hat{\bm n}
\nonumber \\
&=& \!\! -\frac{4\pi}{5}
h_{\rm R0} \Theta_2^{{\rm T}\ast}(\eta_0-\chi)
\int e^{-iK\chi\cos\theta} \,_qY_{2,-2}(\hat{\bm n})
\,_qY_{\ell m}^\ast(\hat{\bm n})\,d^2\hat{\bm n}
\nonumber \\
&=& \!\! -\frac{(4\pi)^{3/2}}{5}
h_{\rm R0} \Theta_2^{{\rm T}\ast}(\eta_0-\chi) \delta_{m,-2} \sum_{\ell'=0}^\infty \sqrt{2\ell'+1}
(-i)^{\ell'}\,j_{\ell'}(K\chi) \int\,_0Y_{\ell'0}(\hat{\bm n})
\,_qY_{2,-2}(\hat{\bm n})
\,_qY_{\ell,-2}^\ast(\hat{\bm n})\,d^2\hat{\bm n}
\nonumber \\
&=& \!\! -4\pi \sqrt{\frac{2\ell+1}5}(-1)^q
h_{\rm R0} \Theta_2^{{\rm T}\ast}(\eta_0-\chi) \delta_{m,-2} \sum_{\ell'=0}^\infty (2\ell'+1)
(-i)^{\ell'}\,j_{\ell'}(K\chi)
\threej{\ell'}{2}{\ell}{0}{q}{-q} \threej{\ell'}{2}{\ell}{0}{-2}{2}
.\end{eqnarray}
Under the transformation $q\leftrightarrow-q$, this changes sign if $\ell'-\ell$ is odd and remains the same if $\ell'-\ell$ is even; thus for the $B$-mode, only the $\ell'-\ell$ odd terms contribute (see Eq.~\ref{eq:EB}). The triangle inequality restricts $|\ell'-\ell|\le2$, so the sum then reduces to $\ell'=\ell\pm 1$. Substituting in the Wigner $3j$ symbols yields for $q>0$:
\begin{equation}
b^B_{q\ell m}(\chi)
= -\frac{2\pi\, (-i)^\ell \delta_{m,-2} }{\sqrt{5(2\ell+1)}}
h_{\rm R0} \Theta_2^{{\rm T}\ast}(\eta_0-\chi) 
\Bigl[
(-1)^q \sqrt{(\ell+2)(\ell+\bar q)} \,j_{\ell-1}(K\chi)
- \sqrt{(\ell-1)(\ell+1-\bar q)} \,j_{\ell+1}(K\chi) \Bigr]
,\end{equation}
\end{widetext}
where we have defined $\bar q \equiv (-1)^qq$. Use of the rules for combining spherical Bessel functions \cite{1972hmfw.book.....A} allows the further simplifications:
\begin{equation}
b^B_{q\ell m}(\chi) = -2\pi \frac{\delta_{m,-2}}{i^\ell} \sqrt{\frac{2\ell+1}5} h_{\rm R0} \Theta_2^{{\rm T}\ast}(\eta_0-\chi) f_{q\ell}(K\chi),
\label{eq:Btrans}
\end{equation}
where we have defined the functions
\begin{equation}
f_{1\ell}(x) = \sqrt{(\ell-1)(\ell+2)} \,\frac{j_\ell(x)}{x}
\end{equation}
and
\begin{equation}
f_{2\ell}(x) = j'_\ell(x) + 2\frac{j_\ell(x)}{x}.
\end{equation}
These functions are always real, and we have $f_{11}(x)=0$.

It remains to express the $B$-mode power spectrum of the remote quadrupole components. This requires us to obtain the product of two $b^B_{q\ell m}(\chi)$s and average over the direction of the plane wave; this is equivalent to summing over $m$ and dividing by $2\ell+1$. Thus for a plane wave in a random direction, we find
\begin{eqnarray}
C^{Bq,Bq'}_\ell(\chi,\chi')
\!\! &=& \!\!
\frac1{2\ell+1}\sum_{m=-\ell}^\ell \langle b^{B\ast}_{q\ell m}(\chi) b^B_{q'\ell m}(\chi') \rangle
\nonumber \\
\!\! &=& \!\!
\frac{4\pi^2}5 |h_{\rm R0}|^2 \Theta^{{\rm T}}_2(\eta_0-\chi)\Theta^{{\rm T}\ast}_2(\eta_0-\chi')
\nonumber \\ && \times
f_{q\ell}(K\chi)f_{q'\ell}(K\chi').
\end{eqnarray}
If we finally replace the plane wave with a statistical distribution of gravitational waves, we find
\begin{eqnarray}
C^{Bq,Bq'}_\ell(\chi,\chi')
\!\! &=& \!\!
\frac{8\pi^2}5 \int_0^\infty \Theta^{{\rm T}}_2(\eta_0-\chi)\Theta^{{\rm T}\ast}_2(\eta_0-\chi')
\nonumber \\ && \times
f_{q\ell}(K\chi)f_{q'\ell}(K\chi') \Delta^2_h(K) \frac{dK}K,
\label{eq:Bpower}
\end{eqnarray}
where $\Delta^2_h(K)$ is the contribution to the variance of the strain per logarithmic range of $K$ (i.e. $d{\rm Var} h/d\ln K$) per gravitational wave polarization (right or left). A factor of 2 has been inserted to account for the existence of two gravitational wave polarizations.

Note that although the spherical harmonic decomposition of a spin-1 field admits an $\ell=1$ component, the $q=1$ $B$-mode of the remote quadrupole vanishes -- i.e. $C^{B1,B1}_\ell(\chi,\chi')=0$ -- because $f_{11}(x)=0$. This is mathematically expected because there is no $\ell=1$ gravitational wave mode.

\subsection{Incorporation of the tensor transfer function}

We now also need the tensor transfer function $\Theta^{\rm T}_2(\eta)$. Fortunately, in the matter-dominated era, well after recombination, there is an analytic solution for this. The strain amplitude has the simple time dependence
\begin{equation}
h_{\rm R}(\eta) = h_{\rm R0} \frac{3j_1(K\eta)}{K\eta}.
\end{equation}
The tensor transfer function is then given by evaluating the temperature quadrupole at the origin at time $\eta$ using the line-of-sight expression for the photon temperature perturbation \cite{1996ApJ...469..437S, 1997PhRvD..56..596H}; in what follows, we assume the temperature perturbation due to the gravitational wave is built up from the time of recombination $\eta_\ast$ to the time $\eta$ in question. We work in terms of the real-space temperature perturbation $\Theta(\mu',\phi')$, where $\mu'=\cos\theta'$ is the direction cosine of the photon's trajectory:
\begin{eqnarray}
\Theta^{\rm T}_2(\eta) \!\!\!\!\!\!\!\!\!\!\!\! && \nonumber \\
\!\! &=& \!\! -\frac{5\sqrt6}{16\pi h_{\rm R0}} \int (1-\mu'{^2})e^{-2i\phi'}\Theta(\mu',\phi')\, d\mu'\,d\phi'
\nonumber \\
\!\! &=& \!\! -\frac{5\sqrt6}{16\pi h_{\rm R0}} \int (1-\mu'{^2})e^{-2i\phi'} \Bigl[
\int_{\eta_\ast}^\eta (1-\mu'{^2}) e^{2i\phi'}
\nonumber \\ && \times \frac{-\dot h_{\rm R}(\eta')}{2\sqrt2} e^{iK\mu'(\eta'-\eta)}\,d\eta' 
\Bigr] \, d\mu'\,d\phi'
\nonumber \\
\!\! &=& \!\! \frac{5\sqrt3}{16 h_{\rm R0}}
\int_{\eta_\ast}^\eta \left[ \int_{-1}^1 (1-\mu'{^2})^2 e^{iK\mu'(\eta'-\eta)} d\mu' \right]
 \dot h_{\rm R}(\eta') d\eta'
\nonumber \\
\!\! &=& \!\! \frac{5\sqrt3}{16 h_{\rm R0}}
\int_{\eta_\ast}^\eta \frac{16 j_2(K(\eta-\eta'))}{[K(\eta-\eta')]^2}
 \dot h_{\rm R}(\eta') \,d\eta'
\nonumber \\
\!\! &=& \!\! \frac{5\sqrt3}{16 h_{\rm R0}}
\int_{\eta_\ast}^\eta \frac{16 j_2(K(\eta-\eta'))}{[K(\eta-\eta')]^2}
h_{\rm R0}\frac{-3j_2(K\eta')}{\eta'} \,d\eta'
\nonumber \\
\!\! &=& \!\! -15\sqrt3
\int_{\eta_\ast}^\eta \frac{j_2(K(\eta-\eta')) j_2(K\eta')}{K^2(\eta-\eta')^2\eta'} \,d\eta'
.\label{eq:Th2}
\end{eqnarray}
Equation~(\ref{eq:Th2}) is an integral form for the tensor transfer function; it is straightforward to compute. With the help of Eq.~(\ref{eq:Bpower}), the general remote quadrupole $B$-mode power spectrum for tensors can be obtained.

\subsection{Sensitivity to tensor-to-scalar ratio}

The uncertainty in the tensor-to-scalar ratio can be forecast using Fisher matrix techniques. In general, if there is a Gaussian-distributed data vector ${\bm d}$ with covariance ${\bf C}$, then the Fisher approximation for the uncertainty in the tensor-to-scalar ratio $r$ is
\begin{equation}
\sigma_r^{-2} = \frac12 {\rm Tr}\,\left( {\bf C}^{-1}\frac{\partial\bf C}{\partial r}{\bf C}^{-1}\frac{\partial\bf C}{\partial r} \right).
\label{eq:F1}
\end{equation}
In our case, we will write as the data vector ${\bm d}$ the sequence of $B$-mode moments $b^{Bq}_{\ell m}(\chi)$: up to some $\ell_{\rm max}$, the number of such moments is $N_d = 2N_z(\ell_{\rm max}^2-4)$, where $N_z$ is the number of redshift slices and $\ell_{\rm max}^2-4 = \sum_{\ell=2}^{\ell_{\rm max}} (2\ell+1)$ is the number of multipoles. In harmonic space, for uniform full-sky coverage, ${\bf C}$ is thus an $N_d\times N_d$ matrix that is block-diagonal with $2N_z\times 2N_z$ blocks; the block corresponding to multipole $\ell$ will be denoted ${\bf C}_{(\ell)}$ and is repeated $2\ell+1$ times. We may thus write Eq.~(\ref{eq:F1}) as
\begin{equation}
\sigma_r^{-2} = \frac{f_{\rm deg}^2}2 \sum_{\ell=2}^{\ell_{\rm max}} (2\ell+1) {\rm Tr}\,\left[ {\bf C}_{(\ell)}^{-1}\frac{\partial\bf C_{(\ell)}}{\partial r}{\bf C}_{(\ell)}^{-1}\frac{\partial\bf C_{(\ell)}}{\partial r} \right].
\label{eq:F2}
\end{equation}
Here $f_{\rm deg}$ is a degradation factor due to reduced sky coverage. In CMB forecasts, it is often assumed that the information content scales with the sky coverage $f_{\rm sky}$, in which case $f_{\rm deg} = f_{\rm sky}^{1/2}$. This is only an approximation however \cite{1997ApJ...480...72K} and is generally valid only for sky coverage $\Delta\theta \ge 2\pi/\Delta\ell$, where $\Delta\ell$ is the width of the features in $\ell$-space under consideration. Since the $B$-mode spectrum peaks at the largest scales, this is only marginally true; forecasts for the reionization $B$-mode that evaluate the cut-sky matrix inversion have shown a factor $f_{\rm deg}\sim f_{\rm sky}^2$ for Galactic Plane cuts with $f_{\rm sky}>0.7$ \cite{2005PhRvD..72l3006A}. In this paper, we consider only observations of the full sky minus the Galactic Plane with an assumed $f_{\rm deg}=0.5$, and stress that Eq.~(\ref{eq:F2}) for $\sigma_r$ is uncertain at the factor of $\sim 2$ level even for this case. 

\begin{figure*}
\includegraphics[trim=0cm 0cm 0cm 0cm, clip=true, width=0.9\textwidth]{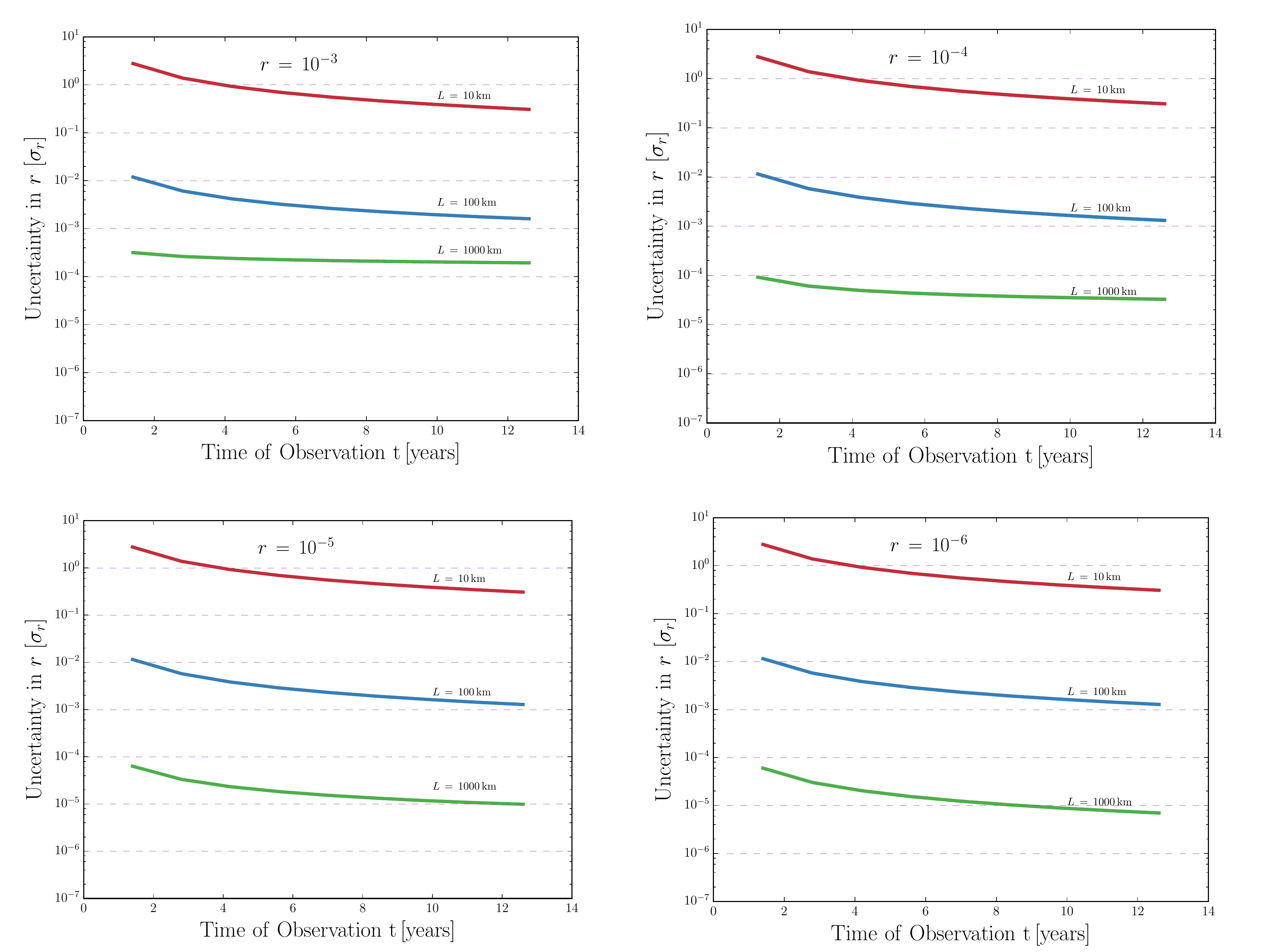}
\caption{Forecasts for $\sigma_{r}$ for different FFTT telescope configurations. The parameters used to make these forecasts are described in Fig.~\ref{fig:reion} \&~\ref{fig:temp} 
and in Section~\ref{ss:refexperiments}. 
For the given Lyman-$\alpha$ flux model the values of weights  $w_1$ and $w_2$ peak around $z\sim 19.5$ as shown in Fig.~\ref{fig:w_z}. For our forecasts we consider a shell with $z_{\rm min}=18$ and $z_{\rm max}=23$. Note that the live observation time quoted here will be shorter than the wall-clock time of the survey.}
\label{fig:sigmar}
\end{figure*}

The matrix ${\bf C}_{(\ell)}$ can be broken up into signal ${\bf S}_{(\ell)}$ and noise ${\bf N}_{(\ell)}$. The noise power spectrum is diagonal in $z$-space:
\begin{equation}
N_{(\ell) q z, q'z'} = \frac{1}{w_q [\chi(z)]^2\Delta\chi}\delta_{qq'} \delta_{zz'},
\end{equation}
where $\Delta\chi = c\Delta z/H(z)$ is the width of the redshift slice and $\chi(z)$ is the comoving distance. (The denominator is the conversion from sr on the sky to Mpc$^3$ of comoving volume). The signal matrix is
\begin{equation}
S_{(\ell) qz,q'z'} = C_\ell^{Bq,Bq'}[\chi(z),\chi(z')],
\end{equation}
which is proportional to the tensor-to-scalar ratio $r$. We thus have $\partial{\bf C}_{(\ell)}/\partial r = {\bf S}_{(\ell)}/r$, which is independent of $r$.

Finally, we need the relation between $\Delta_h^2(k)$ and $r$. This is
\begin{equation}
\Delta_h^2(k) = r \Delta_\zeta^2(k) = 2.4\times 10^{-9}r.
\end{equation}

In Fig.~\ref{fig:sigmar}, we plot the forecasts for $\sigma_{r}$ for different fiducial values of $r$, for 
different FFTT configurations. We choose the pre-reionization Lyman-$\alpha$ flux model described in Sec.~\ref{ss:refexperiments}; for this fiducial model the values of $w_1$ and $w_2$ peak around $z\sim 19.5$. The observation time $\tau$ entering the expression for noise in Eq.~(\ref{eq:noise}) is the time for observing a given portion of the sky that is above the horizon of a given location. We note that this is different from the \textit{total live observation time} $t_{\rm {obs}}$ which is longer than $\tau$. Here $t_{\rm {obs}}$ is longer by a factor equal to the fraction of
the day that a given survey region is above the horizon and is related to $\tau$ via
\be
t_{\rm {obs}} = \tau \f{\Omega_{\rm{total}}}{\Omega_{\rm{instant}}}.
\ee
A FFTT can in principle observe the entire sky above the horizon at a given instant, corresponding to $\Omega_{\rm{instant}}= 2\pi$sr. However, the image quality degrades near the horizon and the effective   $\Omega_{\rm{instant}}= \pi$ sr. For $f_{\rm sky}\sim 0.7$ the corresponding $\Omega_{\rm{total}} = 2.8\pi$ sr (note that achieving this $f_{\rm sky}$ will need two experiments -- one in the northern and one in the southern hemisphere). Fig.~\ref{fig:sigmar} shows our forecasts in terms of the observation time $t_{\rm {obs}}$. We note that there is a third time-scale in our experiments which is the wall-clock time. Practically, an experiment is unlikely to be on-line for the entirety of a survey, and wall-clock time will thus be longer than $t_{\rm {obs}}$. The wall-clock time determines the total duration of a survey.

\section{Foregrounds}
\label{sec:foregrounds}

Foreground contamination by Galactic and extragalactic sources poses the most serious challenge to detecting the cosmological 21 cm temperature and circular polarization signals. Broadband Galactic and extragalactic foregrounds at low-frequencies are expected to be approximately four orders of magnitude larger than the cosmological temperature signal, and their removal has been the subject of extensive study. Broadly, the approaches for foreground removal involve using both the angular structure of foregrounds, and the spectral smoothness of synchrotron and free-free radiation (as compared to the highly structured cosmological signal) to distinguish them from the cosmological signal~\cite{2002ApJ...564..576D, 2004MNRAS.355.1053D, 2004ApJ...608..622Z, 2003MNRAS.346..871O, 2012MNRAS.419.3491L}.

Linear polarization of the redshifted 21cm radiation has been examined by Babich \& Loeb~\cite{2005ApJ...635....1B}. They considered the intrinsic polarization of the 21 cm line due to Thomson scattering during reionization, leading to a 21 cm $E$-mode signal. This signal is expected  to be completely scrambled up by Faraday rotation, although De \& Tashiro~\cite{2014PhRvD..89l3002D} concluded that extremely accurate Galactic rotation measure maps might allow one to reconstruct the intrinsic linear polarization signal. 

Circularly polarized
foregrounds at low-frequencies, relevant for our technique, 
are not as well-understood. King and Lubin~\cite{2016PhRvD..94b3501K} created foreground maps of circular polarization induced by Galactic magnetic fields 
in the GHz frequency range (relevant for CMB observations) and more recently En{\ss}lin et.~al.~\citep{2017arXiv170608539E} have created predicted Galactic circular polarization maps based on synchrotron templates at 408 MHz (see also ~\cite{2012A&A...542A..93O}). 
In this section we examine potential 
foregrounds that could contaminate the measurement of the 
cosmological 21 cm circular polarization signal relevant to our method.

There are two broad mechanisms that can contaminate the
cosmological circular polarization signal: the intrinsic circular polarization of
galactic or extragalactic foreground sources, and that
generated during propagation through the 
interstellar/intergalactic medium. The former is expected to
be spectrally smooth and could potentially be removed 
using spectral smoothness arguments described earlier. 
The circular polarization induced due to propagation effects can 
lead of confusion with the cosmic signal, since it depends 
on the spatial structure of the ISM/IGM, and may have a complicated frequency dependence due to Faraday rotation. As such,
it is important to estimate the amplitude and approximate angular
structure of these foregrounds in order to assess the feasibility
of our technique.

Circularly polarized foregrounds could in principle spoil our measurement in one of two ways. One would be if the circularly polarized foregrounds were correlated with the total intensity with a quadrupolar spatial/spectral pattern such as to mimic a cosmological signal. We discuss in each case whether we expect this to be an issue. The other would be if the circularly polarized foregrounds did not have such a pattern, but were so bright as to effectively add noise to the measured $TV$ correlation and prevent the remote CMB quadrupole estimator from reaching the theoretical thermal noise limit. We can understand this ``foreground noise'' problem if we  consider the $TV$ cross spectrum as a function of wavenumber, 
\be
\Delta_{TV}^2(k) \equiv \f{k^3}{2\pi^2} P_{TV}(k),
\ee
and recall its uncertainty:
\begin{equation}
\sigma[\Delta^{2}_{TV}(k)]
= \sqrt{\f{1}{N_{\mathrm{modes}}} \Delta_{TT,\mathrm{tot}}^{2}(k)  \Delta_{VV,\mathrm{tot}}^{2}(k)},
\end{equation}
where  $N_{\mathrm{modes}}$ is the number of modes probed, and $\Delta_{VV,\mathrm{tot}}^{2}(k)$ is the sum of the intrinsic cosmological signal $\Delta_{VV,\mathrm{cosmo}}^{2}(k)$, the instrument noise $\Delta_{VV,\mathrm{noise}}^{2}(k)$, and the foregrounds $\Delta_{\mathrm{fore}}^{2}(k)$. We have assumed here that the foregrounds for 21 cm temperature have been successfully removed using techniques described in the literature.
As discussed in Section~\ref{ss:subfisher}, $\Delta_{VV,\mathrm{noise}}^{2}(k)\approx \Delta_{TT,\mathrm{noise}}^{2}(k)$ and from Figure~\ref{fig:powerspectra} we see that $\Delta_{VV,\mathrm{noise}}^{2}(k)\gg \Delta_{VV}^{2}(k)$. Hence the ``foreground noise'' contribution to
$\sigma[\Delta^{2}_{TV}(k)]$ depends on the relative magnitude of $\Delta_{VV,\mathrm{fore}}^{2}(k)$
and $\Delta_{VV,\mathrm{noise}}^{2}(k)$.
 
In this section we make order-of-magnitude estimates of $\Delta_{VV,\mathrm{fore}}^{2}(k)$ 
due to the synchrotron emission from the galaxy and extragalactic point sources. These foregrounds
turn out to be the dominant foregrounds but we argue that they can be removed because of their 
spectral smoothness in frequency space. We also
estimate $\Delta_{VV,\mathrm{fore}}^{2}(k)$ due to propagation effects through the ISM. 
These foregrounds are expected to have features correlated to structures in the ISM
and are not spectrally smooth. However, we show that these foregrounds are not likely to
be important for our proposed method.

\subsection{Spectrally Smooth Circular Polarization from Synchrotron}
\label{subsec:synchrotron}

The synchrotron radiation from ultra-relativistic electrons
in the interstellar medium is the strongest source
of foregrounds at low-frequencies~\cite{1982A&AS...47....1H}. 
It is strongly linearly polarized.
At low frequencies in {\em linear} polarization, even a spatially smooth signal is mixed to small angular scales by Faraday rotation, leading to typical  fluctuating signals of a few Kelvin. This signal has been constrained or observed with many instruments at frequencies $<200$ MHz
\cite{2009A&A...500..965B, 2009MNRAS.399..181P, 2013ApJ...771..105B, 2014A&A...568A.101J, 2015arXiv150205072M, 2015A&A...583A.137J}.
In both cases the limits on the Stokes $I$ parameter were $\Delta_I\lesssim10$ K over the range of angular scales probed. (The spatially smooth component can be much brighter.)

Synchrotron radiation is expected to have a small fraction of 
circularly polarization. The circular polarization in synchrotron radio emission
has been observed in 
quasars~\cite{1975AuJPh..28..325R}, AGNs~\cite{1998Natur.395..457W,1999AJ....118.1942H},
and the galactic center~\cite{1999ApJ...526L..85S}. While the degree of circular polarization in these sources is not completely well-understood, it is believed to arise from a combination of intrinsic circular polarization of synchrotron radiation and propagation effects in a plasma~\cite{2000PhRvE..62.4177M, 2000ApJ...545..798M}.

The degree of circular polarization of Galactic synchrotron has not yet been measured, but we can make rough estimates of the strength of this foreground using measured limits on the Stokes $I$ parameter. 
For an electron with Lorentz factor $\gamma$ gyrating around a 
field line at an angle $\theta$ to the line of sight, the degree of circular polarization
observed, to the first order in $\gamma$~\cite{1968ApJ...154..499L, 2015MNRAS.450..533D},
\be
\f{V}{I} \approx \cot\theta \lf \f{\nu_{g}}{\nu}\rt^{1/2} \approx \gamma^{-1} \cot \theta 
\ee
where $\nu_{g}= (eB)/(2\pi \gamma m_{e} c)$ is the gyromagnetic frequency.

The typical Lorentz factor of relativistic electrons that lead to synchrotron radiation
in the frequency range $\nu_{\mathrm{radio}}\sim50-150$MHz is
\be
\gamma = \sqrt{\f{2\pi m_{e} c \nu_{\mathrm{radio}}}{e B_{\mathrm{gal}}}} \changetext{\sim 400},
\ee
where we take the typical magnetic field in the ISM to be \changetext{$B_{\mathrm{gal}}\sim 6\mu$G \cite{2015ASSL..407..483H}}.

Since $\Delta_I\lesssim10\,$K, the typical circular polarization signal from relativistic electrons in the galaxy is expected to be \changetext{$\Delta_V\approx 0.03\,$K} in temperature units, and the typical value of \changetext{$\Delta_{VV,\mathrm{sync}}^{2}(k)\approx10^{-3}\,\mathrm{K}^2$}.

\begin{figure}
\includegraphics[trim=0cm 0cm 0cm 0cm, clip=true, width=0.5\textwidth]{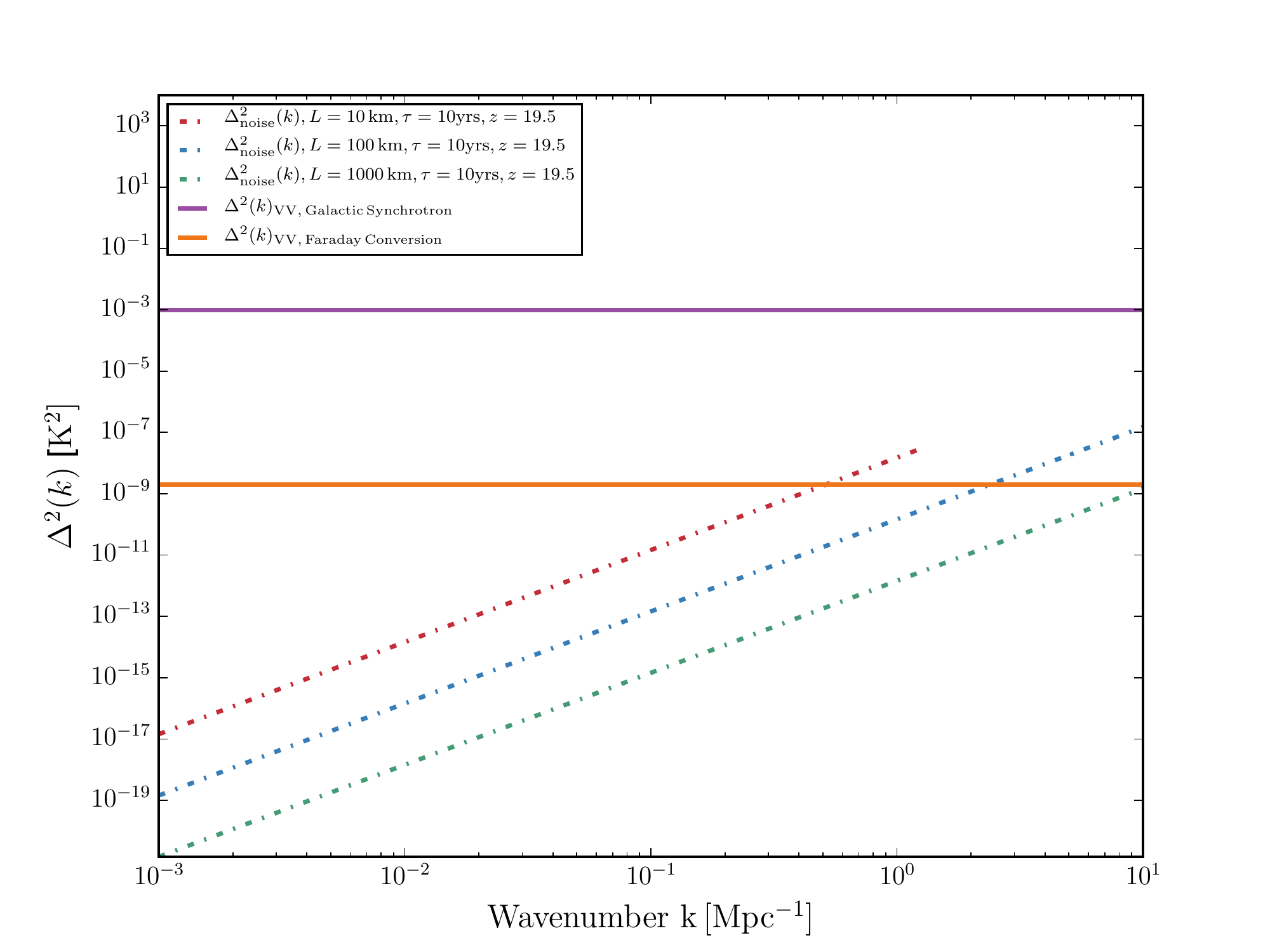}
\caption{Order of magnitude of expected foregrounds for the circular polarization signal from Galactic synchrotron (purple line) and due to Faraday rotation through the ISM (orange line) compared against the noise
power spectra expected for for three different configurations of FFTTs.}
\label{fig:foregrounds}
\end{figure}

As seen in Fig.~\ref{fig:foregrounds}, $\Delta_{VV,\mathrm{sync}}^{2}(k)$ is many orders of magnitude larger than $\Delta_{VV,\mathrm{noise}}^{2}(k)$, and is the most dominant foreground for our method. Moreover, since the sign of the $TV$ correlation depends on the direction of the magnetic field (toward or away from the observer), and the Galactic magnetic field has a large-scale coherent component, we expect significant $TV$ correlations even averaged over a patch of many tens of degrees. However, this synchrotron circular polarization signal is spectrally smooth and hence the same foreground removal techniques applied to total intensity should be applicable~\cite{2012MNRAS.419.3491L}. In particular, it is confined to modes with $k_\parallel \approx 0$.

\subsection{Circular Polarization Foregrounds from Faraday Rotation}
\label{ss:faraday}

Faraday rotation of linearly polarized light through a closed
plasma interconverts $Q$ and $U$ Stokes parameters
but does not lead to generation of Stokes $V$, 
to the first order in the galactic magnetic field $B_{\mathrm{gal}}$. 
However, in the next order in $B_{\mathrm{gal}}$, Faraday rotation can 
lead to an ``leaking" of Stokes $Q$ and $U$ to produce
Stokes $V$~\cite{1969SvA....13..396S}.

The Galactic synchrotron radiation is expected to have a high degree
of linear polarization and the leakage of power from Stokes $Q$ and $U$
to $V$, due to propagation through the cold plasma in the ISM can result in a 
CP foreground. Unlike the intrinsic CP signal discussed in Section~\ref{subsec:synchrotron},
this signal is not smooth in frequency space. 
The signal is expected to trace structures in the ISM and, if it has significant amplitude, can potentially mimic the cosmological signal. 
In this section we estimate the angular power spectrum of the CP
signal due to propagation effects in the Galaxy.

Consider the polarization of radiation
that is propagating through a cold plasma.
The transfer equation for the radiation propagating
along the $z$ direction is given by
\begin{eqnarray}
\frac{dQ}{dz} &=& - 2 \frac\omega c [ n_U V - n_V U ],\nonumber
\\ \frac{dU}{dz} &=& - 2 \frac\omega c [ n_V Q - n_Q V],~~{\rm and}\nonumber
\\ \frac{dV}{dz} &=&  2 \frac\omega c [ n_U Q- n_Q U],
\end{eqnarray}
where $n_{U}$, $n_{Q}$, and $n_{V}$ are the real anti-symmetric
components of the refractive index tensor $n_{ij}$
and are given by
\bea
\nonumber
n_{U}&=& -  \f{\pi n_e e^2} {m_e \omega^2} \left(\f{e}{m_e \omega c}\right)^2 2B_{x}B_{y},\\
\nonumber
n_{Q}&=&  - \f{\pi n_e e^2} {m_e \omega^2} \left(\f{e}{m_e \omega c}\right)^2 (B_{y}^2 -B_{x}^2),~~{\rm and}\\
n_{V} &=&    \f{\pi n_ee^2} {m_e \omega^2} \f{e}{m_e \omega c} B_z.
\label{eq:NBF}
\eea
The circular polarization produced by propagation through a medium is then the integral
\be 
V = \int \f{2\omega}{c} (n_{Q} U -n_{U} Q)\, dz.
\label{eq:V44}
\ee
To estimate the order of magnitude of $V$, we need estimates of the birefringence coefficients $(n_Q,n_U)$; the linear polarization $(Q,U)$; the path length through the ISM; and the coherence length $z_{\rm coh}$ over which the integrand retains the same sign. \changetext{To estimate the order of magnitude of the integrand in Eq.~(\ref{eq:V44}), we consider a magnetic field in the diagonal direction ($B_x=B_y=B_z=B/\sqrt3$). Then the birefringence is in the $n_U$ component and
\begin{eqnarray}
\frac{dV}{dz} &\sim&
\f{2\omega}{c} n_{U} Q
\nonumber \\ &\sim &
\f{2\omega}{c} \f{2\pi n_e e^2} {3m_e \omega^2} \left(\f{e}{m_e \omega c}\right)^2 B^2 \sqrt{\Delta^2_{QQ,\rm sync}}
\nonumber \\ &\sim &
 \f{4\pi e^4 } {3m_e^3 \omega^3 c^3} n_e B^2 \sqrt{\Delta^2_{QQ,\rm sync}},
\label{eq:integrand}
\end{eqnarray}
where we have used $n_U$ from Eq.~(\ref{eq:NBF}) and taken typical linear polarization $\sqrt{\Delta^2_{QQ,\rm sync}}$. The variance of the integral, Eq.~(\ref{eq:V44}), should then be the incoherent sum of $L_{\rm gal}/z_{\rm coh}$ individual segments:
\begin{equation}
\Delta^2_{VV,\rm Faraday} \sim \frac{L_{\rm gal}}{z_{\rm coh}}\left(\frac{dV}{dz}\,z_{\rm coh}\right)^2,
\end{equation}
which -- using Eq.~(\ref{eq:integrand}) -- simplifies to
\be
\Delta_{VV,\mathrm{Faraday}}^{2}
 \sim \lf \f{4\pi e^4}{3m_e^3 \omega^3 c^3}\rt^2 n_{e}^2 L_{\mathrm{gal}} B^4 z_{\mathrm{coh}}  \Delta_{QQ,\mathrm{sync}}^{2} .
\label{eq:V2}
\ee}
The coherence length $z_{\rm coh}$ could be set by either de-correlation of $(n_Q,n_U)$ or of $(Q,U)$; the latter occurs on a distance scale of order a Faraday rotation cycle. \changetext{We take as our estimate the distance for a rotation of $(Q,U)$ by $\pi/4$, so that if $Q$ is maximal at position $z$ it crosses zero at $z+z_{\rm coh}$. Then:
\begin{equation}
z_{\rm coh} \sim \frac{\pi c}{4\omega|n_V|}
\sim \frac{\sqrt3\,m_e^2c^2\omega^2}{4 n_e e^3 B}
\end{equation}
(recall that $B_z\sim B/\sqrt3$). Plugging this into Eq.~(\ref{eq:V2}) gives
\begin{equation}
\Delta_{VV,\mathrm{Faraday}}^{2}
 \sim \frac{4\pi^2 e^5 B^3 n_e}{3\sqrt3\,m_e^4 \omega^4 c^4} L_{\mathrm{gal}} \Delta_{QQ,\mathrm{sync}}^{2} .
\label{eq:V2X}
\end{equation}}
For order-of-magnitude purposes, we take $\Delta_{QQ,\mathrm{sync}}^{2}\sim 10 \,\mathrm{K}^2$ (the order of magnitude of recent detections or upper limits), a path length of \changetext{$0.95$ kpc corresponding to electron scale height of the Milky Way thick disk inferred from the NE2001 model~\cite{2002astro.ph..7156C}}, a Galactic magnetic field of \changetext{$B\sim 6\,\mu$G}, and an electron density \changetext{$n_e\sim 0.035\,\mathrm{cm}^{-3}$~\cite{2002astro.ph..7156C}. We then estimate
$|n_V|\approx 2\times 10^{-17}$,
$|n_U|\approx 6\times 10^{-24}$,
$z_{\rm coh}\approx 1\,$pc, and
$\Delta_{VV,\mathrm{Faraday}}^{2}\approx 2\times 10^{-9}\,\mathrm{K}^2$ at $\nu=69$ MHz (corresponding to $z=19.5$)}. As seen in Fig.~\ref{fig:foregrounds} the circular polarization foreground due to Faraday rotation is \changetext{lower} than the noise power spectra of the proposed experimental setups \changetext{for $\ge90$\%\ of the accessible Fourier modes (recall that ${\bm k}$-space volume is proportional to $k_{\rm max}^3$)}. Note that we have not determined the peak angular scale for this foreground; since the Galactic magnetic field and hence $(n_Q,n_U)$ exhibit large-scale coherence, we expect the circular polarization induced by $Q,U\rightarrow V$ conversion to trace the same angular scales as linear polarization.

Since at low frequencies the linear polarization has been rotated through many cycles, we expect a very weak correlation of $(Q,U)$ (and hence $V$) with the total synchrotron intensity.

\subsection{Extragalactic Point Sources}

\begin{figure}
\includegraphics[trim=0cm 0cm 0cm 0cm, clip=true, width=0.5\textwidth]{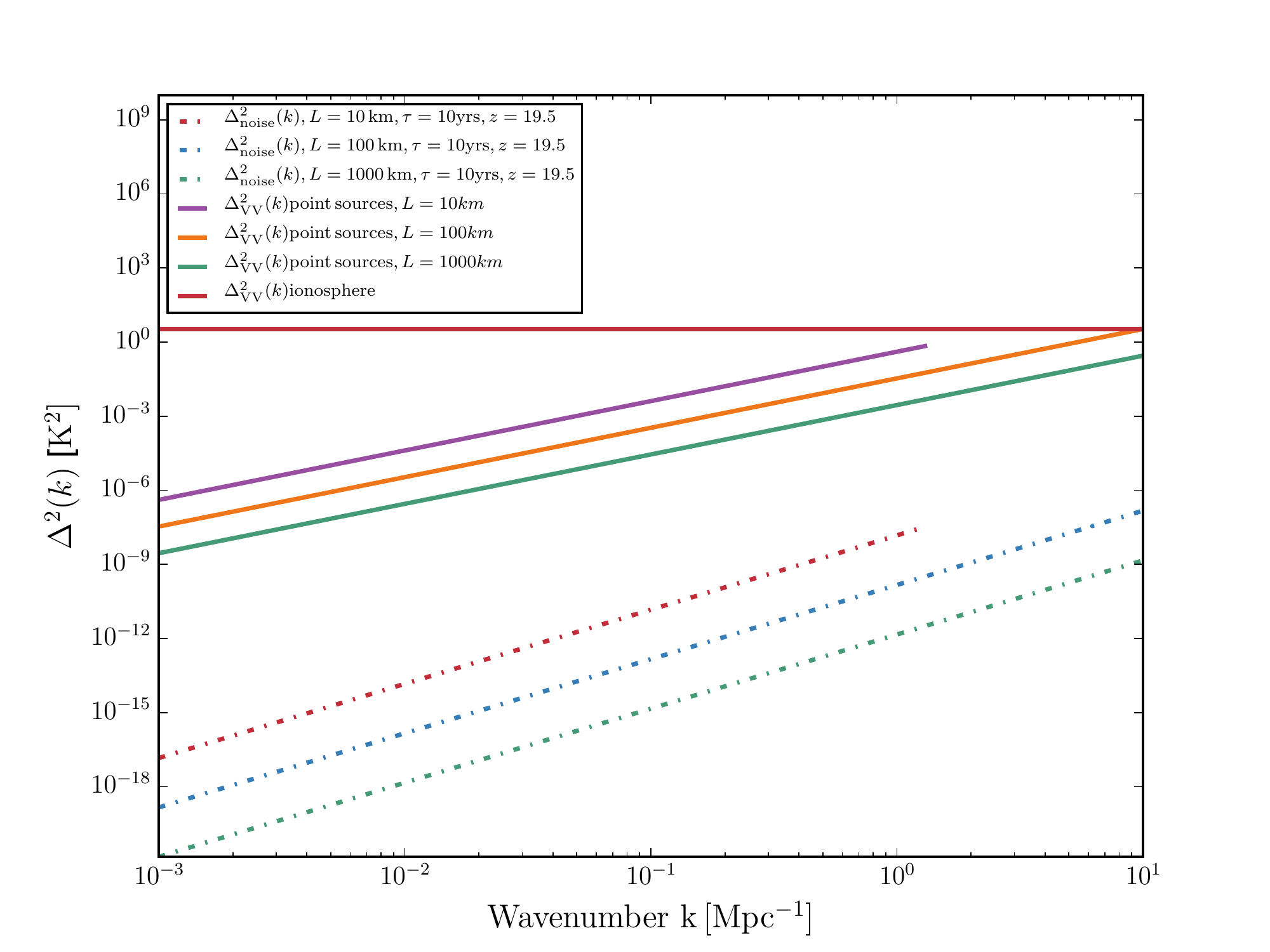}
\caption{Order of magnitude of expected foregrounds for the circular polarization signal from unresolved point sources (purple, orange, and solid green line)and Faraday rotation due to the ionosphere (solid red line) compared against the noise power spectra expected for three different configurations of FFTTs.}
\label{fig:point_sources}
\end{figure}

After Galactic synchrotron, unresolved, extragalactic 
point sources are expected to be one of the most challenging foregrounds for 21 cm tomography~\cite{2009MNRAS.394.1575L, 2004MNRAS.355.1053D}. An interferometer is usually characterized by a classical confusion limit, defined as having one source above the threshold flux $S_c$ per $m$ synthesized beams. Then the threshold flux density $S_c$ is defined  such that $m \times 1.13 \, \, \theta^2 \, \,N(S_{c}) = 1$. Here $N(s)$ is the number density of sources above flux density $s$ and $\theta$ is the FWHM of each synthesized beam. Here we assume that $N(s)=As^{\beta}$ which implies,
\be
\label{eq:confusion_limit}
S_c = (1.13m A)^{-1/\beta} \theta^{-2/\beta}
\ee
Here we consider the classical confusion limit calculated at $74$MHz for the VLSS survey which gives $A = 1.14 $, $\beta = -1.3$, and $m = 12.9$ as calculated by Cohen~\cite{LWA}, and the units of flux are in Jansky and beam size is in degrees.

We can detect and remove point sources of flux $S$ from a survey if the thermal noise of the survey is much less than $S$ and if the source has a flux density much greater than $S_c$. Sources with flux density less than $S_c$ will lead to a confusion noise even in the limit of infinite integration time. In this section we assume that the resolved point sources have been removed using standard techniques and estimate the noise contribution to $\Delta_{VV,\mathrm{fore}}^{2}(k)$ due to unresolved point sources, for different configurations of FFTTs. 
To estimate the foreground contribution due to unresolved point sources we need the classical confusion for low-frequency radio experiments. Here we consider the confusion limit calculation based on the VLSS sky survey at 75 MHz \cite{LWA} given by Eq.(~\ref{eq:confusion_limit}). For a FFTT the beam size is $\theta\sim \lambda/L$ where $L$ is the side length. For observations around 68 MHz and for FFTT side lengths of $10, 100, 1000$km the beam size corresponds to $\theta = 0.025 \, \, , 0.0025 \,\, \& \,\, 0.00025$ degrees respectively. The corresponding confusion limits are $S_c = 3\times 10^{-2} \, \, , 8.6\times 10^{-4} \,\, \& \,\, 2.5\times 10^{-5}$ respectively. The contribution to the temperature power spectrum due to unresolved point sources (flux less than $S_c$ per beam) is
\be
\Delta^2_{\rm{TT}}= \f{l^2}{2\pi}C_{l}^{\rm{TT}}\approx  \f{l^2}{2\pi}\lf\f{\lambda^2}{2k_{\mathrm{B}}}\rt^2   \int^{S_c}_0 S^2 \f{dN}{dS} dS,
\ee
where $\Delta^2_{\rm{TT}}$ is the total power per log$k$, $N(S)$ is the number density of sources above a flux $S$ and $l = k(1+z)D_{A}(z)$. We use a power law source count function,
$N(S)=A S^{\alpha}$ where $A=  1.14$ and $\alpha= -1.3$ \cite{LWA}. 
The point source foreground at low frequencies is dominated by synchrotron emission from radio-loud galaxies and AGNs~\cite{2012MNRAS.426.3295G, 2004MNRAS.355.1053D}; the circular polarization foreground due to the confused background of point sources is given by
\be
\Delta_{VV,\mathrm{fore}}^{2}\approx  \left( \frac VI \right)^2 \Delta_{TT,\mathrm{fore}}^{2}.
\ee
The measured circular polarization fraction for typical radio-loud AGNs is $V/I\sim 10^{-4}$ at $4.9$ GHz~\cite{2000MNRAS.319..484R}. Note that these measurements are dominated by the brightest radio-galaxies while the point sources dominating the foregrounds we are interested in are likely to be much fainted. The fractional circular polarization for blazars are expected to be much higher (e.g.~\cite{2000MNRAS.319..484R}) but these blazars are not likely to dominate the unresolved point source background.

Assuming the circular polarization of radio galaxies is dominated by synchrotron, the degree of circular polarization at low frequencies (relevant to our estimates) can be determined by scaling $V/I \propto \gamma^{-1} \propto \nu^{-1/2}$, so at 68 MHz $V/I$ is a factor of 8.5 larger than at 4.9 GHz. We plot $\Delta_{VV,\mathrm{fore}}^{2}$ for different configurations of the FFTT in Fig.~\ref{fig:point_sources}. 
The synchrotron emission from point sources is expected to vary smoothly in frequency space, whereas the redshifted 21 cm signal varies
rapidly in frequency space (similar to the galactic synchrotron signal). This is a similar situation to 21 cm temperature, and similar techniques should be applicable~\cite{2006ApJ...650..529W, 2012MNRAS.419.3491L}.

We note that the sign of the circular polarization of a point source is determined by its internal magnetic field structure, so our result for $\Delta^2_{VV,\rm fore}$ is not affected by source clustering so long as the sign of $V$ is independent for each source. Furthermore, under these circumstances, there is no systematic contribution to $TV$, only a source of excess noise in $VV$.

\subsection{Atmospheric Effects}

Radio propagation through the Earth's atmosphere is one of the key
calibration challenges in low-frequency radio astronomy. At low
frequencies ($\nu\leq 200$ MHz), propagation effects through the ionosphere 
become dominant. The physics of the propagation of the radio waves
through a magnetized ionosphere is well understood. There are two primary 
effects at play after the polarization-dependent geometrical refraction by the ionosphere is removed. First, propagation through a turbulent
ionosphere leads to stochastic interferometric visibilities, which contribute to an additional
``scintillation noise'' to the measurement of the power spectrum (e.g.\ \cite{2016MNRAS.458.3099V}). This scintillation noise
can be larger than the thermal noise associated with low-frequency radio experiments. 

Second, and most directly relevant here, is the inter-conversion of the polarization Stokes parameters ($Q$, $U$, $V$) and hence the generation of additional circular polarization signal due to Faraday rotation as discussed in Section~\ref{ss:faraday}. This mechanism for generating Stokes $V$ is much more significant in the Earth's ionosphere than in the ISM since the magnetic field $B$ is much larger (generation of $V$ depends on $B^2$ times column density, unlike regular Faraday rotation that depends on $B$ times column density). Since again at low frequencies the ionosphere can result in $\gtrsim 1$ cycle of Faraday rotation, we use Eq.~(\ref{eq:V2X}), with low-frequency linear polarization of order $\Delta_{QQ,\mathrm{gal}} \sim 10$\,K (see discussion in Section \ref{subsec:synchrotron}).
The typical electron density in the F-layer of the ionosphere is $n_e\sim 10^5 \,\mathrm{cm}^{-3}$,
the magnetic field is $B\sim 0.5$ G, and the typical path length is $L\sim500$ km. At $1+z\sim 20$ this leads to an expected circular polarization signal of $\Delta_{VV,\mathrm{atm}}^2\approx 1\,$K. We plot
the expected order of magnitude of $\Delta_{VV,\mathrm{atm}}^{2}$ due to atmospheric Faraday rotation against the noise power spectra in Fig.~\ref{fig:point_sources}. As evident from the figure, the is likely to be the most challenging foreground for low-frequency circular polarization studies. 

Calibration and correction of Faraday rotation distorted low-frequency measurements has been extensively studied in the literature, particularly in the context of ongoing 21 cm experiments. For primordial gravitational wave detection, such techniques would clearly have to be pushed many orders of magnitude beyond the present state of the art. In any case, the ionosphere represents perhaps the greatest foreground challenge to cosmological circular polarization studies.

\section{Discussion}
\label{sec:discussion}

In Paper I of this series, we showed that the remote CMB quadrupole during the pre-reionization epoch leads to a small circular polarization of the emitted 21 cm radiation.  In this paper we showed that measurement of the temperature-circular polarization cross-spectrum $P_{TV}(k)$ allows us to measure the remote quadrupole of the CMB. The remote quadrupole field at high redshifts can then be decomposed into $E$ and $B$ modes, and we showed that measurement of the $B$ modes of this field can help us measure the tensor-to-scalar ratio $r$. We showed that, given the fiducial model for pre-reionization physics, a Fast Fourier Transform Telescope (FFTT) with side length $100$ km can achieve $\sigma(r)\sim 4\times 10^{-3}$ after ten years of observation while a FFTT of side length $1000$ km can achieve $\sigma(r)\sim 10^{-5}$ after ten years of observation time. 

One of the key results of this paper is that the sensitivity to measuring the remote CMB quadrupole is sensitive to the measurement of the modes with largest wavenumber (corresponding to the longest baselines in an interferometric experiment). For the fiducial model of pre-reionization physics considered in our paper, Fig.~\ref{fig:w_z} implies that the method is most sensitive around $z\approx 19.5$, i.e.\ the time at which Lyman-$\alpha$ coupling becomes efficient in the fiducial model. Fig.~\ref{fig:sigmar} shows the sensitivity to measuring the tensor-to-scalar ratio $r$ as a function of the side length of a FFTT, and for different observation times.

Our forecasts depend on assumptions made about
the pre-reionization history of the universe, in particular 
on the rate of depolarization of the ground state of hydrogen through Lyman-$\alpha$ pumping, which is proportional to the mean Lyman-$\alpha$ flux. The parameters for the fiducial model that we consider for our sensitivity calculation are shown in Fig.~\ref{fig:reion}. We note however that the Lyman-$\alpha$ flux at the redshifts of interest is completely unconstrained observationally; the ``optimal'' window in redshift, when $x_\alpha\sim {\cal O}(1)$, would be earlier (later) if the Lyman-$\alpha$ flux is higher (lower). We note that lower Lyman-$\alpha$ flux would be advantageous for our method, since it places the transition at higher observed frequency where the foregrounds are less severe.

Another assumption in our technique is that the magnetic fields during the Dark Ages
are below the ``saturation limit" as described in Ref.~\cite{2017PhRvD..95h3011G}. A saturated magnetic
field has a strength such that the precession of hydrogen atoms in 
in the hyperfine excited state is much faster
than the decay (natural or stimulated) of the excited state. If the magnetic field is above
the saturation limit, then the circular polarization signal will be suppressed.
However most conventional models for magnetic fields during the Dark Ages
predict unsaturated magnetic fields. A constraint on magnetic field
strength during the Dark Ages as described in Refs.~\cite{2014arXiv1410.2250V, 2017PhRvD..95h3011G} will thus be crucial before embarking on an experiment that uses the technique described in this paper. 

To contrast our results to existing bounds on the tensor-to-scalar ratio, we note that the current upper bounds on $r$ from the combination of the CMB $B$-mode and other observables are $r<0.07$ (95\%\ CL) \cite{2016PhRvL.116c1302B}. Next generation ``Stage-4"  CMB experiments have a goal of probing $r\leq 0.002$ \cite{2016arXiv161002743A}, although several challenges remain in dealing with systematic effects.

Other authors have proposed techniques to detect inflationary gravitational waves that, while futuristic, have the potential to confirm a CMB detection, probe another range of scales, and/or improve sensitivity to $r$. Some of these techniques are based on conventional large-scale structure observables \cite{2003PhRvL..91b1301D, 2012PhRvD..85l3540A, 2012PhRvD..86h3513S, 2014PhRvD..89h3507S, 2014PhRvD..90d3527C}, although the surveys required even to detect $r\sim 0.07$ are very futuristic and many run up against cosmic variance limitations. Direct detection of the high-wavenumber gravitational waves with a network of space-based laser interferometers has been studied \cite{2006CQGra..23.2435C, 2011CQGra..28i4011K}.

The techniques most comparable to this work are other proposals using the enormous number of modes in redshifted 21 cm radiation. While the foreground-to-signal ratio is much higher for 21 cm experiments than for the CMB, the 21 cm measurement is a line measurement against a continuum foreground (as opposed to the continuum CMB signal) and so the ultimate factor by which foregrounds can be suppressed in analysis could be much larger.
Masui \& Pen~\cite{2010PhRvL.105p1302M} proposed using the large number of Fourier modes available in a 21 cm survey to measure the intrinsic distortion of large scale structure due to inflationary gravitational waves. For a FFTT with $L=100$ km their technique can detect $r\sim10^{-3}$ which is similar to our forecasts. Book et~al.~\cite{2012PhRvL.108u1301B} proposed using the weak lensing of the 21 cm intensity fluctuations by gravitational waves to put bounds on $r$. This involves the measurement of the 21 cm power spectrum up to very small angular scales; to reach $r\sim 10^{-3}$, they would need to probe to $l_{\mathrm{max}}\sim 10^5$, requiring an array size of $L\gtrsim 100$\ km.

The method proposed in this series is the first to make use of the 21 cm circular polarization signal (in cross-correlation with temperature: $TV$). It is also very futuristic, in the sense of requiring $L\sim 100$ km radio arrays (or $\sim 5\times 10^8$ antennas). However, the foregrounds in circular polarization are much fainter than in brightness temperature, so our method for measuring $r$ may turn out to be less problematic than methods based on the local anisotropy of the temperature power spectrum. In any case, the radio arrays that could implement the $TT$ methods \cite{2010PhRvL.105p1302M, 2012PhRvL.108u1301B} are likely similar to what one would need for $TV$, so the techniques could be used to cross-check each other.

While the experimental setup required for the circular polarization method is very futuristic, it illustrates the rich array of physical processes and diagnostics that are in principle available in 21 cm surveys. Given the long-term interest in detecting inflationary gravitational waves, we hope that this idea will serve both to further motivate the goal of the ultimate 21 cm cosmology experiment, and to inspire additional work on novel applications.

\begin{acknowledgments}
AM and CH are supported by the U.S. Department of Energy. CH is also supported by the David \& Lucile Packard Foundation, the Simons Foundation, and NASA. We thank Olivier Dor\'{e}, Michael Eastwood, Ashish Goel, and Tejaswi Venumadhav for useful discussions during the project.
\end{acknowledgments}

\vspace{-12pt}
\bibliographystyle{apsrev4-1}
\bibliography{detectability}

\begin{thebibliography}{87}%
\makeatletter
\providecommand \@ifxundefined [1]{%
 \@ifx{#1\undefined}
}%
\providecommand \@ifnum [1]{%
 \ifnum #1\expandafter \@firstoftwo
 \else \expandafter \@secondoftwo
 \fi
}%
\providecommand \@ifx [1]{%
 \ifx #1\expandafter \@firstoftwo
 \else \expandafter \@secondoftwo
 \fi
}%
\providecommand \natexlab [1]{#1}%
\providecommand \enquote  [1]{``#1''}%
\providecommand \bibnamefont  [1]{#1}%
\providecommand \bibfnamefont [1]{#1}%
\providecommand \citenamefont [1]{#1}%
\providecommand \href@noop [0]{\@secondoftwo}%
\providecommand \href [0]{\begingroup \@sanitize@url \@href}%
\providecommand \@href[1]{\@@startlink{#1}\@@href}%
\providecommand \@@href[1]{\endgroup#1\@@endlink}%
\providecommand \@sanitize@url [0]{\catcode `\\12\catcode `\$12\catcode
  `\&12\catcode `\#12\catcode `\^12\catcode `\_12\catcode `\%12\relax}%
\providecommand \@@startlink[1]{}%
\providecommand \@@endlink[0]{}%
\providecommand \url  [0]{\begingroup\@sanitize@url \@url }%
\providecommand \@url [1]{\endgroup\@href {#1}{\urlprefix }}%
\providecommand \urlprefix  [0]{URL }%
\providecommand \Eprint [0]{\href }%
\providecommand \doibase [0]{http://dx.doi.org/}%
\providecommand \selectlanguage [0]{\@gobble}%
\providecommand \bibinfo  [0]{\@secondoftwo}%
\providecommand \bibfield  [0]{\@secondoftwo}%
\providecommand \translation [1]{[#1]}%
\providecommand \BibitemOpen [0]{}%
\providecommand \bibitemStop [0]{}%
\providecommand \bibitemNoStop [0]{.\EOS\space}%
\providecommand \EOS [0]{\spacefactor3000\relax}%
\providecommand \BibitemShut  [1]{\csname bibitem#1\endcsname}%
\let\auto@bib@innerbib\@empty
\bibitem [{\citenamefont {{Guth}}(1981)}]{1981PhRvD..23..347G}%
  \BibitemOpen
  \bibfield  {author} {\bibinfo {author} {\bibfnamefont {A.~H.}\ \bibnamefont
  {{Guth}}},\ }\href {\doibase 10.1103/PhysRevD.23.347} {\bibfield  {journal}
  {\bibinfo  {journal} {\prd}\ }\textbf {\bibinfo {volume} {23}},\ \bibinfo
  {pages} {347} (\bibinfo {year} {1981})}\BibitemShut {NoStop}%
\bibitem [{\citenamefont {{Guth}}\ and\ \citenamefont
  {{Pi}}(1982)}]{1982PhRvL..49.1110G}%
  \BibitemOpen
  \bibfield  {author} {\bibinfo {author} {\bibfnamefont {A.~H.}\ \bibnamefont
  {{Guth}}}\ and\ \bibinfo {author} {\bibfnamefont {S.-Y.}\ \bibnamefont
  {{Pi}}},\ }\href {\doibase 10.1103/PhysRevLett.49.1110} {\bibfield  {journal}
  {\bibinfo  {journal} {Physical Review Letters}\ }\textbf {\bibinfo {volume}
  {49}},\ \bibinfo {pages} {1110} (\bibinfo {year} {1982})}\BibitemShut
  {NoStop}%
\bibitem [{\citenamefont {{Bardeen}}\ \emph {et~al.}(1983)\citenamefont
  {{Bardeen}}, \citenamefont {{Steinhardt}},\ and\ \citenamefont
  {{Turner}}}]{1983PhRvD..28..679B}%
  \BibitemOpen
  \bibfield  {author} {\bibinfo {author} {\bibfnamefont {J.~M.}\ \bibnamefont
  {{Bardeen}}}, \bibinfo {author} {\bibfnamefont {P.~J.}\ \bibnamefont
  {{Steinhardt}}}, \ and\ \bibinfo {author} {\bibfnamefont {M.~S.}\
  \bibnamefont {{Turner}}},\ }\href {\doibase 10.1103/PhysRevD.28.679}
  {\bibfield  {journal} {\bibinfo  {journal} {\prd}\ }\textbf {\bibinfo
  {volume} {28}},\ \bibinfo {pages} {679} (\bibinfo {year} {1983})}\BibitemShut
  {NoStop}%
\bibitem [{\citenamefont {{Hawking}}(1982)}]{1982PhLB..115..295H}%
  \BibitemOpen
  \bibfield  {author} {\bibinfo {author} {\bibfnamefont {S.~W.}\ \bibnamefont
  {{Hawking}}},\ }\href {\doibase 10.1016/0370-2693(82)90373-2} {\bibfield
  {journal} {\bibinfo  {journal} {Physics Letters B}\ }\textbf {\bibinfo
  {volume} {115}},\ \bibinfo {pages} {295} (\bibinfo {year}
  {1982})}\BibitemShut {NoStop}%
\bibitem [{\citenamefont {{Linde}}(1982)}]{1982PhLB..108..389L}%
  \BibitemOpen
  \bibfield  {author} {\bibinfo {author} {\bibfnamefont {A.~D.}\ \bibnamefont
  {{Linde}}},\ }\href {\doibase 10.1016/0370-2693(82)91219-9} {\bibfield
  {journal} {\bibinfo  {journal} {Physics Letters B}\ }\textbf {\bibinfo
  {volume} {108}},\ \bibinfo {pages} {389} (\bibinfo {year}
  {1982})}\BibitemShut {NoStop}%
\bibitem [{\citenamefont {{Mukhanov}}\ and\ \citenamefont
  {{Chibisov}}(1981)}]{1981ZhPmR..33..549M}%
  \BibitemOpen
  \bibfield  {author} {\bibinfo {author} {\bibfnamefont {V.~F.}\ \bibnamefont
  {{Mukhanov}}}\ and\ \bibinfo {author} {\bibfnamefont {G.~V.}\ \bibnamefont
  {{Chibisov}}},\ }\href@noop {} {\bibfield  {journal} {\bibinfo  {journal}
  {ZhETF Pisma Redaktsiiu}\ }\textbf {\bibinfo {volume} {33}},\ \bibinfo
  {pages} {549} (\bibinfo {year} {1981})}\BibitemShut {NoStop}%
\bibitem [{\citenamefont {{Planck Collaboration}}\ \emph
  {et~al.}(2016{\natexlab{a}})\citenamefont {{Planck Collaboration}},
  \citenamefont {{Ade}}, \citenamefont {{Aghanim}}, \citenamefont {{Arnaud}},
  \citenamefont {{Arroja}}, \citenamefont {{Ashdown}}, \citenamefont
  {{Aumont}}, \citenamefont {{Baccigalupi}}, \citenamefont {{Ballardini}},
  \citenamefont {{Banday}},\ and\ \citenamefont
  {et~al.}}]{2016A&A...594A..20P}%
  \BibitemOpen
  \bibfield  {author} {\bibinfo {author} {\bibnamefont {{Planck
  Collaboration}}}, \bibinfo {author} {\bibfnamefont {P.~A.~R.}\ \bibnamefont
  {{Ade}}}, \bibinfo {author} {\bibfnamefont {N.}~\bibnamefont {{Aghanim}}},
  \bibinfo {author} {\bibfnamefont {M.}~\bibnamefont {{Arnaud}}}, \bibinfo
  {author} {\bibfnamefont {F.}~\bibnamefont {{Arroja}}}, \bibinfo {author}
  {\bibfnamefont {M.}~\bibnamefont {{Ashdown}}}, \bibinfo {author}
  {\bibfnamefont {J.}~\bibnamefont {{Aumont}}}, \bibinfo {author}
  {\bibfnamefont {C.}~\bibnamefont {{Baccigalupi}}}, \bibinfo {author}
  {\bibfnamefont {M.}~\bibnamefont {{Ballardini}}}, \bibinfo {author}
  {\bibfnamefont {A.~J.}\ \bibnamefont {{Banday}}}, \ and\ \bibinfo {author}
  {\bibnamefont {et~al.}},\ }\href {\doibase 10.1051/0004-6361/201525898}
  {\bibfield  {journal} {\bibinfo  {journal} {\aap}\ }\textbf {\bibinfo
  {volume} {594}},\ \bibinfo {eid} {A20} (\bibinfo {year}
  {2016}{\natexlab{a}})},\ \Eprint {http://arxiv.org/abs/1502.02114}
  {arXiv:1502.02114} \BibitemShut {NoStop}%
\bibitem [{\citenamefont {{Planck Collaboration}}\ \emph
  {et~al.}(2016{\natexlab{b}})\citenamefont {{Planck Collaboration}},
  \citenamefont {{Ade}}, \citenamefont {{Aghanim}}, \citenamefont {{Arnaud}},
  \citenamefont {{Arroja}}, \citenamefont {{Ashdown}}, \citenamefont
  {{Aumont}}, \citenamefont {{Baccigalupi}}, \citenamefont {{Ballardini}},
  \citenamefont {{Banday}},\ and\ \citenamefont
  {et~al.}}]{2016A&A...594A..17P}%
  \BibitemOpen
  \bibfield  {author} {\bibinfo {author} {\bibnamefont {{Planck
  Collaboration}}}, \bibinfo {author} {\bibfnamefont {P.~A.~R.}\ \bibnamefont
  {{Ade}}}, \bibinfo {author} {\bibfnamefont {N.}~\bibnamefont {{Aghanim}}},
  \bibinfo {author} {\bibfnamefont {M.}~\bibnamefont {{Arnaud}}}, \bibinfo
  {author} {\bibfnamefont {F.}~\bibnamefont {{Arroja}}}, \bibinfo {author}
  {\bibfnamefont {M.}~\bibnamefont {{Ashdown}}}, \bibinfo {author}
  {\bibfnamefont {J.}~\bibnamefont {{Aumont}}}, \bibinfo {author}
  {\bibfnamefont {C.}~\bibnamefont {{Baccigalupi}}}, \bibinfo {author}
  {\bibfnamefont {M.}~\bibnamefont {{Ballardini}}}, \bibinfo {author}
  {\bibfnamefont {A.~J.}\ \bibnamefont {{Banday}}}, \ and\ \bibinfo {author}
  {\bibnamefont {et~al.}},\ }\href {\doibase 10.1051/0004-6361/201525836}
  {\bibfield  {journal} {\bibinfo  {journal} {\aap}\ }\textbf {\bibinfo
  {volume} {594}},\ \bibinfo {eid} {A17} (\bibinfo {year}
  {2016}{\natexlab{b}})},\ \Eprint {http://arxiv.org/abs/1502.01592}
  {arXiv:1502.01592} \BibitemShut {NoStop}%
\bibitem [{\citenamefont {{Martin}}(2016)}]{2016ASSP...45...41M}%
  \BibitemOpen
  \bibfield  {author} {\bibinfo {author} {\bibfnamefont {J.}~\bibnamefont
  {{Martin}}},\ }\href {\doibase 10.1007/978-3-319-44769-8_2} {\bibfield
  {journal} {\bibinfo  {journal} {Astrophysics and Space Science Proceedings}\
  }\textbf {\bibinfo {volume} {45}},\ \bibinfo {pages} {41} (\bibinfo {year}
  {2016})},\ \Eprint {http://arxiv.org/abs/1502.05733} {arXiv:1502.05733}
  \BibitemShut {NoStop}%
\bibitem [{\citenamefont {{Abbott}}\ and\ \citenamefont
  {{Wise}}(1984)}]{1984NuPhB.244..541A}%
  \BibitemOpen
  \bibfield  {author} {\bibinfo {author} {\bibfnamefont {L.~F.}\ \bibnamefont
  {{Abbott}}}\ and\ \bibinfo {author} {\bibfnamefont {M.~B.}\ \bibnamefont
  {{Wise}}},\ }\href {\doibase 10.1016/0550-3213(84)90329-8} {\bibfield
  {journal} {\bibinfo  {journal} {Nuclear Physics B}\ }\textbf {\bibinfo
  {volume} {244}},\ \bibinfo {pages} {541} (\bibinfo {year}
  {1984})}\BibitemShut {NoStop}%
\bibitem [{\citenamefont {{Rubakov}}\ \emph {et~al.}(1982)\citenamefont
  {{Rubakov}}, \citenamefont {{Sazhin}},\ and\ \citenamefont
  {{Veryaskin}}}]{1982PhLB..115..189R}%
  \BibitemOpen
  \bibfield  {author} {\bibinfo {author} {\bibfnamefont {V.~A.}\ \bibnamefont
  {{Rubakov}}}, \bibinfo {author} {\bibfnamefont {M.~V.}\ \bibnamefont
  {{Sazhin}}}, \ and\ \bibinfo {author} {\bibfnamefont {A.~V.}\ \bibnamefont
  {{Veryaskin}}},\ }\href {\doibase 10.1016/0370-2693(82)90641-4} {\bibfield
  {journal} {\bibinfo  {journal} {Physics Letters B}\ }\textbf {\bibinfo
  {volume} {115}},\ \bibinfo {pages} {189} (\bibinfo {year}
  {1982})}\BibitemShut {NoStop}%
\bibitem [{\citenamefont {{Fabbri}}\ and\ \citenamefont
  {{Pollock}}(1983)}]{1983PhLB..125..445F}%
  \BibitemOpen
  \bibfield  {author} {\bibinfo {author} {\bibfnamefont {R.}~\bibnamefont
  {{Fabbri}}}\ and\ \bibinfo {author} {\bibfnamefont {M.~D.}\ \bibnamefont
  {{Pollock}}},\ }\href {\doibase 10.1016/0370-2693(83)91322-9} {\bibfield
  {journal} {\bibinfo  {journal} {Physics Letters B}\ }\textbf {\bibinfo
  {volume} {125}},\ \bibinfo {pages} {445} (\bibinfo {year}
  {1983})}\BibitemShut {NoStop}%
\bibitem [{\citenamefont {{Starobinski{\v i}}}(1979)}]{1979JETPL..30..682S}%
  \BibitemOpen
  \bibfield  {author} {\bibinfo {author} {\bibfnamefont {A.~A.}\ \bibnamefont
  {{Starobinski{\v i}}}},\ }\href@noop {} {\bibfield  {journal} {\bibinfo
  {journal} {Soviet Journal of Experimental and Theoretical Physics Letters}\
  }\textbf {\bibinfo {volume} {30}},\ \bibinfo {pages} {682} (\bibinfo {year}
  {1979})}\BibitemShut {NoStop}%
\bibitem [{\citenamefont {{Kamionkowski}}\ \emph
  {et~al.}(1997{\natexlab{a}})\citenamefont {{Kamionkowski}}, \citenamefont
  {{Kosowsky}},\ and\ \citenamefont {{Stebbins}}}]{1997PhRvD..55.7368K}%
  \BibitemOpen
  \bibfield  {author} {\bibinfo {author} {\bibfnamefont {M.}~\bibnamefont
  {{Kamionkowski}}}, \bibinfo {author} {\bibfnamefont {A.}~\bibnamefont
  {{Kosowsky}}}, \ and\ \bibinfo {author} {\bibfnamefont {A.}~\bibnamefont
  {{Stebbins}}},\ }\href {\doibase 10.1103/PhysRevD.55.7368} {\bibfield
  {journal} {\bibinfo  {journal} {\prd}\ }\textbf {\bibinfo {volume} {55}},\
  \bibinfo {pages} {7368} (\bibinfo {year} {1997}{\natexlab{a}})},\ \Eprint
  {http://arxiv.org/abs/astro-ph/9611125} {astro-ph/9611125} \BibitemShut
  {NoStop}%
\bibitem [{\citenamefont {{Kamionkowski}}\ \emph
  {et~al.}(1997{\natexlab{b}})\citenamefont {{Kamionkowski}}, \citenamefont
  {{Kosowsky}},\ and\ \citenamefont {{Stebbins}}}]{1997PhRvL..78.2058K}%
  \BibitemOpen
  \bibfield  {author} {\bibinfo {author} {\bibfnamefont {M.}~\bibnamefont
  {{Kamionkowski}}}, \bibinfo {author} {\bibfnamefont {A.}~\bibnamefont
  {{Kosowsky}}}, \ and\ \bibinfo {author} {\bibfnamefont {A.}~\bibnamefont
  {{Stebbins}}},\ }\href {\doibase 10.1103/PhysRevLett.78.2058} {\bibfield
  {journal} {\bibinfo  {journal} {Physical Review Letters}\ }\textbf {\bibinfo
  {volume} {78}},\ \bibinfo {pages} {2058} (\bibinfo {year}
  {1997}{\natexlab{b}})},\ \Eprint {http://arxiv.org/abs/astro-ph/9609132}
  {astro-ph/9609132} \BibitemShut {NoStop}%
\bibitem [{\citenamefont {{Seljak}}(1997)}]{1997ApJ...482....6S}%
  \BibitemOpen
  \bibfield  {author} {\bibinfo {author} {\bibfnamefont {U.}~\bibnamefont
  {{Seljak}}},\ }\href {\doibase 10.1086/304123} {\bibfield  {journal}
  {\bibinfo  {journal} {\apj}\ }\textbf {\bibinfo {volume} {482}},\ \bibinfo
  {pages} {6} (\bibinfo {year} {1997})},\ \Eprint
  {http://arxiv.org/abs/astro-ph/9608131} {astro-ph/9608131} \BibitemShut
  {NoStop}%
\bibitem [{\citenamefont {{Seljak}}\ and\ \citenamefont
  {{Zaldarriaga}}(1997)}]{1997PhRvL..78.2054S}%
  \BibitemOpen
  \bibfield  {author} {\bibinfo {author} {\bibfnamefont {U.}~\bibnamefont
  {{Seljak}}}\ and\ \bibinfo {author} {\bibfnamefont {M.}~\bibnamefont
  {{Zaldarriaga}}},\ }\href {\doibase 10.1103/PhysRevLett.78.2054} {\bibfield
  {journal} {\bibinfo  {journal} {Physical Review Letters}\ }\textbf {\bibinfo
  {volume} {78}},\ \bibinfo {pages} {2054} (\bibinfo {year} {1997})},\ \Eprint
  {http://arxiv.org/abs/astro-ph/9609169} {astro-ph/9609169} \BibitemShut
  {NoStop}%
\bibitem [{\citenamefont {{Zaldarriaga}}\ and\ \citenamefont
  {{Seljak}}(1997)}]{1997PhRvD..55.1830Z}%
  \BibitemOpen
  \bibfield  {author} {\bibinfo {author} {\bibfnamefont {M.}~\bibnamefont
  {{Zaldarriaga}}}\ and\ \bibinfo {author} {\bibfnamefont {U.}~\bibnamefont
  {{Seljak}}},\ }\href {\doibase 10.1103/PhysRevD.55.1830} {\bibfield
  {journal} {\bibinfo  {journal} {\prd}\ }\textbf {\bibinfo {volume} {55}},\
  \bibinfo {pages} {1830} (\bibinfo {year} {1997})},\ \Eprint
  {http://arxiv.org/abs/astro-ph/9609170} {astro-ph/9609170} \BibitemShut
  {NoStop}%
\bibitem [{\citenamefont {{Essinger-Hileman}}\ \emph
  {et~al.}(2010)\citenamefont {{Essinger-Hileman}}, \citenamefont {{Appel}},
  \citenamefont {{Beall}}, \citenamefont {{Cho}}, \citenamefont {{Fowler}},
  \citenamefont {{Halpern}}, \citenamefont {{Hasselfield}}, \citenamefont
  {{Irwin}}, \citenamefont {{Marriage}}, \citenamefont {{Niemack}},
  \citenamefont {{Page}}, \citenamefont {{Parker}}, \citenamefont {{Pufu}},
  \citenamefont {{Staggs}}, \citenamefont {{Stryzak}}, \citenamefont
  {{Visnjic}}, \citenamefont {{Yoon}},\ and\ \citenamefont
  {{Zhao}}}]{2010arXiv1008.3915E}%
  \BibitemOpen
  \bibfield  {author} {\bibinfo {author} {\bibfnamefont {T.}~\bibnamefont
  {{Essinger-Hileman}}}, \bibinfo {author} {\bibfnamefont {J.~W.}\ \bibnamefont
  {{Appel}}}, \bibinfo {author} {\bibfnamefont {J.~A.}\ \bibnamefont
  {{Beall}}}, \bibinfo {author} {\bibfnamefont {H.~M.}\ \bibnamefont {{Cho}}},
  \bibinfo {author} {\bibfnamefont {J.}~\bibnamefont {{Fowler}}}, \bibinfo
  {author} {\bibfnamefont {M.}~\bibnamefont {{Halpern}}}, \bibinfo {author}
  {\bibfnamefont {M.}~\bibnamefont {{Hasselfield}}}, \bibinfo {author}
  {\bibfnamefont {K.~D.}\ \bibnamefont {{Irwin}}}, \bibinfo {author}
  {\bibfnamefont {T.~A.}\ \bibnamefont {{Marriage}}}, \bibinfo {author}
  {\bibfnamefont {M.~D.}\ \bibnamefont {{Niemack}}}, \bibinfo {author}
  {\bibfnamefont {L.}~\bibnamefont {{Page}}}, \bibinfo {author} {\bibfnamefont
  {L.~P.}\ \bibnamefont {{Parker}}}, \bibinfo {author} {\bibfnamefont
  {S.}~\bibnamefont {{Pufu}}}, \bibinfo {author} {\bibfnamefont {S.~T.}\
  \bibnamefont {{Staggs}}}, \bibinfo {author} {\bibfnamefont {O.}~\bibnamefont
  {{Stryzak}}}, \bibinfo {author} {\bibfnamefont {C.}~\bibnamefont
  {{Visnjic}}}, \bibinfo {author} {\bibfnamefont {K.~W.}\ \bibnamefont
  {{Yoon}}}, \ and\ \bibinfo {author} {\bibfnamefont {Y.}~\bibnamefont
  {{Zhao}}},\ }\href@noop {} {\bibfield  {journal} {\bibinfo  {journal} {ArXiv
  e-prints}\ } (\bibinfo {year} {2010})},\ \Eprint
  {http://arxiv.org/abs/1008.3915} {arXiv:1008.3915 [astro-ph.IM]} \BibitemShut
  {NoStop}%
\bibitem [{\citenamefont {{Naess}}\ \emph {et~al.}(2014)\citenamefont
  {{Naess}}, \citenamefont {{Hasselfield}}, \citenamefont {{McMahon}},
  \citenamefont {{Niemack}}, \citenamefont {{Addison}}, \citenamefont {{Ade}},
  \citenamefont {{Allison}}, \citenamefont {{Amiri}}, \citenamefont
  {{Battaglia}}, \citenamefont {{Beall}}, \citenamefont {{de Bernardis}},
  \citenamefont {{Bond}}, \citenamefont {{Britton}}, \citenamefont
  {{Calabrese}}, \citenamefont {{Cho}}, \citenamefont {{Coughlin}},
  \citenamefont {{Crichton}}, \citenamefont {{Das}}, \citenamefont {{Datta}},
  \citenamefont {{Devlin}}, \citenamefont {{Dicker}}, \citenamefont
  {{Dunkley}}, \citenamefont {{D{\"u}nner}}, \citenamefont {{Fowler}},
  \citenamefont {{Fox}}, \citenamefont {{Gallardo}}, \citenamefont {{Grace}},
  \citenamefont {{Gralla}}, \citenamefont {{Hajian}}, \citenamefont
  {{Halpern}}, \citenamefont {{Henderson}}, \citenamefont {{Hill}},
  \citenamefont {{Hilton}}, \citenamefont {{Hilton}}, \citenamefont {{Hincks}},
  \citenamefont {{Hlozek}}, \citenamefont {{Ho}}, \citenamefont {{Hubmayr}},
  \citenamefont {{Huffenberger}}, \citenamefont {{Hughes}}, \citenamefont
  {{Infante}}, \citenamefont {{Irwin}}, \citenamefont {{Jackson}},
  \citenamefont {{Muya Kasanda}}, \citenamefont {{Klein}}, \citenamefont
  {{Koopman}}, \citenamefont {{Kosowsky}}, \citenamefont {{Li}}, \citenamefont
  {{Louis}}, \citenamefont {{Lungu}}, \citenamefont {{Madhavacheril}},
  \citenamefont {{Marriage}}, \citenamefont {{Maurin}}, \citenamefont
  {{Menanteau}}, \citenamefont {{Moodley}}, \citenamefont {{Munson}},
  \citenamefont {{Newburgh}}, \citenamefont {{Nibarger}}, \citenamefont
  {{Nolta}}, \citenamefont {{Page}}, \citenamefont {{Pappas}}, \citenamefont
  {{Partridge}}, \citenamefont {{Rojas}}, \citenamefont {{Schmitt}},
  \citenamefont {{Sehgal}}, \citenamefont {{Sherwin}}, \citenamefont
  {{Sievers}}, \citenamefont {{Simon}}, \citenamefont {{Spergel}},
  \citenamefont {{Staggs}}, \citenamefont {{Switzer}}, \citenamefont
  {{Thornton}}, \citenamefont {{Trac}}, \citenamefont {{Tucker}}, \citenamefont
  {{Uehara}}, \citenamefont {{Van Engelen}}, \citenamefont {{Ward}},\ and\
  \citenamefont {{Wollack}}}]{2014JCAP...10..007N}%
  \BibitemOpen
  \bibfield  {author} {\bibinfo {author} {\bibfnamefont {S.}~\bibnamefont
  {{Naess}}}, \bibinfo {author} {\bibfnamefont {M.}~\bibnamefont
  {{Hasselfield}}}, \bibinfo {author} {\bibfnamefont {J.}~\bibnamefont
  {{McMahon}}}, \bibinfo {author} {\bibfnamefont {M.~D.}\ \bibnamefont
  {{Niemack}}}, \bibinfo {author} {\bibfnamefont {G.~E.}\ \bibnamefont
  {{Addison}}}, \bibinfo {author} {\bibfnamefont {P.~A.~R.}\ \bibnamefont
  {{Ade}}}, \bibinfo {author} {\bibfnamefont {R.}~\bibnamefont {{Allison}}},
  \bibinfo {author} {\bibfnamefont {M.}~\bibnamefont {{Amiri}}}, \bibinfo
  {author} {\bibfnamefont {N.}~\bibnamefont {{Battaglia}}}, \bibinfo {author}
  {\bibfnamefont {J.~A.}\ \bibnamefont {{Beall}}}, \bibinfo {author}
  {\bibfnamefont {F.}~\bibnamefont {{de Bernardis}}}, \bibinfo {author}
  {\bibfnamefont {J.~R.}\ \bibnamefont {{Bond}}}, \bibinfo {author}
  {\bibfnamefont {J.}~\bibnamefont {{Britton}}}, \bibinfo {author}
  {\bibfnamefont {E.}~\bibnamefont {{Calabrese}}}, \bibinfo {author}
  {\bibfnamefont {H.-m.}\ \bibnamefont {{Cho}}}, \bibinfo {author}
  {\bibfnamefont {K.}~\bibnamefont {{Coughlin}}}, \bibinfo {author}
  {\bibfnamefont {D.}~\bibnamefont {{Crichton}}}, \bibinfo {author}
  {\bibfnamefont {S.}~\bibnamefont {{Das}}}, \bibinfo {author} {\bibfnamefont
  {R.}~\bibnamefont {{Datta}}}, \bibinfo {author} {\bibfnamefont {M.~J.}\
  \bibnamefont {{Devlin}}}, \bibinfo {author} {\bibfnamefont {S.~R.}\
  \bibnamefont {{Dicker}}}, \bibinfo {author} {\bibfnamefont {J.}~\bibnamefont
  {{Dunkley}}}, \bibinfo {author} {\bibfnamefont {R.}~\bibnamefont
  {{D{\"u}nner}}}, \bibinfo {author} {\bibfnamefont {J.~W.}\ \bibnamefont
  {{Fowler}}}, \bibinfo {author} {\bibfnamefont {A.~E.}\ \bibnamefont {{Fox}}},
  \bibinfo {author} {\bibfnamefont {P.}~\bibnamefont {{Gallardo}}}, \bibinfo
  {author} {\bibfnamefont {E.}~\bibnamefont {{Grace}}}, \bibinfo {author}
  {\bibfnamefont {M.}~\bibnamefont {{Gralla}}}, \bibinfo {author}
  {\bibfnamefont {A.}~\bibnamefont {{Hajian}}}, \bibinfo {author}
  {\bibfnamefont {M.}~\bibnamefont {{Halpern}}}, \bibinfo {author}
  {\bibfnamefont {S.}~\bibnamefont {{Henderson}}}, \bibinfo {author}
  {\bibfnamefont {J.~C.}\ \bibnamefont {{Hill}}}, \bibinfo {author}
  {\bibfnamefont {G.~C.}\ \bibnamefont {{Hilton}}}, \bibinfo {author}
  {\bibfnamefont {M.}~\bibnamefont {{Hilton}}}, \bibinfo {author}
  {\bibfnamefont {A.~D.}\ \bibnamefont {{Hincks}}}, \bibinfo {author}
  {\bibfnamefont {R.}~\bibnamefont {{Hlozek}}}, \bibinfo {author}
  {\bibfnamefont {P.}~\bibnamefont {{Ho}}}, \bibinfo {author} {\bibfnamefont
  {J.}~\bibnamefont {{Hubmayr}}}, \bibinfo {author} {\bibfnamefont {K.~M.}\
  \bibnamefont {{Huffenberger}}}, \bibinfo {author} {\bibfnamefont {J.~P.}\
  \bibnamefont {{Hughes}}}, \bibinfo {author} {\bibfnamefont {L.}~\bibnamefont
  {{Infante}}}, \bibinfo {author} {\bibfnamefont {K.}~\bibnamefont {{Irwin}}},
  \bibinfo {author} {\bibfnamefont {R.}~\bibnamefont {{Jackson}}}, \bibinfo
  {author} {\bibfnamefont {S.}~\bibnamefont {{Muya Kasanda}}}, \bibinfo
  {author} {\bibfnamefont {J.}~\bibnamefont {{Klein}}}, \bibinfo {author}
  {\bibfnamefont {B.}~\bibnamefont {{Koopman}}}, \bibinfo {author}
  {\bibfnamefont {A.}~\bibnamefont {{Kosowsky}}}, \bibinfo {author}
  {\bibfnamefont {D.}~\bibnamefont {{Li}}}, \bibinfo {author} {\bibfnamefont
  {T.}~\bibnamefont {{Louis}}}, \bibinfo {author} {\bibfnamefont
  {M.}~\bibnamefont {{Lungu}}}, \bibinfo {author} {\bibfnamefont
  {M.}~\bibnamefont {{Madhavacheril}}}, \bibinfo {author} {\bibfnamefont
  {T.~A.}\ \bibnamefont {{Marriage}}}, \bibinfo {author} {\bibfnamefont
  {L.}~\bibnamefont {{Maurin}}}, \bibinfo {author} {\bibfnamefont
  {F.}~\bibnamefont {{Menanteau}}}, \bibinfo {author} {\bibfnamefont
  {K.}~\bibnamefont {{Moodley}}}, \bibinfo {author} {\bibfnamefont
  {C.}~\bibnamefont {{Munson}}}, \bibinfo {author} {\bibfnamefont
  {L.}~\bibnamefont {{Newburgh}}}, \bibinfo {author} {\bibfnamefont
  {J.}~\bibnamefont {{Nibarger}}}, \bibinfo {author} {\bibfnamefont {M.~R.}\
  \bibnamefont {{Nolta}}}, \bibinfo {author} {\bibfnamefont {L.~A.}\
  \bibnamefont {{Page}}}, \bibinfo {author} {\bibfnamefont {C.}~\bibnamefont
  {{Pappas}}}, \bibinfo {author} {\bibfnamefont {B.}~\bibnamefont
  {{Partridge}}}, \bibinfo {author} {\bibfnamefont {F.}~\bibnamefont
  {{Rojas}}}, \bibinfo {author} {\bibfnamefont {B.~L.}\ \bibnamefont
  {{Schmitt}}}, \bibinfo {author} {\bibfnamefont {N.}~\bibnamefont {{Sehgal}}},
  \bibinfo {author} {\bibfnamefont {B.~D.}\ \bibnamefont {{Sherwin}}}, \bibinfo
  {author} {\bibfnamefont {J.}~\bibnamefont {{Sievers}}}, \bibinfo {author}
  {\bibfnamefont {S.}~\bibnamefont {{Simon}}}, \bibinfo {author} {\bibfnamefont
  {D.~N.}\ \bibnamefont {{Spergel}}}, \bibinfo {author} {\bibfnamefont {S.~T.}\
  \bibnamefont {{Staggs}}}, \bibinfo {author} {\bibfnamefont {E.~R.}\
  \bibnamefont {{Switzer}}}, \bibinfo {author} {\bibfnamefont {R.}~\bibnamefont
  {{Thornton}}}, \bibinfo {author} {\bibfnamefont {H.}~\bibnamefont {{Trac}}},
  \bibinfo {author} {\bibfnamefont {C.}~\bibnamefont {{Tucker}}}, \bibinfo
  {author} {\bibfnamefont {M.}~\bibnamefont {{Uehara}}}, \bibinfo {author}
  {\bibfnamefont {A.}~\bibnamefont {{Van Engelen}}}, \bibinfo {author}
  {\bibfnamefont {J.~T.}\ \bibnamefont {{Ward}}}, \ and\ \bibinfo {author}
  {\bibfnamefont {E.~J.}\ \bibnamefont {{Wollack}}},\ }\href {\doibase
  10.1088/1475-7516/2014/10/007} {\bibfield  {journal} {\bibinfo  {journal}
  {\jcap}\ }\textbf {\bibinfo {volume} {10}},\ \bibinfo {eid} {007} (\bibinfo
  {year} {2014})},\ \Eprint {http://arxiv.org/abs/1405.5524} {arXiv:1405.5524}
  \BibitemShut {NoStop}%
\bibitem [{\citenamefont {{BICEP2 Collaboration}}\ \emph
  {et~al.}(2014)\citenamefont {{BICEP2 Collaboration}}, \citenamefont {{Ade}},
  \citenamefont {{Aikin}}, \citenamefont {{Amiri}}, \citenamefont {{Barkats}},
  \citenamefont {{Benton}}, \citenamefont {{Bischoff}}, \citenamefont {{Bock}},
  \citenamefont {{Brevik}}, \citenamefont {{Buder}}, \citenamefont {{Bullock}},
  \citenamefont {{Davis}}, \citenamefont {{Day}}, \citenamefont {{Dowell}},
  \citenamefont {{Duband}}, \citenamefont {{Filippini}}, \citenamefont
  {{Fliescher}}, \citenamefont {{Golwala}}, \citenamefont {{Halpern}},
  \citenamefont {{Hasselfield}}, \citenamefont {{Hildebrandt}}, \citenamefont
  {{Hilton}}, \citenamefont {{Irwin}}, \citenamefont {{Karkare}}, \citenamefont
  {{Kaufman}}, \citenamefont {{Keating}}, \citenamefont {{Kernasovskiy}},
  \citenamefont {{Kovac}}, \citenamefont {{Kuo}}, \citenamefont {{Leitch}},
  \citenamefont {{Llombart}}, \citenamefont {{Lueker}}, \citenamefont
  {{Netterfield}}, \citenamefont {{Nguyen}}, \citenamefont {{O'Brient}},
  \citenamefont {{Ogburn}}, \citenamefont {{Orlando}}, \citenamefont {{Pryke}},
  \citenamefont {{Reintsema}}, \citenamefont {{Richter}}, \citenamefont
  {{Schwarz}}, \citenamefont {{Sheehy}}, \citenamefont {{Staniszewski}},
  \citenamefont {{Story}}, \citenamefont {{Sudiwala}}, \citenamefont {{Teply}},
  \citenamefont {{Tolan}}, \citenamefont {{Turner}}, \citenamefont {{Vieregg}},
  \citenamefont {{Wilson}}, \citenamefont {{Wong}},\ and\ \citenamefont
  {{Yoon}}}]{2014ApJ...792...62B}%
  \BibitemOpen
  \bibfield  {author} {\bibinfo {author} {\bibnamefont {{BICEP2
  Collaboration}}}, \bibinfo {author} {\bibfnamefont {P.~A.~R.}\ \bibnamefont
  {{Ade}}}, \bibinfo {author} {\bibfnamefont {R.~W.}\ \bibnamefont {{Aikin}}},
  \bibinfo {author} {\bibfnamefont {M.}~\bibnamefont {{Amiri}}}, \bibinfo
  {author} {\bibfnamefont {D.}~\bibnamefont {{Barkats}}}, \bibinfo {author}
  {\bibfnamefont {S.~J.}\ \bibnamefont {{Benton}}}, \bibinfo {author}
  {\bibfnamefont {C.~A.}\ \bibnamefont {{Bischoff}}}, \bibinfo {author}
  {\bibfnamefont {J.~J.}\ \bibnamefont {{Bock}}}, \bibinfo {author}
  {\bibfnamefont {J.~A.}\ \bibnamefont {{Brevik}}}, \bibinfo {author}
  {\bibfnamefont {I.}~\bibnamefont {{Buder}}}, \bibinfo {author} {\bibfnamefont
  {E.}~\bibnamefont {{Bullock}}}, \bibinfo {author} {\bibfnamefont
  {G.}~\bibnamefont {{Davis}}}, \bibinfo {author} {\bibfnamefont {P.~K.}\
  \bibnamefont {{Day}}}, \bibinfo {author} {\bibfnamefont {C.~D.}\ \bibnamefont
  {{Dowell}}}, \bibinfo {author} {\bibfnamefont {L.}~\bibnamefont {{Duband}}},
  \bibinfo {author} {\bibfnamefont {J.~P.}\ \bibnamefont {{Filippini}}},
  \bibinfo {author} {\bibfnamefont {S.}~\bibnamefont {{Fliescher}}}, \bibinfo
  {author} {\bibfnamefont {S.~R.}\ \bibnamefont {{Golwala}}}, \bibinfo {author}
  {\bibfnamefont {M.}~\bibnamefont {{Halpern}}}, \bibinfo {author}
  {\bibfnamefont {M.}~\bibnamefont {{Hasselfield}}}, \bibinfo {author}
  {\bibfnamefont {S.~R.}\ \bibnamefont {{Hildebrandt}}}, \bibinfo {author}
  {\bibfnamefont {G.~C.}\ \bibnamefont {{Hilton}}}, \bibinfo {author}
  {\bibfnamefont {K.~D.}\ \bibnamefont {{Irwin}}}, \bibinfo {author}
  {\bibfnamefont {K.~S.}\ \bibnamefont {{Karkare}}}, \bibinfo {author}
  {\bibfnamefont {J.~P.}\ \bibnamefont {{Kaufman}}}, \bibinfo {author}
  {\bibfnamefont {B.~G.}\ \bibnamefont {{Keating}}}, \bibinfo {author}
  {\bibfnamefont {S.~A.}\ \bibnamefont {{Kernasovskiy}}}, \bibinfo {author}
  {\bibfnamefont {J.~M.}\ \bibnamefont {{Kovac}}}, \bibinfo {author}
  {\bibfnamefont {C.~L.}\ \bibnamefont {{Kuo}}}, \bibinfo {author}
  {\bibfnamefont {E.~M.}\ \bibnamefont {{Leitch}}}, \bibinfo {author}
  {\bibfnamefont {N.}~\bibnamefont {{Llombart}}}, \bibinfo {author}
  {\bibfnamefont {M.}~\bibnamefont {{Lueker}}}, \bibinfo {author}
  {\bibfnamefont {C.~B.}\ \bibnamefont {{Netterfield}}}, \bibinfo {author}
  {\bibfnamefont {H.~T.}\ \bibnamefont {{Nguyen}}}, \bibinfo {author}
  {\bibfnamefont {R.}~\bibnamefont {{O'Brient}}}, \bibinfo {author}
  {\bibfnamefont {R.~W.}\ \bibnamefont {{Ogburn}}, \bibfnamefont {IV}},
  \bibinfo {author} {\bibfnamefont {A.}~\bibnamefont {{Orlando}}}, \bibinfo
  {author} {\bibfnamefont {C.}~\bibnamefont {{Pryke}}}, \bibinfo {author}
  {\bibfnamefont {C.~D.}\ \bibnamefont {{Reintsema}}}, \bibinfo {author}
  {\bibfnamefont {S.}~\bibnamefont {{Richter}}}, \bibinfo {author}
  {\bibfnamefont {R.}~\bibnamefont {{Schwarz}}}, \bibinfo {author}
  {\bibfnamefont {C.~D.}\ \bibnamefont {{Sheehy}}}, \bibinfo {author}
  {\bibfnamefont {Z.~K.}\ \bibnamefont {{Staniszewski}}}, \bibinfo {author}
  {\bibfnamefont {K.~T.}\ \bibnamefont {{Story}}}, \bibinfo {author}
  {\bibfnamefont {R.~V.}\ \bibnamefont {{Sudiwala}}}, \bibinfo {author}
  {\bibfnamefont {G.~P.}\ \bibnamefont {{Teply}}}, \bibinfo {author}
  {\bibfnamefont {J.~E.}\ \bibnamefont {{Tolan}}}, \bibinfo {author}
  {\bibfnamefont {A.~D.}\ \bibnamefont {{Turner}}}, \bibinfo {author}
  {\bibfnamefont {A.~G.}\ \bibnamefont {{Vieregg}}}, \bibinfo {author}
  {\bibfnamefont {P.}~\bibnamefont {{Wilson}}}, \bibinfo {author}
  {\bibfnamefont {C.~L.}\ \bibnamefont {{Wong}}}, \ and\ \bibinfo {author}
  {\bibfnamefont {K.~W.}\ \bibnamefont {{Yoon}}},\ }\href {\doibase
  10.1088/0004-637X/792/1/62} {\bibfield  {journal} {\bibinfo  {journal}
  {\apj}\ }\textbf {\bibinfo {volume} {792}},\ \bibinfo {eid} {62} (\bibinfo
  {year} {2014})},\ \Eprint {http://arxiv.org/abs/1403.4302} {arXiv:1403.4302}
  \BibitemShut {NoStop}%
\bibitem [{\citenamefont {{BICEP2 and Keck Array Collaborations}}\ \emph
  {et~al.}(2015)\citenamefont {{BICEP2 and Keck Array Collaborations}},
  \citenamefont {{Ade}}, \citenamefont {{Ahmed}}, \citenamefont {{Aikin}},
  \citenamefont {{Alexander}}, \citenamefont {{Barkats}}, \citenamefont
  {{Benton}}, \citenamefont {{Bischoff}}, \citenamefont {{Bock}}, \citenamefont
  {{Brevik}}, \citenamefont {{Buder}}, \citenamefont {{Bullock}}, \citenamefont
  {{Buza}}, \citenamefont {{Connors}}, \citenamefont {{Crill}}, \citenamefont
  {{Dowell}}, \citenamefont {{Dvorkin}}, \citenamefont {{Duband}},
  \citenamefont {{Filippini}}, \citenamefont {{Fliescher}}, \citenamefont
  {{Golwala}}, \citenamefont {{Halpern}}, \citenamefont {{Harrison}},
  \citenamefont {{Hasselfield}}, \citenamefont {{Hildebrandt}}, \citenamefont
  {{Hilton}}, \citenamefont {{Hristov}}, \citenamefont {{Hui}}, \citenamefont
  {{Irwin}}, \citenamefont {{Karkare}}, \citenamefont {{Kaufman}},
  \citenamefont {{Keating}}, \citenamefont {{Kefeli}}, \citenamefont
  {{Kernasovskiy}}, \citenamefont {{Kovac}}, \citenamefont {{Kuo}},
  \citenamefont {{Leitch}}, \citenamefont {{Lueker}}, \citenamefont {{Mason}},
  \citenamefont {{Megerian}}, \citenamefont {{Netterfield}}, \citenamefont
  {{Nguyen}}, \citenamefont {{O'Brient}}, \citenamefont {{Ogburn}},
  \citenamefont {{Orlando}}, \citenamefont {{Pryke}}, \citenamefont
  {{Reintsema}}, \citenamefont {{Richter}}, \citenamefont {{Schwarz}},
  \citenamefont {{Sheehy}}, \citenamefont {{Staniszewski}}, \citenamefont
  {{Sudiwala}}, \citenamefont {{Teply}}, \citenamefont {{Thompson}},
  \citenamefont {{Tolan}}, \citenamefont {{Turner}}, \citenamefont {{Vieregg}},
  \citenamefont {{Weber}}, \citenamefont {{Willmert}}, \citenamefont {{Wong}},\
  and\ \citenamefont {{Yoon}}}]{2015ApJ...811..126B}%
  \BibitemOpen
  \bibfield  {author} {\bibinfo {author} {\bibnamefont {{BICEP2 and Keck Array
  Collaborations}}}, \bibinfo {author} {\bibfnamefont {P.~A.~R.}\ \bibnamefont
  {{Ade}}}, \bibinfo {author} {\bibfnamefont {Z.}~\bibnamefont {{Ahmed}}},
  \bibinfo {author} {\bibfnamefont {R.~W.}\ \bibnamefont {{Aikin}}}, \bibinfo
  {author} {\bibfnamefont {K.~D.}\ \bibnamefont {{Alexander}}}, \bibinfo
  {author} {\bibfnamefont {D.}~\bibnamefont {{Barkats}}}, \bibinfo {author}
  {\bibfnamefont {S.~J.}\ \bibnamefont {{Benton}}}, \bibinfo {author}
  {\bibfnamefont {C.~A.}\ \bibnamefont {{Bischoff}}}, \bibinfo {author}
  {\bibfnamefont {J.~J.}\ \bibnamefont {{Bock}}}, \bibinfo {author}
  {\bibfnamefont {J.~A.}\ \bibnamefont {{Brevik}}}, \bibinfo {author}
  {\bibfnamefont {I.}~\bibnamefont {{Buder}}}, \bibinfo {author} {\bibfnamefont
  {E.}~\bibnamefont {{Bullock}}}, \bibinfo {author} {\bibfnamefont
  {V.}~\bibnamefont {{Buza}}}, \bibinfo {author} {\bibfnamefont
  {J.}~\bibnamefont {{Connors}}}, \bibinfo {author} {\bibfnamefont {B.~P.}\
  \bibnamefont {{Crill}}}, \bibinfo {author} {\bibfnamefont {C.~D.}\
  \bibnamefont {{Dowell}}}, \bibinfo {author} {\bibfnamefont {C.}~\bibnamefont
  {{Dvorkin}}}, \bibinfo {author} {\bibfnamefont {L.}~\bibnamefont {{Duband}}},
  \bibinfo {author} {\bibfnamefont {J.~P.}\ \bibnamefont {{Filippini}}},
  \bibinfo {author} {\bibfnamefont {S.}~\bibnamefont {{Fliescher}}}, \bibinfo
  {author} {\bibfnamefont {S.~R.}\ \bibnamefont {{Golwala}}}, \bibinfo {author}
  {\bibfnamefont {M.}~\bibnamefont {{Halpern}}}, \bibinfo {author}
  {\bibfnamefont {S.}~\bibnamefont {{Harrison}}}, \bibinfo {author}
  {\bibfnamefont {M.}~\bibnamefont {{Hasselfield}}}, \bibinfo {author}
  {\bibfnamefont {S.~R.}\ \bibnamefont {{Hildebrandt}}}, \bibinfo {author}
  {\bibfnamefont {G.~C.}\ \bibnamefont {{Hilton}}}, \bibinfo {author}
  {\bibfnamefont {V.~V.}\ \bibnamefont {{Hristov}}}, \bibinfo {author}
  {\bibfnamefont {H.}~\bibnamefont {{Hui}}}, \bibinfo {author} {\bibfnamefont
  {K.~D.}\ \bibnamefont {{Irwin}}}, \bibinfo {author} {\bibfnamefont {K.~S.}\
  \bibnamefont {{Karkare}}}, \bibinfo {author} {\bibfnamefont {J.~P.}\
  \bibnamefont {{Kaufman}}}, \bibinfo {author} {\bibfnamefont {B.~G.}\
  \bibnamefont {{Keating}}}, \bibinfo {author} {\bibfnamefont {S.}~\bibnamefont
  {{Kefeli}}}, \bibinfo {author} {\bibfnamefont {S.~A.}\ \bibnamefont
  {{Kernasovskiy}}}, \bibinfo {author} {\bibfnamefont {J.~M.}\ \bibnamefont
  {{Kovac}}}, \bibinfo {author} {\bibfnamefont {C.~L.}\ \bibnamefont {{Kuo}}},
  \bibinfo {author} {\bibfnamefont {E.~M.}\ \bibnamefont {{Leitch}}}, \bibinfo
  {author} {\bibfnamefont {M.}~\bibnamefont {{Lueker}}}, \bibinfo {author}
  {\bibfnamefont {P.}~\bibnamefont {{Mason}}}, \bibinfo {author} {\bibfnamefont
  {K.~G.}\ \bibnamefont {{Megerian}}}, \bibinfo {author} {\bibfnamefont
  {C.~B.}\ \bibnamefont {{Netterfield}}}, \bibinfo {author} {\bibfnamefont
  {H.~T.}\ \bibnamefont {{Nguyen}}}, \bibinfo {author} {\bibfnamefont
  {R.}~\bibnamefont {{O'Brient}}}, \bibinfo {author} {\bibfnamefont {R.~W.}\
  \bibnamefont {{Ogburn}}, \bibfnamefont {IV}}, \bibinfo {author}
  {\bibfnamefont {A.}~\bibnamefont {{Orlando}}}, \bibinfo {author}
  {\bibfnamefont {C.}~\bibnamefont {{Pryke}}}, \bibinfo {author} {\bibfnamefont
  {C.~D.}\ \bibnamefont {{Reintsema}}}, \bibinfo {author} {\bibfnamefont
  {S.}~\bibnamefont {{Richter}}}, \bibinfo {author} {\bibfnamefont
  {R.}~\bibnamefont {{Schwarz}}}, \bibinfo {author} {\bibfnamefont {C.~D.}\
  \bibnamefont {{Sheehy}}}, \bibinfo {author} {\bibfnamefont {Z.~K.}\
  \bibnamefont {{Staniszewski}}}, \bibinfo {author} {\bibfnamefont {R.~V.}\
  \bibnamefont {{Sudiwala}}}, \bibinfo {author} {\bibfnamefont {G.~P.}\
  \bibnamefont {{Teply}}}, \bibinfo {author} {\bibfnamefont {K.~L.}\
  \bibnamefont {{Thompson}}}, \bibinfo {author} {\bibfnamefont {J.~E.}\
  \bibnamefont {{Tolan}}}, \bibinfo {author} {\bibfnamefont {A.~D.}\
  \bibnamefont {{Turner}}}, \bibinfo {author} {\bibfnamefont {A.~G.}\
  \bibnamefont {{Vieregg}}}, \bibinfo {author} {\bibfnamefont {A.~C.}\
  \bibnamefont {{Weber}}}, \bibinfo {author} {\bibfnamefont {J.}~\bibnamefont
  {{Willmert}}}, \bibinfo {author} {\bibfnamefont {C.~L.}\ \bibnamefont
  {{Wong}}}, \ and\ \bibinfo {author} {\bibfnamefont {K.~W.}\ \bibnamefont
  {{Yoon}}},\ }\href {\doibase 10.1088/0004-637X/811/2/126} {\bibfield
  {journal} {\bibinfo  {journal} {\apj}\ }\textbf {\bibinfo {volume} {811}},\
  \bibinfo {eid} {126} (\bibinfo {year} {2015})},\ \Eprint
  {http://arxiv.org/abs/1502.00643} {arXiv:1502.00643} \BibitemShut {NoStop}%
\bibitem [{\citenamefont {{Arnold}}\ \emph {et~al.}(2014)\citenamefont
  {{Arnold}}, \citenamefont {{Stebor}}, \citenamefont {{Ade}}, \citenamefont
  {{Akiba}}, \citenamefont {{Anthony}}, \citenamefont {{Atlas}}, \citenamefont
  {{Barron}}, \citenamefont {{Bender}}, \citenamefont {{Boettger}},
  \citenamefont {{Borrill}}, \citenamefont {{Chapman}}, \citenamefont
  {{Chinone}}, \citenamefont {{Cukierman}}, \citenamefont {{Dobbs}},
  \citenamefont {{Elleflot}}, \citenamefont {{Errard}}, \citenamefont
  {{Fabbian}}, \citenamefont {{Feng}}, \citenamefont {{Gilbert}}, \citenamefont
  {{Goeckner-Wald}}, \citenamefont {{Halverson}}, \citenamefont {{Hasegawa}},
  \citenamefont {{Hattori}}, \citenamefont {{Hazumi}}, \citenamefont
  {{Holzapfel}}, \citenamefont {{Hori}}, \citenamefont {{Inoue}}, \citenamefont
  {{Jaehnig}}, \citenamefont {{Jaffe}}, \citenamefont {{Katayama}},
  \citenamefont {{Keating}}, \citenamefont {{Kermish}}, \citenamefont
  {{Keskitalo}}, \citenamefont {{Kisner}}, \citenamefont {{Le Jeune}},
  \citenamefont {{Lee}}, \citenamefont {{Leitch}}, \citenamefont {{Linder}},
  \citenamefont {{Matsuda}}, \citenamefont {{Matsumura}}, \citenamefont
  {{Meng}}, \citenamefont {{Miller}}, \citenamefont {{Morii}}, \citenamefont
  {{Myers}}, \citenamefont {{Navaroli}}, \citenamefont {{Nishino}},
  \citenamefont {{Okamura}}, \citenamefont {{Paar}}, \citenamefont {{Peloton}},
  \citenamefont {{Poletti}}, \citenamefont {{Raum}}, \citenamefont {{Rebeiz}},
  \citenamefont {{Reichardt}}, \citenamefont {{Richards}}, \citenamefont
  {{Ross}}, \citenamefont {{Rotermund}}, \citenamefont {{Schenck}},
  \citenamefont {{Sherwin}}, \citenamefont {{Shirley}}, \citenamefont
  {{Sholl}}, \citenamefont {{Siritanasak}}, \citenamefont {{Smecher}},
  \citenamefont {{Steinbach}}, \citenamefont {{Stompor}}, \citenamefont
  {{Suzuki}}, \citenamefont {{Suzuki}}, \citenamefont {{Takada}}, \citenamefont
  {{Takakura}}, \citenamefont {{Tomaru}}, \citenamefont {{Wilson}},
  \citenamefont {{Yadav}},\ and\ \citenamefont {{Zahn}}}]{2014SPIE.9153E..1FA}%
  \BibitemOpen
  \bibfield  {author} {\bibinfo {author} {\bibfnamefont {K.}~\bibnamefont
  {{Arnold}}}, \bibinfo {author} {\bibfnamefont {N.}~\bibnamefont {{Stebor}}},
  \bibinfo {author} {\bibfnamefont {P.~A.~R.}\ \bibnamefont {{Ade}}}, \bibinfo
  {author} {\bibfnamefont {Y.}~\bibnamefont {{Akiba}}}, \bibinfo {author}
  {\bibfnamefont {A.~E.}\ \bibnamefont {{Anthony}}}, \bibinfo {author}
  {\bibfnamefont {M.}~\bibnamefont {{Atlas}}}, \bibinfo {author} {\bibfnamefont
  {D.}~\bibnamefont {{Barron}}}, \bibinfo {author} {\bibfnamefont
  {A.}~\bibnamefont {{Bender}}}, \bibinfo {author} {\bibfnamefont
  {D.}~\bibnamefont {{Boettger}}}, \bibinfo {author} {\bibfnamefont
  {J.}~\bibnamefont {{Borrill}}}, \bibinfo {author} {\bibfnamefont
  {S.}~\bibnamefont {{Chapman}}}, \bibinfo {author} {\bibfnamefont
  {Y.}~\bibnamefont {{Chinone}}}, \bibinfo {author} {\bibfnamefont
  {A.}~\bibnamefont {{Cukierman}}}, \bibinfo {author} {\bibfnamefont
  {M.}~\bibnamefont {{Dobbs}}}, \bibinfo {author} {\bibfnamefont
  {T.}~\bibnamefont {{Elleflot}}}, \bibinfo {author} {\bibfnamefont
  {J.}~\bibnamefont {{Errard}}}, \bibinfo {author} {\bibfnamefont
  {G.}~\bibnamefont {{Fabbian}}}, \bibinfo {author} {\bibfnamefont
  {C.}~\bibnamefont {{Feng}}}, \bibinfo {author} {\bibfnamefont
  {A.}~\bibnamefont {{Gilbert}}}, \bibinfo {author} {\bibfnamefont
  {N.}~\bibnamefont {{Goeckner-Wald}}}, \bibinfo {author} {\bibfnamefont
  {N.~W.}\ \bibnamefont {{Halverson}}}, \bibinfo {author} {\bibfnamefont
  {M.}~\bibnamefont {{Hasegawa}}}, \bibinfo {author} {\bibfnamefont
  {K.}~\bibnamefont {{Hattori}}}, \bibinfo {author} {\bibfnamefont
  {M.}~\bibnamefont {{Hazumi}}}, \bibinfo {author} {\bibfnamefont {W.~L.}\
  \bibnamefont {{Holzapfel}}}, \bibinfo {author} {\bibfnamefont
  {Y.}~\bibnamefont {{Hori}}}, \bibinfo {author} {\bibfnamefont
  {Y.}~\bibnamefont {{Inoue}}}, \bibinfo {author} {\bibfnamefont {G.~C.}\
  \bibnamefont {{Jaehnig}}}, \bibinfo {author} {\bibfnamefont {A.~H.}\
  \bibnamefont {{Jaffe}}}, \bibinfo {author} {\bibfnamefont {N.}~\bibnamefont
  {{Katayama}}}, \bibinfo {author} {\bibfnamefont {B.}~\bibnamefont
  {{Keating}}}, \bibinfo {author} {\bibfnamefont {Z.}~\bibnamefont
  {{Kermish}}}, \bibinfo {author} {\bibfnamefont {R.}~\bibnamefont
  {{Keskitalo}}}, \bibinfo {author} {\bibfnamefont {T.}~\bibnamefont
  {{Kisner}}}, \bibinfo {author} {\bibfnamefont {M.}~\bibnamefont {{Le
  Jeune}}}, \bibinfo {author} {\bibfnamefont {A.~T.}\ \bibnamefont {{Lee}}},
  \bibinfo {author} {\bibfnamefont {E.~M.}\ \bibnamefont {{Leitch}}}, \bibinfo
  {author} {\bibfnamefont {E.}~\bibnamefont {{Linder}}}, \bibinfo {author}
  {\bibfnamefont {F.}~\bibnamefont {{Matsuda}}}, \bibinfo {author}
  {\bibfnamefont {T.}~\bibnamefont {{Matsumura}}}, \bibinfo {author}
  {\bibfnamefont {X.}~\bibnamefont {{Meng}}}, \bibinfo {author} {\bibfnamefont
  {N.~J.}\ \bibnamefont {{Miller}}}, \bibinfo {author} {\bibfnamefont
  {H.}~\bibnamefont {{Morii}}}, \bibinfo {author} {\bibfnamefont {M.~J.}\
  \bibnamefont {{Myers}}}, \bibinfo {author} {\bibfnamefont {M.}~\bibnamefont
  {{Navaroli}}}, \bibinfo {author} {\bibfnamefont {H.}~\bibnamefont
  {{Nishino}}}, \bibinfo {author} {\bibfnamefont {T.}~\bibnamefont
  {{Okamura}}}, \bibinfo {author} {\bibfnamefont {H.}~\bibnamefont {{Paar}}},
  \bibinfo {author} {\bibfnamefont {J.}~\bibnamefont {{Peloton}}}, \bibinfo
  {author} {\bibfnamefont {D.}~\bibnamefont {{Poletti}}}, \bibinfo {author}
  {\bibfnamefont {C.}~\bibnamefont {{Raum}}}, \bibinfo {author} {\bibfnamefont
  {G.}~\bibnamefont {{Rebeiz}}}, \bibinfo {author} {\bibfnamefont {C.~L.}\
  \bibnamefont {{Reichardt}}}, \bibinfo {author} {\bibfnamefont {P.~L.}\
  \bibnamefont {{Richards}}}, \bibinfo {author} {\bibfnamefont
  {C.}~\bibnamefont {{Ross}}}, \bibinfo {author} {\bibfnamefont {K.~M.}\
  \bibnamefont {{Rotermund}}}, \bibinfo {author} {\bibfnamefont {D.~E.}\
  \bibnamefont {{Schenck}}}, \bibinfo {author} {\bibfnamefont {B.~D.}\
  \bibnamefont {{Sherwin}}}, \bibinfo {author} {\bibfnamefont {I.}~\bibnamefont
  {{Shirley}}}, \bibinfo {author} {\bibfnamefont {M.}~\bibnamefont {{Sholl}}},
  \bibinfo {author} {\bibfnamefont {P.}~\bibnamefont {{Siritanasak}}}, \bibinfo
  {author} {\bibfnamefont {G.}~\bibnamefont {{Smecher}}}, \bibinfo {author}
  {\bibfnamefont {B.}~\bibnamefont {{Steinbach}}}, \bibinfo {author}
  {\bibfnamefont {R.}~\bibnamefont {{Stompor}}}, \bibinfo {author}
  {\bibfnamefont {A.}~\bibnamefont {{Suzuki}}}, \bibinfo {author}
  {\bibfnamefont {J.}~\bibnamefont {{Suzuki}}}, \bibinfo {author}
  {\bibfnamefont {S.}~\bibnamefont {{Takada}}}, \bibinfo {author}
  {\bibfnamefont {S.}~\bibnamefont {{Takakura}}}, \bibinfo {author}
  {\bibfnamefont {T.}~\bibnamefont {{Tomaru}}}, \bibinfo {author}
  {\bibfnamefont {B.}~\bibnamefont {{Wilson}}}, \bibinfo {author}
  {\bibfnamefont {A.}~\bibnamefont {{Yadav}}}, \ and\ \bibinfo {author}
  {\bibfnamefont {O.}~\bibnamefont {{Zahn}}},\ }in\ \href {\doibase
  10.1117/12.2057332} {\emph {\bibinfo {booktitle} {Millimeter, Submillimeter,
  and Far-Infrared Detectors and Instrumentation for Astronomy VII}}},\
  \bibinfo {series} {\procspie}, Vol.\ \bibinfo {volume} {9153}\ (\bibinfo
  {year} {2014})\ p.\ \bibinfo {pages} {91531F}\BibitemShut {NoStop}%
\bibitem [{\citenamefont {{Abazajian}}\ \emph {et~al.}(2016)\citenamefont
  {{Abazajian}}, \citenamefont {{Adshead}}, \citenamefont {{Ahmed}},
  \citenamefont {{Allen}}, \citenamefont {{Alonso}}, \citenamefont {{Arnold}},
  \citenamefont {{Baccigalupi}}, \citenamefont {{Bartlett}}, \citenamefont
  {{Battaglia}}, \citenamefont {{Benson}}, \citenamefont {{Bischoff}},
  \citenamefont {{Borrill}}, \citenamefont {{Buza}}, \citenamefont
  {{Calabrese}}, \citenamefont {{Caldwell}}, \citenamefont {{Carlstrom}},
  \citenamefont {{Chang}}, \citenamefont {{Crawford}}, \citenamefont
  {{Cyr-Racine}}, \citenamefont {{De Bernardis}}, \citenamefont {{de Haan}},
  \citenamefont {{di Serego Alighieri}}, \citenamefont {{Dunkley}},
  \citenamefont {{Dvorkin}}, \citenamefont {{Errard}}, \citenamefont
  {{Fabbian}}, \citenamefont {{Feeney}}, \citenamefont {{Ferraro}},
  \citenamefont {{Filippini}}, \citenamefont {{Flauger}}, \citenamefont
  {{Fuller}}, \citenamefont {{Gluscevic}}, \citenamefont {{Green}},
  \citenamefont {{Grin}}, \citenamefont {{Grohs}}, \citenamefont {{Henning}},
  \citenamefont {{Hill}}, \citenamefont {{Hlozek}}, \citenamefont {{Holder}},
  \citenamefont {{Holzapfel}}, \citenamefont {{Hu}}, \citenamefont
  {{Huffenberger}}, \citenamefont {{Keskitalo}}, \citenamefont {{Knox}},
  \citenamefont {{Kosowsky}}, \citenamefont {{Kovac}}, \citenamefont
  {{Kovetz}}, \citenamefont {{Kuo}}, \citenamefont {{Kusaka}}, \citenamefont
  {{Le Jeune}}, \citenamefont {{Lee}}, \citenamefont {{Lilley}}, \citenamefont
  {{Loverde}}, \citenamefont {{Madhavacheril}}, \citenamefont {{Mantz}},
  \citenamefont {{Marsh}}, \citenamefont {{McMahon}}, \citenamefont
  {{Meerburg}}, \citenamefont {{Meyers}}, \citenamefont {{Miller}},
  \citenamefont {{Munoz}}, \citenamefont {{Nguyen}}, \citenamefont {{Niemack}},
  \citenamefont {{Peloso}}, \citenamefont {{Peloton}}, \citenamefont
  {{Pogosian}}, \citenamefont {{Pryke}}, \citenamefont {{Raveri}},
  \citenamefont {{Reichardt}}, \citenamefont {{Rocha}}, \citenamefont
  {{Rotti}}, \citenamefont {{Schaan}}, \citenamefont {{Schmittfull}},
  \citenamefont {{Scott}}, \citenamefont {{Sehgal}}, \citenamefont
  {{Shandera}}, \citenamefont {{Sherwin}}, \citenamefont {{Smith}},
  \citenamefont {{Sorbo}}, \citenamefont {{Starkman}}, \citenamefont {{Story}},
  \citenamefont {{van Engelen}}, \citenamefont {{Vieira}}, \citenamefont
  {{Watson}}, \citenamefont {{Whitehorn}},\ and\ \citenamefont {{Kimmy
  Wu}}}]{2016arXiv161002743A}%
  \BibitemOpen
  \bibfield  {author} {\bibinfo {author} {\bibfnamefont {K.~N.}\ \bibnamefont
  {{Abazajian}}}, \bibinfo {author} {\bibfnamefont {P.}~\bibnamefont
  {{Adshead}}}, \bibinfo {author} {\bibfnamefont {Z.}~\bibnamefont {{Ahmed}}},
  \bibinfo {author} {\bibfnamefont {S.~W.}\ \bibnamefont {{Allen}}}, \bibinfo
  {author} {\bibfnamefont {D.}~\bibnamefont {{Alonso}}}, \bibinfo {author}
  {\bibfnamefont {K.~S.}\ \bibnamefont {{Arnold}}}, \bibinfo {author}
  {\bibfnamefont {C.}~\bibnamefont {{Baccigalupi}}}, \bibinfo {author}
  {\bibfnamefont {J.~G.}\ \bibnamefont {{Bartlett}}}, \bibinfo {author}
  {\bibfnamefont {N.}~\bibnamefont {{Battaglia}}}, \bibinfo {author}
  {\bibfnamefont {B.~A.}\ \bibnamefont {{Benson}}}, \bibinfo {author}
  {\bibfnamefont {C.~A.}\ \bibnamefont {{Bischoff}}}, \bibinfo {author}
  {\bibfnamefont {J.}~\bibnamefont {{Borrill}}}, \bibinfo {author}
  {\bibfnamefont {V.}~\bibnamefont {{Buza}}}, \bibinfo {author} {\bibfnamefont
  {E.}~\bibnamefont {{Calabrese}}}, \bibinfo {author} {\bibfnamefont
  {R.}~\bibnamefont {{Caldwell}}}, \bibinfo {author} {\bibfnamefont {J.~E.}\
  \bibnamefont {{Carlstrom}}}, \bibinfo {author} {\bibfnamefont {C.~L.}\
  \bibnamefont {{Chang}}}, \bibinfo {author} {\bibfnamefont {T.~M.}\
  \bibnamefont {{Crawford}}}, \bibinfo {author} {\bibfnamefont {F.-Y.}\
  \bibnamefont {{Cyr-Racine}}}, \bibinfo {author} {\bibfnamefont
  {F.}~\bibnamefont {{De Bernardis}}}, \bibinfo {author} {\bibfnamefont
  {T.}~\bibnamefont {{de Haan}}}, \bibinfo {author} {\bibfnamefont
  {S.}~\bibnamefont {{di Serego Alighieri}}}, \bibinfo {author} {\bibfnamefont
  {J.}~\bibnamefont {{Dunkley}}}, \bibinfo {author} {\bibfnamefont
  {C.}~\bibnamefont {{Dvorkin}}}, \bibinfo {author} {\bibfnamefont
  {J.}~\bibnamefont {{Errard}}}, \bibinfo {author} {\bibfnamefont
  {G.}~\bibnamefont {{Fabbian}}}, \bibinfo {author} {\bibfnamefont
  {S.}~\bibnamefont {{Feeney}}}, \bibinfo {author} {\bibfnamefont
  {S.}~\bibnamefont {{Ferraro}}}, \bibinfo {author} {\bibfnamefont {J.~P.}\
  \bibnamefont {{Filippini}}}, \bibinfo {author} {\bibfnamefont
  {R.}~\bibnamefont {{Flauger}}}, \bibinfo {author} {\bibfnamefont {G.~M.}\
  \bibnamefont {{Fuller}}}, \bibinfo {author} {\bibfnamefont {V.}~\bibnamefont
  {{Gluscevic}}}, \bibinfo {author} {\bibfnamefont {D.}~\bibnamefont
  {{Green}}}, \bibinfo {author} {\bibfnamefont {D.}~\bibnamefont {{Grin}}},
  \bibinfo {author} {\bibfnamefont {E.}~\bibnamefont {{Grohs}}}, \bibinfo
  {author} {\bibfnamefont {J.~W.}\ \bibnamefont {{Henning}}}, \bibinfo {author}
  {\bibfnamefont {J.~C.}\ \bibnamefont {{Hill}}}, \bibinfo {author}
  {\bibfnamefont {R.}~\bibnamefont {{Hlozek}}}, \bibinfo {author}
  {\bibfnamefont {G.}~\bibnamefont {{Holder}}}, \bibinfo {author}
  {\bibfnamefont {W.}~\bibnamefont {{Holzapfel}}}, \bibinfo {author}
  {\bibfnamefont {W.}~\bibnamefont {{Hu}}}, \bibinfo {author} {\bibfnamefont
  {K.~M.}\ \bibnamefont {{Huffenberger}}}, \bibinfo {author} {\bibfnamefont
  {R.}~\bibnamefont {{Keskitalo}}}, \bibinfo {author} {\bibfnamefont
  {L.}~\bibnamefont {{Knox}}}, \bibinfo {author} {\bibfnamefont
  {A.}~\bibnamefont {{Kosowsky}}}, \bibinfo {author} {\bibfnamefont
  {J.}~\bibnamefont {{Kovac}}}, \bibinfo {author} {\bibfnamefont {E.~D.}\
  \bibnamefont {{Kovetz}}}, \bibinfo {author} {\bibfnamefont {C.-L.}\
  \bibnamefont {{Kuo}}}, \bibinfo {author} {\bibfnamefont {A.}~\bibnamefont
  {{Kusaka}}}, \bibinfo {author} {\bibfnamefont {M.}~\bibnamefont {{Le
  Jeune}}}, \bibinfo {author} {\bibfnamefont {A.~T.}\ \bibnamefont {{Lee}}},
  \bibinfo {author} {\bibfnamefont {M.}~\bibnamefont {{Lilley}}}, \bibinfo
  {author} {\bibfnamefont {M.}~\bibnamefont {{Loverde}}}, \bibinfo {author}
  {\bibfnamefont {M.~S.}\ \bibnamefont {{Madhavacheril}}}, \bibinfo {author}
  {\bibfnamefont {A.}~\bibnamefont {{Mantz}}}, \bibinfo {author} {\bibfnamefont
  {D.~J.~E.}\ \bibnamefont {{Marsh}}}, \bibinfo {author} {\bibfnamefont
  {J.}~\bibnamefont {{McMahon}}}, \bibinfo {author} {\bibfnamefont {P.~D.}\
  \bibnamefont {{Meerburg}}}, \bibinfo {author} {\bibfnamefont
  {J.}~\bibnamefont {{Meyers}}}, \bibinfo {author} {\bibfnamefont {A.~D.}\
  \bibnamefont {{Miller}}}, \bibinfo {author} {\bibfnamefont {J.~B.}\
  \bibnamefont {{Munoz}}}, \bibinfo {author} {\bibfnamefont {H.~N.}\
  \bibnamefont {{Nguyen}}}, \bibinfo {author} {\bibfnamefont {M.~D.}\
  \bibnamefont {{Niemack}}}, \bibinfo {author} {\bibfnamefont {M.}~\bibnamefont
  {{Peloso}}}, \bibinfo {author} {\bibfnamefont {J.}~\bibnamefont {{Peloton}}},
  \bibinfo {author} {\bibfnamefont {L.}~\bibnamefont {{Pogosian}}}, \bibinfo
  {author} {\bibfnamefont {C.}~\bibnamefont {{Pryke}}}, \bibinfo {author}
  {\bibfnamefont {M.}~\bibnamefont {{Raveri}}}, \bibinfo {author}
  {\bibfnamefont {C.~L.}\ \bibnamefont {{Reichardt}}}, \bibinfo {author}
  {\bibfnamefont {G.}~\bibnamefont {{Rocha}}}, \bibinfo {author} {\bibfnamefont
  {A.}~\bibnamefont {{Rotti}}}, \bibinfo {author} {\bibfnamefont
  {E.}~\bibnamefont {{Schaan}}}, \bibinfo {author} {\bibfnamefont {M.~M.}\
  \bibnamefont {{Schmittfull}}}, \bibinfo {author} {\bibfnamefont
  {D.}~\bibnamefont {{Scott}}}, \bibinfo {author} {\bibfnamefont
  {N.}~\bibnamefont {{Sehgal}}}, \bibinfo {author} {\bibfnamefont
  {S.}~\bibnamefont {{Shandera}}}, \bibinfo {author} {\bibfnamefont {B.~D.}\
  \bibnamefont {{Sherwin}}}, \bibinfo {author} {\bibfnamefont {T.~L.}\
  \bibnamefont {{Smith}}}, \bibinfo {author} {\bibfnamefont {L.}~\bibnamefont
  {{Sorbo}}}, \bibinfo {author} {\bibfnamefont {G.~D.}\ \bibnamefont
  {{Starkman}}}, \bibinfo {author} {\bibfnamefont {K.~T.}\ \bibnamefont
  {{Story}}}, \bibinfo {author} {\bibfnamefont {A.}~\bibnamefont {{van
  Engelen}}}, \bibinfo {author} {\bibfnamefont {J.~D.}\ \bibnamefont
  {{Vieira}}}, \bibinfo {author} {\bibfnamefont {S.}~\bibnamefont {{Watson}}},
  \bibinfo {author} {\bibfnamefont {N.}~\bibnamefont {{Whitehorn}}}, \ and\
  \bibinfo {author} {\bibfnamefont {W.~L.}\ \bibnamefont {{Kimmy Wu}}},\
  }\href@noop {} {\bibfield  {journal} {\bibinfo  {journal} {ArXiv e-prints}\ }
  (\bibinfo {year} {2016})},\ \Eprint {http://arxiv.org/abs/1610.02743}
  {arXiv:1610.02743} \BibitemShut {NoStop}%
\bibitem [{\citenamefont {{BICEP2 Collaboration}}\ \emph
  {et~al.}(2016)\citenamefont {{BICEP2 Collaboration}}, \citenamefont {{Keck
  Array Collaboration}}, \citenamefont {{Ade}}, \citenamefont {{Ahmed}},
  \citenamefont {{Aikin}}, \citenamefont {{Alexander}}, \citenamefont
  {{Barkats}}, \citenamefont {{Benton}}, \citenamefont {{Bischoff}},
  \citenamefont {{Bock}}, \citenamefont {{Bowens-Rubin}}, \citenamefont
  {{Brevik}}, \citenamefont {{Buder}}, \citenamefont {{Bullock}}, \citenamefont
  {{Buza}}, \citenamefont {{Connors}}, \citenamefont {{Crill}}, \citenamefont
  {{Duband}}, \citenamefont {{Dvorkin}}, \citenamefont {{Filippini}},
  \citenamefont {{Fliescher}}, \citenamefont {{Grayson}}, \citenamefont
  {{Halpern}}, \citenamefont {{Harrison}}, \citenamefont {{Hilton}},
  \citenamefont {{Hui}}, \citenamefont {{Irwin}}, \citenamefont {{Karkare}},
  \citenamefont {{Karpel}}, \citenamefont {{Kaufman}}, \citenamefont
  {{Keating}}, \citenamefont {{Kefeli}}, \citenamefont {{Kernasovskiy}},
  \citenamefont {{Kovac}}, \citenamefont {{Kuo}}, \citenamefont {{Leitch}},
  \citenamefont {{Lueker}}, \citenamefont {{Megerian}}, \citenamefont
  {{Netterfield}}, \citenamefont {{Nguyen}}, \citenamefont {{O'Brient}},
  \citenamefont {{Ogburn}}, \citenamefont {{Orlando}}, \citenamefont {{Pryke}},
  \citenamefont {{Richter}}, \citenamefont {{Schwarz}}, \citenamefont
  {{Sheehy}}, \citenamefont {{Staniszewski}}, \citenamefont {{Steinbach}},
  \citenamefont {{Sudiwala}}, \citenamefont {{Teply}}, \citenamefont
  {{Thompson}}, \citenamefont {{Tolan}}, \citenamefont {{Tucker}},
  \citenamefont {{Turner}}, \citenamefont {{Vieregg}}, \citenamefont {{Weber}},
  \citenamefont {{Wiebe}}, \citenamefont {{Willmert}}, \citenamefont {{Wong}},
  \citenamefont {{Wu}},\ and\ \citenamefont {{Yoon}}}]{2016PhRvL.116c1302B}%
  \BibitemOpen
  \bibfield  {author} {\bibinfo {author} {\bibnamefont {{BICEP2
  Collaboration}}}, \bibinfo {author} {\bibnamefont {{Keck Array
  Collaboration}}}, \bibinfo {author} {\bibfnamefont {P.~A.~R.}\ \bibnamefont
  {{Ade}}}, \bibinfo {author} {\bibfnamefont {Z.}~\bibnamefont {{Ahmed}}},
  \bibinfo {author} {\bibfnamefont {R.~W.}\ \bibnamefont {{Aikin}}}, \bibinfo
  {author} {\bibfnamefont {K.~D.}\ \bibnamefont {{Alexander}}}, \bibinfo
  {author} {\bibfnamefont {D.}~\bibnamefont {{Barkats}}}, \bibinfo {author}
  {\bibfnamefont {S.~J.}\ \bibnamefont {{Benton}}}, \bibinfo {author}
  {\bibfnamefont {C.~A.}\ \bibnamefont {{Bischoff}}}, \bibinfo {author}
  {\bibfnamefont {J.~J.}\ \bibnamefont {{Bock}}}, \bibinfo {author}
  {\bibfnamefont {R.}~\bibnamefont {{Bowens-Rubin}}}, \bibinfo {author}
  {\bibfnamefont {J.~A.}\ \bibnamefont {{Brevik}}}, \bibinfo {author}
  {\bibfnamefont {I.}~\bibnamefont {{Buder}}}, \bibinfo {author} {\bibfnamefont
  {E.}~\bibnamefont {{Bullock}}}, \bibinfo {author} {\bibfnamefont
  {V.}~\bibnamefont {{Buza}}}, \bibinfo {author} {\bibfnamefont
  {J.}~\bibnamefont {{Connors}}}, \bibinfo {author} {\bibfnamefont {B.~P.}\
  \bibnamefont {{Crill}}}, \bibinfo {author} {\bibfnamefont {L.}~\bibnamefont
  {{Duband}}}, \bibinfo {author} {\bibfnamefont {C.}~\bibnamefont {{Dvorkin}}},
  \bibinfo {author} {\bibfnamefont {J.~P.}\ \bibnamefont {{Filippini}}},
  \bibinfo {author} {\bibfnamefont {S.}~\bibnamefont {{Fliescher}}}, \bibinfo
  {author} {\bibfnamefont {J.}~\bibnamefont {{Grayson}}}, \bibinfo {author}
  {\bibfnamefont {M.}~\bibnamefont {{Halpern}}}, \bibinfo {author}
  {\bibfnamefont {S.}~\bibnamefont {{Harrison}}}, \bibinfo {author}
  {\bibfnamefont {G.~C.}\ \bibnamefont {{Hilton}}}, \bibinfo {author}
  {\bibfnamefont {H.}~\bibnamefont {{Hui}}}, \bibinfo {author} {\bibfnamefont
  {K.~D.}\ \bibnamefont {{Irwin}}}, \bibinfo {author} {\bibfnamefont {K.~S.}\
  \bibnamefont {{Karkare}}}, \bibinfo {author} {\bibfnamefont {E.}~\bibnamefont
  {{Karpel}}}, \bibinfo {author} {\bibfnamefont {J.~P.}\ \bibnamefont
  {{Kaufman}}}, \bibinfo {author} {\bibfnamefont {B.~G.}\ \bibnamefont
  {{Keating}}}, \bibinfo {author} {\bibfnamefont {S.}~\bibnamefont {{Kefeli}}},
  \bibinfo {author} {\bibfnamefont {S.~A.}\ \bibnamefont {{Kernasovskiy}}},
  \bibinfo {author} {\bibfnamefont {J.~M.}\ \bibnamefont {{Kovac}}}, \bibinfo
  {author} {\bibfnamefont {C.~L.}\ \bibnamefont {{Kuo}}}, \bibinfo {author}
  {\bibfnamefont {E.~M.}\ \bibnamefont {{Leitch}}}, \bibinfo {author}
  {\bibfnamefont {M.}~\bibnamefont {{Lueker}}}, \bibinfo {author}
  {\bibfnamefont {K.~G.}\ \bibnamefont {{Megerian}}}, \bibinfo {author}
  {\bibfnamefont {C.~B.}\ \bibnamefont {{Netterfield}}}, \bibinfo {author}
  {\bibfnamefont {H.~T.}\ \bibnamefont {{Nguyen}}}, \bibinfo {author}
  {\bibfnamefont {R.}~\bibnamefont {{O'Brient}}}, \bibinfo {author}
  {\bibfnamefont {R.~W.}\ \bibnamefont {{Ogburn}}}, \bibinfo {author}
  {\bibfnamefont {A.}~\bibnamefont {{Orlando}}}, \bibinfo {author}
  {\bibfnamefont {C.}~\bibnamefont {{Pryke}}}, \bibinfo {author} {\bibfnamefont
  {S.}~\bibnamefont {{Richter}}}, \bibinfo {author} {\bibfnamefont
  {R.}~\bibnamefont {{Schwarz}}}, \bibinfo {author} {\bibfnamefont {C.~D.}\
  \bibnamefont {{Sheehy}}}, \bibinfo {author} {\bibfnamefont {Z.~K.}\
  \bibnamefont {{Staniszewski}}}, \bibinfo {author} {\bibfnamefont
  {B.}~\bibnamefont {{Steinbach}}}, \bibinfo {author} {\bibfnamefont {R.~V.}\
  \bibnamefont {{Sudiwala}}}, \bibinfo {author} {\bibfnamefont {G.~P.}\
  \bibnamefont {{Teply}}}, \bibinfo {author} {\bibfnamefont {K.~L.}\
  \bibnamefont {{Thompson}}}, \bibinfo {author} {\bibfnamefont {J.~E.}\
  \bibnamefont {{Tolan}}}, \bibinfo {author} {\bibfnamefont {C.}~\bibnamefont
  {{Tucker}}}, \bibinfo {author} {\bibfnamefont {A.~D.}\ \bibnamefont
  {{Turner}}}, \bibinfo {author} {\bibfnamefont {A.~G.}\ \bibnamefont
  {{Vieregg}}}, \bibinfo {author} {\bibfnamefont {A.~C.}\ \bibnamefont
  {{Weber}}}, \bibinfo {author} {\bibfnamefont {D.~V.}\ \bibnamefont
  {{Wiebe}}}, \bibinfo {author} {\bibfnamefont {J.}~\bibnamefont {{Willmert}}},
  \bibinfo {author} {\bibfnamefont {C.~L.}\ \bibnamefont {{Wong}}}, \bibinfo
  {author} {\bibfnamefont {W.~L.~K.}\ \bibnamefont {{Wu}}}, \ and\ \bibinfo
  {author} {\bibfnamefont {K.~W.}\ \bibnamefont {{Yoon}}},\ }\href {\doibase
  10.1103/PhysRevLett.116.031302} {\bibfield  {journal} {\bibinfo  {journal}
  {Physical Review Letters}\ }\textbf {\bibinfo {volume} {116}},\ \bibinfo
  {eid} {031302} (\bibinfo {year} {2016})},\ \Eprint
  {http://arxiv.org/abs/1510.09217} {arXiv:1510.09217} \BibitemShut {NoStop}%
\bibitem [{\citenamefont {{Tegmark}}\ and\ \citenamefont
  {{Zaldarriaga}}(2009)}]{2009PhRvD..79h3530T}%
  \BibitemOpen
  \bibfield  {author} {\bibinfo {author} {\bibfnamefont {M.}~\bibnamefont
  {{Tegmark}}}\ and\ \bibinfo {author} {\bibfnamefont {M.}~\bibnamefont
  {{Zaldarriaga}}},\ }\href {\doibase 10.1103/PhysRevD.79.083530} {\bibfield
  {journal} {\bibinfo  {journal} {\prd}\ }\textbf {\bibinfo {volume} {79}},\
  \bibinfo {eid} {083530} (\bibinfo {year} {2009})},\ \Eprint
  {http://arxiv.org/abs/0805.4414} {arXiv:0805.4414} \BibitemShut {NoStop}%
\bibitem [{\citenamefont {{Furlanetto}}\ \emph {et~al.}(2006)\citenamefont
  {{Furlanetto}}, \citenamefont {{Oh}},\ and\ \citenamefont
  {{Briggs}}}]{2006PhR...433..181F}%
  \BibitemOpen
  \bibfield  {author} {\bibinfo {author} {\bibfnamefont {S.~R.}\ \bibnamefont
  {{Furlanetto}}}, \bibinfo {author} {\bibfnamefont {S.~P.}\ \bibnamefont
  {{Oh}}}, \ and\ \bibinfo {author} {\bibfnamefont {F.~H.}\ \bibnamefont
  {{Briggs}}},\ }\href {\doibase 10.1016/j.physrep.2006.08.002} {\bibfield
  {journal} {\bibinfo  {journal} {\physrep}\ }\textbf {\bibinfo {volume}
  {433}},\ \bibinfo {pages} {181} (\bibinfo {year} {2006})},\ \Eprint
  {http://arxiv.org/abs/astro-ph/0608032} {astro-ph/0608032} \BibitemShut
  {NoStop}%
\bibitem [{\citenamefont {{Venumadhav}}\ \emph {et~al.}(2014)\citenamefont
  {{Venumadhav}}, \citenamefont {{Oklopcic}}, \citenamefont {{Gluscevic}},
  \citenamefont {{Mishra}},\ and\ \citenamefont
  {{Hirata}}}]{2014arXiv1410.2250V}%
  \BibitemOpen
  \bibfield  {author} {\bibinfo {author} {\bibfnamefont {T.}~\bibnamefont
  {{Venumadhav}}}, \bibinfo {author} {\bibfnamefont {A.}~\bibnamefont
  {{Oklopcic}}}, \bibinfo {author} {\bibfnamefont {V.}~\bibnamefont
  {{Gluscevic}}}, \bibinfo {author} {\bibfnamefont {A.}~\bibnamefont
  {{Mishra}}}, \ and\ \bibinfo {author} {\bibfnamefont {C.~M.}\ \bibnamefont
  {{Hirata}}},\ }\href@noop {} {\bibfield  {journal} {\bibinfo  {journal}
  {ArXiv e-prints}\ } (\bibinfo {year} {2014})},\ \Eprint
  {http://arxiv.org/abs/1410.2250} {arXiv:1410.2250} \BibitemShut {NoStop}%
\bibitem [{\citenamefont {{Gluscevic}}\ \emph {et~al.}(2017)\citenamefont
  {{Gluscevic}}, \citenamefont {{Venumadhav}}, \citenamefont {{Fang}},
  \citenamefont {{Hirata}}, \citenamefont {{Oklop{\v c}i{\'c}}},\ and\
  \citenamefont {{Mishra}}}]{2017PhRvD..95h3011G}%
  \BibitemOpen
  \bibfield  {author} {\bibinfo {author} {\bibfnamefont {V.}~\bibnamefont
  {{Gluscevic}}}, \bibinfo {author} {\bibfnamefont {T.}~\bibnamefont
  {{Venumadhav}}}, \bibinfo {author} {\bibfnamefont {X.}~\bibnamefont
  {{Fang}}}, \bibinfo {author} {\bibfnamefont {C.}~\bibnamefont {{Hirata}}},
  \bibinfo {author} {\bibfnamefont {A.}~\bibnamefont {{Oklop{\v c}i{\'c}}}}, \
  and\ \bibinfo {author} {\bibfnamefont {A.}~\bibnamefont {{Mishra}}},\ }\href
  {\doibase 10.1103/PhysRevD.95.083011} {\bibfield  {journal} {\bibinfo
  {journal} {\prd}\ }\textbf {\bibinfo {volume} {95}},\ \bibinfo {eid} {083011}
  (\bibinfo {year} {2017})},\ \Eprint {http://arxiv.org/abs/1604.06327}
  {arXiv:1604.06327} \BibitemShut {NoStop}%
\bibitem [{\citenamefont {{Kamionkowski}}\ and\ \citenamefont
  {{Loeb}}(1997)}]{1997PhRvD..56.4511K}%
  \BibitemOpen
  \bibfield  {author} {\bibinfo {author} {\bibfnamefont {M.}~\bibnamefont
  {{Kamionkowski}}}\ and\ \bibinfo {author} {\bibfnamefont {A.}~\bibnamefont
  {{Loeb}}},\ }\href {\doibase 10.1103/PhysRevD.56.4511} {\bibfield  {journal}
  {\bibinfo  {journal} {\prd}\ }\textbf {\bibinfo {volume} {56}},\ \bibinfo
  {pages} {4511} (\bibinfo {year} {1997})},\ \Eprint
  {http://arxiv.org/abs/astro-ph/9703118} {astro-ph/9703118} \BibitemShut
  {NoStop}%
\bibitem [{\citenamefont {{Cooray}}\ and\ \citenamefont
  {{Baumann}}(2003)}]{2003PhRvD..67f3505C}%
  \BibitemOpen
  \bibfield  {author} {\bibinfo {author} {\bibfnamefont {A.}~\bibnamefont
  {{Cooray}}}\ and\ \bibinfo {author} {\bibfnamefont {D.}~\bibnamefont
  {{Baumann}}},\ }\href {\doibase 10.1103/PhysRevD.67.063505} {\bibfield
  {journal} {\bibinfo  {journal} {\prd}\ }\textbf {\bibinfo {volume} {67}},\
  \bibinfo {eid} {063505} (\bibinfo {year} {2003})},\ \Eprint
  {http://arxiv.org/abs/astro-ph/0211095} {astro-ph/0211095} \BibitemShut
  {NoStop}%
\bibitem [{\citenamefont {{Dor{\'e}}}\ \emph {et~al.}(2004)\citenamefont
  {{Dor{\'e}}}, \citenamefont {{Holder}},\ and\ \citenamefont
  {{Loeb}}}]{2004ApJ...612...81D}%
  \BibitemOpen
  \bibfield  {author} {\bibinfo {author} {\bibfnamefont {O.}~\bibnamefont
  {{Dor{\'e}}}}, \bibinfo {author} {\bibfnamefont {G.~P.}\ \bibnamefont
  {{Holder}}}, \ and\ \bibinfo {author} {\bibfnamefont {A.}~\bibnamefont
  {{Loeb}}},\ }\href {\doibase 10.1086/422496} {\bibfield  {journal} {\bibinfo
  {journal} {\apj}\ }\textbf {\bibinfo {volume} {612}},\ \bibinfo {pages} {81}
  (\bibinfo {year} {2004})},\ \Eprint {http://arxiv.org/abs/astro-ph/0309281}
  {astro-ph/0309281} \BibitemShut {NoStop}%
\bibitem [{\citenamefont {{Skordis}}\ and\ \citenamefont
  {{Silk}}(2004)}]{2004astro.ph..2474S}%
  \BibitemOpen
  \bibfield  {author} {\bibinfo {author} {\bibfnamefont {C.}~\bibnamefont
  {{Skordis}}}\ and\ \bibinfo {author} {\bibfnamefont {J.}~\bibnamefont
  {{Silk}}},\ }\href@noop {} {\bibfield  {journal} {\bibinfo  {journal} {ArXiv
  Astrophysics e-prints}\ } (\bibinfo {year} {2004})},\ \Eprint
  {http://arxiv.org/abs/astro-ph/0402474} {astro-ph/0402474} \BibitemShut
  {NoStop}%
\bibitem [{\citenamefont {{Portsmouth}}(2004)}]{2004PhRvD..70f3504P}%
  \BibitemOpen
  \bibfield  {author} {\bibinfo {author} {\bibfnamefont {J.}~\bibnamefont
  {{Portsmouth}}},\ }\href {\doibase 10.1103/PhysRevD.70.063504} {\bibfield
  {journal} {\bibinfo  {journal} {\prd}\ }\textbf {\bibinfo {volume} {70}},\
  \bibinfo {eid} {063504} (\bibinfo {year} {2004})},\ \Eprint
  {http://arxiv.org/abs/astro-ph/0402173} {astro-ph/0402173} \BibitemShut
  {NoStop}%
\bibitem [{\citenamefont {{Bunn}}(2006)}]{2006PhRvD..73l3517B}%
  \BibitemOpen
  \bibfield  {author} {\bibinfo {author} {\bibfnamefont {E.~F.}\ \bibnamefont
  {{Bunn}}},\ }\href {\doibase 10.1103/PhysRevD.73.123517} {\bibfield
  {journal} {\bibinfo  {journal} {\prd}\ }\textbf {\bibinfo {volume} {73}},\
  \bibinfo {eid} {123517} (\bibinfo {year} {2006})},\ \Eprint
  {http://arxiv.org/abs/astro-ph/0603271} {astro-ph/0603271} \BibitemShut
  {NoStop}%
\bibitem [{\citenamefont {{Alizadeh}}\ and\ \citenamefont
  {{Hirata}}(2012)}]{2012PhRvD..85l3540A}%
  \BibitemOpen
  \bibfield  {author} {\bibinfo {author} {\bibfnamefont {E.}~\bibnamefont
  {{Alizadeh}}}\ and\ \bibinfo {author} {\bibfnamefont {C.~M.}\ \bibnamefont
  {{Hirata}}},\ }\href {\doibase 10.1103/PhysRevD.85.123540} {\bibfield
  {journal} {\bibinfo  {journal} {\prd}\ }\textbf {\bibinfo {volume} {85}},\
  \bibinfo {eid} {123540} (\bibinfo {year} {2012})},\ \Eprint
  {http://arxiv.org/abs/1201.5374} {arXiv:1201.5374 [astro-ph.CO]} \BibitemShut
  {NoStop}%
\bibitem [{\citenamefont {{Hu}}(2001)}]{2001PhRvD..64h3005H}%
  \BibitemOpen
  \bibfield  {author} {\bibinfo {author} {\bibfnamefont {W.}~\bibnamefont
  {{Hu}}},\ }\href {\doibase 10.1103/PhysRevD.64.083005} {\bibfield  {journal}
  {\bibinfo  {journal} {\prd}\ }\textbf {\bibinfo {volume} {64}},\ \bibinfo
  {eid} {083005} (\bibinfo {year} {2001})},\ \Eprint
  {http://arxiv.org/abs/astro-ph/0105117} {astro-ph/0105117} \BibitemShut
  {NoStop}%
\bibitem [{\citenamefont {{Mesinger}}\ \emph {et~al.}(2011)\citenamefont
  {{Mesinger}}, \citenamefont {{Furlanetto}},\ and\ \citenamefont
  {{Cen}}}]{2011MNRAS.411..955M}%
  \BibitemOpen
  \bibfield  {author} {\bibinfo {author} {\bibfnamefont {A.}~\bibnamefont
  {{Mesinger}}}, \bibinfo {author} {\bibfnamefont {S.}~\bibnamefont
  {{Furlanetto}}}, \ and\ \bibinfo {author} {\bibfnamefont {R.}~\bibnamefont
  {{Cen}}},\ }\href {\doibase 10.1111/j.1365-2966.2010.17731.x} {\bibfield
  {journal} {\bibinfo  {journal} {\mnras}\ }\textbf {\bibinfo {volume} {411}},\
  \bibinfo {pages} {955} (\bibinfo {year} {2011})},\ \Eprint
  {http://arxiv.org/abs/1003.3878} {arXiv:1003.3878} \BibitemShut {NoStop}%
\bibitem [{\citenamefont {{Planck Collaboration}}\ \emph
  {et~al.}(2016{\natexlab{c}})\citenamefont {{Planck Collaboration}},
  \citenamefont {{Ade}}, \citenamefont {{Aghanim}}, \citenamefont {{Arnaud}},
  \citenamefont {{Ashdown}}, \citenamefont {{Aumont}}, \citenamefont
  {{Baccigalupi}}, \citenamefont {{Banday}}, \citenamefont {{Barreiro}},
  \citenamefont {{Bartlett}},\ and\ \citenamefont
  {et~al.}}]{2016A&A...594A..13P}%
  \BibitemOpen
  \bibfield  {author} {\bibinfo {author} {\bibnamefont {{Planck
  Collaboration}}}, \bibinfo {author} {\bibfnamefont {P.~A.~R.}\ \bibnamefont
  {{Ade}}}, \bibinfo {author} {\bibfnamefont {N.}~\bibnamefont {{Aghanim}}},
  \bibinfo {author} {\bibfnamefont {M.}~\bibnamefont {{Arnaud}}}, \bibinfo
  {author} {\bibfnamefont {M.}~\bibnamefont {{Ashdown}}}, \bibinfo {author}
  {\bibfnamefont {J.}~\bibnamefont {{Aumont}}}, \bibinfo {author}
  {\bibfnamefont {C.}~\bibnamefont {{Baccigalupi}}}, \bibinfo {author}
  {\bibfnamefont {A.~J.}\ \bibnamefont {{Banday}}}, \bibinfo {author}
  {\bibfnamefont {R.~B.}\ \bibnamefont {{Barreiro}}}, \bibinfo {author}
  {\bibfnamefont {J.~G.}\ \bibnamefont {{Bartlett}}}, \ and\ \bibinfo {author}
  {\bibnamefont {et~al.}},\ }\href {\doibase 10.1051/0004-6361/201525830}
  {\bibfield  {journal} {\bibinfo  {journal} {\aap}\ }\textbf {\bibinfo
  {volume} {594}},\ \bibinfo {eid} {A13} (\bibinfo {year}
  {2016}{\natexlab{c}})},\ \Eprint {http://arxiv.org/abs/1502.01589}
  {arXiv:1502.01589} \BibitemShut {NoStop}%
\bibitem [{\citenamefont {{Baumann}}(2009)}]{2009arXiv0907.5424B}%
  \BibitemOpen
  \bibfield  {author} {\bibinfo {author} {\bibfnamefont {D.}~\bibnamefont
  {{Baumann}}},\ }\href@noop {} {\bibfield  {journal} {\bibinfo  {journal}
  {ArXiv e-prints}\ } (\bibinfo {year} {2009})},\ \Eprint
  {http://arxiv.org/abs/0907.5424} {arXiv:0907.5424 [hep-th]} \BibitemShut
  {NoStop}%
\bibitem [{\citenamefont {{Abramowitz}}\ and\ \citenamefont
  {{Stegun}}(1972)}]{1972hmfw.book.....A}%
  \BibitemOpen
  \bibfield  {author} {\bibinfo {author} {\bibfnamefont {M.}~\bibnamefont
  {{Abramowitz}}}\ and\ \bibinfo {author} {\bibfnamefont {I.~A.}\ \bibnamefont
  {{Stegun}}},\ }\href@noop {} {\emph {\bibinfo {title} {Handbook of
  Mathematical Functions, New York: Dover, 1972}}}\ (\bibinfo {year}
  {1972})\BibitemShut {NoStop}%
\bibitem [{\citenamefont {{Seljak}}\ and\ \citenamefont
  {{Zaldarriaga}}(1996)}]{1996ApJ...469..437S}%
  \BibitemOpen
  \bibfield  {author} {\bibinfo {author} {\bibfnamefont {U.}~\bibnamefont
  {{Seljak}}}\ and\ \bibinfo {author} {\bibfnamefont {M.}~\bibnamefont
  {{Zaldarriaga}}},\ }\href {\doibase 10.1086/177793} {\bibfield  {journal}
  {\bibinfo  {journal} {\apj}\ }\textbf {\bibinfo {volume} {469}},\ \bibinfo
  {pages} {437} (\bibinfo {year} {1996})},\ \Eprint
  {http://arxiv.org/abs/astro-ph/9603033} {astro-ph/9603033} \BibitemShut
  {NoStop}%
\bibitem [{\citenamefont {{Hu}}\ and\ \citenamefont
  {{White}}(1997)}]{1997PhRvD..56..596H}%
  \BibitemOpen
  \bibfield  {author} {\bibinfo {author} {\bibfnamefont {W.}~\bibnamefont
  {{Hu}}}\ and\ \bibinfo {author} {\bibfnamefont {M.}~\bibnamefont {{White}}},\
  }\href {\doibase 10.1103/PhysRevD.56.596} {\bibfield  {journal} {\bibinfo
  {journal} {\prd}\ }\textbf {\bibinfo {volume} {56}},\ \bibinfo {pages} {596}
  (\bibinfo {year} {1997})},\ \Eprint {http://arxiv.org/abs/astro-ph/9702170}
  {astro-ph/9702170} \BibitemShut {NoStop}%
\bibitem [{\citenamefont {{Knox}}(1997)}]{1997ApJ...480...72K}%
  \BibitemOpen
  \bibfield  {author} {\bibinfo {author} {\bibfnamefont {L.}~\bibnamefont
  {{Knox}}},\ }\href {\doibase 10.1086/303959} {\bibfield  {journal} {\bibinfo
  {journal} {\apj}\ }\textbf {\bibinfo {volume} {480}},\ \bibinfo {pages} {72}
  (\bibinfo {year} {1997})},\ \Eprint {http://arxiv.org/abs/astro-ph/9606066}
  {astro-ph/9606066} \BibitemShut {NoStop}%
\bibitem [{\citenamefont {{Amarie}}\ \emph {et~al.}(2005)\citenamefont
  {{Amarie}}, \citenamefont {{Hirata}},\ and\ \citenamefont
  {{Seljak}}}]{2005PhRvD..72l3006A}%
  \BibitemOpen
  \bibfield  {author} {\bibinfo {author} {\bibfnamefont {M.}~\bibnamefont
  {{Amarie}}}, \bibinfo {author} {\bibfnamefont {C.}~\bibnamefont {{Hirata}}},
  \ and\ \bibinfo {author} {\bibfnamefont {U.}~\bibnamefont {{Seljak}}},\
  }\href {\doibase 10.1103/PhysRevD.72.123006} {\bibfield  {journal} {\bibinfo
  {journal} {\prd}\ }\textbf {\bibinfo {volume} {72}},\ \bibinfo {eid} {123006}
  (\bibinfo {year} {2005})},\ \Eprint {http://arxiv.org/abs/astro-ph/0508293}
  {astro-ph/0508293} \BibitemShut {NoStop}%
\bibitem [{\citenamefont {{Di Matteo}}\ \emph {et~al.}(2002)\citenamefont {{Di
  Matteo}}, \citenamefont {{Perna}}, \citenamefont {{Abel}},\ and\
  \citenamefont {{Rees}}}]{2002ApJ...564..576D}%
  \BibitemOpen
  \bibfield  {author} {\bibinfo {author} {\bibfnamefont {T.}~\bibnamefont {{Di
  Matteo}}}, \bibinfo {author} {\bibfnamefont {R.}~\bibnamefont {{Perna}}},
  \bibinfo {author} {\bibfnamefont {T.}~\bibnamefont {{Abel}}}, \ and\ \bibinfo
  {author} {\bibfnamefont {M.~J.}\ \bibnamefont {{Rees}}},\ }\href {\doibase
  10.1086/324293} {\bibfield  {journal} {\bibinfo  {journal} {\apj}\ }\textbf
  {\bibinfo {volume} {564}},\ \bibinfo {pages} {576} (\bibinfo {year}
  {2002})},\ \Eprint {http://arxiv.org/abs/astro-ph/0109241} {astro-ph/0109241}
  \BibitemShut {NoStop}%
\bibitem [{\citenamefont {{Di Matteo}}\ \emph {et~al.}(2004)\citenamefont {{Di
  Matteo}}, \citenamefont {{Ciardi}},\ and\ \citenamefont
  {{Miniati}}}]{2004MNRAS.355.1053D}%
  \BibitemOpen
  \bibfield  {author} {\bibinfo {author} {\bibfnamefont {T.}~\bibnamefont {{Di
  Matteo}}}, \bibinfo {author} {\bibfnamefont {B.}~\bibnamefont {{Ciardi}}}, \
  and\ \bibinfo {author} {\bibfnamefont {F.}~\bibnamefont {{Miniati}}},\ }\href
  {\doibase 10.1111/j.1365-2966.2004.08443.x} {\bibfield  {journal} {\bibinfo
  {journal} {\mnras}\ }\textbf {\bibinfo {volume} {355}},\ \bibinfo {pages}
  {1053} (\bibinfo {year} {2004})},\ \Eprint
  {http://arxiv.org/abs/astro-ph/0402322} {astro-ph/0402322} \BibitemShut
  {NoStop}%
\bibitem [{\citenamefont {{Zaldarriaga}}\ \emph {et~al.}(2004)\citenamefont
  {{Zaldarriaga}}, \citenamefont {{Furlanetto}},\ and\ \citenamefont
  {{Hernquist}}}]{2004ApJ...608..622Z}%
  \BibitemOpen
  \bibfield  {author} {\bibinfo {author} {\bibfnamefont {M.}~\bibnamefont
  {{Zaldarriaga}}}, \bibinfo {author} {\bibfnamefont {S.~R.}\ \bibnamefont
  {{Furlanetto}}}, \ and\ \bibinfo {author} {\bibfnamefont {L.}~\bibnamefont
  {{Hernquist}}},\ }\href {\doibase 10.1086/386327} {\bibfield  {journal}
  {\bibinfo  {journal} {\apj}\ }\textbf {\bibinfo {volume} {608}},\ \bibinfo
  {pages} {622} (\bibinfo {year} {2004})},\ \Eprint
  {http://arxiv.org/abs/astro-ph/0311514} {astro-ph/0311514} \BibitemShut
  {NoStop}%
\bibitem [{\citenamefont {{Oh}}\ and\ \citenamefont
  {{Mack}}(2003)}]{2003MNRAS.346..871O}%
  \BibitemOpen
  \bibfield  {author} {\bibinfo {author} {\bibfnamefont {S.~P.}\ \bibnamefont
  {{Oh}}}\ and\ \bibinfo {author} {\bibfnamefont {K.~J.}\ \bibnamefont
  {{Mack}}},\ }\href {\doibase 10.1111/j.1365-2966.2003.07133.x} {\bibfield
  {journal} {\bibinfo  {journal} {\mnras}\ }\textbf {\bibinfo {volume} {346}},\
  \bibinfo {pages} {871} (\bibinfo {year} {2003})},\ \Eprint
  {http://arxiv.org/abs/astro-ph/0302099} {astro-ph/0302099} \BibitemShut
  {NoStop}%
\bibitem [{\citenamefont {{Liu}}\ and\ \citenamefont
  {{Tegmark}}(2012)}]{2012MNRAS.419.3491L}%
  \BibitemOpen
  \bibfield  {author} {\bibinfo {author} {\bibfnamefont {A.}~\bibnamefont
  {{Liu}}}\ and\ \bibinfo {author} {\bibfnamefont {M.}~\bibnamefont
  {{Tegmark}}},\ }\href {\doibase 10.1111/j.1365-2966.2011.19989.x} {\bibfield
  {journal} {\bibinfo  {journal} {\mnras}\ }\textbf {\bibinfo {volume} {419}},\
  \bibinfo {pages} {3491} (\bibinfo {year} {2012})},\ \Eprint
  {http://arxiv.org/abs/1106.0007} {arXiv:1106.0007 [astro-ph.CO]} \BibitemShut
  {NoStop}%
\bibitem [{\citenamefont {{Babich}}\ and\ \citenamefont
  {{Loeb}}(2005)}]{2005ApJ...635....1B}%
  \BibitemOpen
  \bibfield  {author} {\bibinfo {author} {\bibfnamefont {D.}~\bibnamefont
  {{Babich}}}\ and\ \bibinfo {author} {\bibfnamefont {A.}~\bibnamefont
  {{Loeb}}},\ }\href {\doibase 10.1086/497297} {\bibfield  {journal} {\bibinfo
  {journal} {\apj}\ }\textbf {\bibinfo {volume} {635}},\ \bibinfo {pages} {1}
  (\bibinfo {year} {2005})},\ \Eprint {http://arxiv.org/abs/astro-ph/0505358}
  {astro-ph/0505358} \BibitemShut {NoStop}%
\bibitem [{\citenamefont {{De}}\ and\ \citenamefont
  {{Tashiro}}(2014)}]{2014PhRvD..89l3002D}%
  \BibitemOpen
  \bibfield  {author} {\bibinfo {author} {\bibfnamefont {S.}~\bibnamefont
  {{De}}}\ and\ \bibinfo {author} {\bibfnamefont {H.}~\bibnamefont
  {{Tashiro}}},\ }\href {\doibase 10.1103/PhysRevD.89.123002} {\bibfield
  {journal} {\bibinfo  {journal} {\prd}\ }\textbf {\bibinfo {volume} {89}},\
  \bibinfo {eid} {123002} (\bibinfo {year} {2014})},\ \Eprint
  {http://arxiv.org/abs/1307.3584} {arXiv:1307.3584} \BibitemShut {NoStop}%
\bibitem [{\citenamefont {{King}}\ and\ \citenamefont
  {{Lubin}}(2016)}]{2016PhRvD..94b3501K}%
  \BibitemOpen
  \bibfield  {author} {\bibinfo {author} {\bibfnamefont {S.}~\bibnamefont
  {{King}}}\ and\ \bibinfo {author} {\bibfnamefont {P.}~\bibnamefont
  {{Lubin}}},\ }\href {\doibase 10.1103/PhysRevD.94.023501} {\bibfield
  {journal} {\bibinfo  {journal} {\prd}\ }\textbf {\bibinfo {volume} {94}},\
  \bibinfo {eid} {023501} (\bibinfo {year} {2016})},\ \Eprint
  {http://arxiv.org/abs/1606.04112} {arXiv:1606.04112} \BibitemShut {NoStop}%
\bibitem [{\citenamefont {{En{\ss}lin}}\ \emph {et~al.}(2017)\citenamefont
  {{En{\ss}lin}}, \citenamefont {{Hutschenreuter}}, \citenamefont {{Vacca}},\
  and\ \citenamefont {{Oppermann}}}]{2017arXiv170608539E}%
  \BibitemOpen
  \bibfield  {author} {\bibinfo {author} {\bibfnamefont {T.~A.}\ \bibnamefont
  {{En{\ss}lin}}}, \bibinfo {author} {\bibfnamefont {S.}~\bibnamefont
  {{Hutschenreuter}}}, \bibinfo {author} {\bibfnamefont {V.}~\bibnamefont
  {{Vacca}}}, \ and\ \bibinfo {author} {\bibfnamefont {N.}~\bibnamefont
  {{Oppermann}}},\ }\href@noop {} {\bibfield  {journal} {\bibinfo  {journal}
  {ArXiv e-prints}\ } (\bibinfo {year} {2017})},\ \Eprint
  {http://arxiv.org/abs/1706.08539} {arXiv:1706.08539 [astro-ph.HE]}
  \BibitemShut {NoStop}%
\bibitem [{\citenamefont {{Oppermann}}\ \emph {et~al.}(2012)\citenamefont
  {{Oppermann}}, \citenamefont {{Junklewitz}}, \citenamefont {{Robbers}},
  \citenamefont {{Bell}}, \citenamefont {{En{\ss}lin}}, \citenamefont
  {{Bonafede}}, \citenamefont {{Braun}}, \citenamefont {{Brown}}, \citenamefont
  {{Clarke}}, \citenamefont {{Feain}}, \citenamefont {{Gaensler}},
  \citenamefont {{Hammond}}, \citenamefont {{Harvey-Smith}}, \citenamefont
  {{Heald}}, \citenamefont {{Johnston-Hollitt}}, \citenamefont {{Klein}},
  \citenamefont {{Kronberg}}, \citenamefont {{Mao}}, \citenamefont
  {{McClure-Griffiths}}, \citenamefont {{O'Sullivan}}, \citenamefont
  {{Pratley}}, \citenamefont {{Robishaw}}, \citenamefont {{Roy}}, \citenamefont
  {{Schnitzeler}}, \citenamefont {{Sotomayor-Beltran}}, \citenamefont
  {{Stevens}}, \citenamefont {{Stil}}, \citenamefont {{Sunstrum}},
  \citenamefont {{Tanna}}, \citenamefont {{Taylor}},\ and\ \citenamefont {{Van
  Eck}}}]{2012A&A...542A..93O}%
  \BibitemOpen
  \bibfield  {author} {\bibinfo {author} {\bibfnamefont {N.}~\bibnamefont
  {{Oppermann}}}, \bibinfo {author} {\bibfnamefont {H.}~\bibnamefont
  {{Junklewitz}}}, \bibinfo {author} {\bibfnamefont {G.}~\bibnamefont
  {{Robbers}}}, \bibinfo {author} {\bibfnamefont {M.~R.}\ \bibnamefont
  {{Bell}}}, \bibinfo {author} {\bibfnamefont {T.~A.}\ \bibnamefont
  {{En{\ss}lin}}}, \bibinfo {author} {\bibfnamefont {A.}~\bibnamefont
  {{Bonafede}}}, \bibinfo {author} {\bibfnamefont {R.}~\bibnamefont {{Braun}}},
  \bibinfo {author} {\bibfnamefont {J.~C.}\ \bibnamefont {{Brown}}}, \bibinfo
  {author} {\bibfnamefont {T.~E.}\ \bibnamefont {{Clarke}}}, \bibinfo {author}
  {\bibfnamefont {I.~J.}\ \bibnamefont {{Feain}}}, \bibinfo {author}
  {\bibfnamefont {B.~M.}\ \bibnamefont {{Gaensler}}}, \bibinfo {author}
  {\bibfnamefont {A.}~\bibnamefont {{Hammond}}}, \bibinfo {author}
  {\bibfnamefont {L.}~\bibnamefont {{Harvey-Smith}}}, \bibinfo {author}
  {\bibfnamefont {G.}~\bibnamefont {{Heald}}}, \bibinfo {author} {\bibfnamefont
  {M.}~\bibnamefont {{Johnston-Hollitt}}}, \bibinfo {author} {\bibfnamefont
  {U.}~\bibnamefont {{Klein}}}, \bibinfo {author} {\bibfnamefont {P.~P.}\
  \bibnamefont {{Kronberg}}}, \bibinfo {author} {\bibfnamefont {S.~A.}\
  \bibnamefont {{Mao}}}, \bibinfo {author} {\bibfnamefont {N.~M.}\ \bibnamefont
  {{McClure-Griffiths}}}, \bibinfo {author} {\bibfnamefont {S.~P.}\
  \bibnamefont {{O'Sullivan}}}, \bibinfo {author} {\bibfnamefont
  {L.}~\bibnamefont {{Pratley}}}, \bibinfo {author} {\bibfnamefont
  {T.}~\bibnamefont {{Robishaw}}}, \bibinfo {author} {\bibfnamefont
  {S.}~\bibnamefont {{Roy}}}, \bibinfo {author} {\bibfnamefont {D.~H.~F.~M.}\
  \bibnamefont {{Schnitzeler}}}, \bibinfo {author} {\bibfnamefont
  {C.}~\bibnamefont {{Sotomayor-Beltran}}}, \bibinfo {author} {\bibfnamefont
  {J.}~\bibnamefont {{Stevens}}}, \bibinfo {author} {\bibfnamefont {J.~M.}\
  \bibnamefont {{Stil}}}, \bibinfo {author} {\bibfnamefont {C.}~\bibnamefont
  {{Sunstrum}}}, \bibinfo {author} {\bibfnamefont {A.}~\bibnamefont {{Tanna}}},
  \bibinfo {author} {\bibfnamefont {A.~R.}\ \bibnamefont {{Taylor}}}, \ and\
  \bibinfo {author} {\bibfnamefont {C.~L.}\ \bibnamefont {{Van Eck}}},\ }\href
  {\doibase 10.1051/0004-6361/201118526} {\bibfield  {journal} {\bibinfo
  {journal} {\aap}\ }\textbf {\bibinfo {volume} {542}},\ \bibinfo {eid} {A93}
  (\bibinfo {year} {2012})},\ \Eprint {http://arxiv.org/abs/1111.6186}
  {arXiv:1111.6186 [astro-ph.GA]} \BibitemShut {NoStop}%
\bibitem [{\citenamefont {{Haslam}}\ \emph {et~al.}(1982)\citenamefont
  {{Haslam}}, \citenamefont {{Salter}}, \citenamefont {{Stoffel}},\ and\
  \citenamefont {{Wilson}}}]{1982A&AS...47....1H}%
  \BibitemOpen
  \bibfield  {author} {\bibinfo {author} {\bibfnamefont {C.~G.~T.}\
  \bibnamefont {{Haslam}}}, \bibinfo {author} {\bibfnamefont {C.~J.}\
  \bibnamefont {{Salter}}}, \bibinfo {author} {\bibfnamefont {H.}~\bibnamefont
  {{Stoffel}}}, \ and\ \bibinfo {author} {\bibfnamefont {W.~E.}\ \bibnamefont
  {{Wilson}}},\ }\href@noop {} {\bibfield  {journal} {\bibinfo  {journal}
  {\aaps}\ }\textbf {\bibinfo {volume} {47}},\ \bibinfo {pages} {1} (\bibinfo
  {year} {1982})}\BibitemShut {NoStop}%
\bibitem [{\citenamefont {{Bernardi}}\ \emph {et~al.}(2009)\citenamefont
  {{Bernardi}}, \citenamefont {{de Bruyn}}, \citenamefont {{Brentjens}},
  \citenamefont {{Ciardi}}, \citenamefont {{Harker}}, \citenamefont
  {{Jeli{\'c}}}, \citenamefont {{Koopmans}}, \citenamefont {{Labropoulos}},
  \citenamefont {{Offringa}}, \citenamefont {{Pandey}}, \citenamefont
  {{Schaye}}, \citenamefont {{Thomas}}, \citenamefont {{Yatawatta}},\ and\
  \citenamefont {{Zaroubi}}}]{2009A&A...500..965B}%
  \BibitemOpen
  \bibfield  {author} {\bibinfo {author} {\bibfnamefont {G.}~\bibnamefont
  {{Bernardi}}}, \bibinfo {author} {\bibfnamefont {A.~G.}\ \bibnamefont {{de
  Bruyn}}}, \bibinfo {author} {\bibfnamefont {M.~A.}\ \bibnamefont
  {{Brentjens}}}, \bibinfo {author} {\bibfnamefont {B.}~\bibnamefont
  {{Ciardi}}}, \bibinfo {author} {\bibfnamefont {G.}~\bibnamefont {{Harker}}},
  \bibinfo {author} {\bibfnamefont {V.}~\bibnamefont {{Jeli{\'c}}}}, \bibinfo
  {author} {\bibfnamefont {L.~V.~E.}\ \bibnamefont {{Koopmans}}}, \bibinfo
  {author} {\bibfnamefont {P.}~\bibnamefont {{Labropoulos}}}, \bibinfo {author}
  {\bibfnamefont {A.}~\bibnamefont {{Offringa}}}, \bibinfo {author}
  {\bibfnamefont {V.~N.}\ \bibnamefont {{Pandey}}}, \bibinfo {author}
  {\bibfnamefont {J.}~\bibnamefont {{Schaye}}}, \bibinfo {author}
  {\bibfnamefont {R.~M.}\ \bibnamefont {{Thomas}}}, \bibinfo {author}
  {\bibfnamefont {S.}~\bibnamefont {{Yatawatta}}}, \ and\ \bibinfo {author}
  {\bibfnamefont {S.}~\bibnamefont {{Zaroubi}}},\ }\href {\doibase
  10.1051/0004-6361/200911627} {\bibfield  {journal} {\bibinfo  {journal}
  {\aap}\ }\textbf {\bibinfo {volume} {500}},\ \bibinfo {pages} {965} (\bibinfo
  {year} {2009})},\ \Eprint {http://arxiv.org/abs/0904.0404} {arXiv:0904.0404}
  \BibitemShut {NoStop}%
\bibitem [{\citenamefont {{Pen}}\ \emph {et~al.}(2009)\citenamefont {{Pen}},
  \citenamefont {{Chang}}, \citenamefont {{Hirata}}, \citenamefont
  {{Peterson}}, \citenamefont {{Roy}}, \citenamefont {{Gupta}}, \citenamefont
  {{Odegova}},\ and\ \citenamefont {{Sigurdson}}}]{2009MNRAS.399..181P}%
  \BibitemOpen
  \bibfield  {author} {\bibinfo {author} {\bibfnamefont {U.-L.}\ \bibnamefont
  {{Pen}}}, \bibinfo {author} {\bibfnamefont {T.-C.}\ \bibnamefont {{Chang}}},
  \bibinfo {author} {\bibfnamefont {C.~M.}\ \bibnamefont {{Hirata}}}, \bibinfo
  {author} {\bibfnamefont {J.~B.}\ \bibnamefont {{Peterson}}}, \bibinfo
  {author} {\bibfnamefont {J.}~\bibnamefont {{Roy}}}, \bibinfo {author}
  {\bibfnamefont {Y.}~\bibnamefont {{Gupta}}}, \bibinfo {author} {\bibfnamefont
  {J.}~\bibnamefont {{Odegova}}}, \ and\ \bibinfo {author} {\bibfnamefont
  {K.}~\bibnamefont {{Sigurdson}}},\ }\href {\doibase
  10.1111/j.1365-2966.2009.14980.x} {\bibfield  {journal} {\bibinfo  {journal}
  {\mnras}\ }\textbf {\bibinfo {volume} {399}},\ \bibinfo {pages} {181}
  (\bibinfo {year} {2009})},\ \Eprint {http://arxiv.org/abs/0807.1056}
  {arXiv:0807.1056} \BibitemShut {NoStop}%
\bibitem [{\citenamefont {{Bernardi}}\ \emph {et~al.}(2013)\citenamefont
  {{Bernardi}}, \citenamefont {{Greenhill}}, \citenamefont {{Mitchell}},
  \citenamefont {{Ord}}, \citenamefont {{Hazelton}}, \citenamefont
  {{Gaensler}}, \citenamefont {{de Oliveira-Costa}}, \citenamefont {{Morales}},
  \citenamefont {{Udaya Shankar}}, \citenamefont {{Subrahmanyan}},
  \citenamefont {{Wayth}}, \citenamefont {{Lenc}}, \citenamefont {{Williams}},
  \citenamefont {{Arcus}}, \citenamefont {{Arora}}, \citenamefont {{Barnes}},
  \citenamefont {{Bowman}}, \citenamefont {{Briggs}}, \citenamefont {{Bunton}},
  \citenamefont {{Cappallo}}, \citenamefont {{Corey}}, \citenamefont
  {{Deshpande}}, \citenamefont {{deSouza}}, \citenamefont {{Emrich}},
  \citenamefont {{Goeke}}, \citenamefont {{Herne}}, \citenamefont {{Hewitt}},
  \citenamefont {{Johnston-Hollitt}}, \citenamefont {{Kaplan}}, \citenamefont
  {{Kasper}}, \citenamefont {{Kincaid}}, \citenamefont {{Koenig}},
  \citenamefont {{Kratzenberg}}, \citenamefont {{Lonsdale}}, \citenamefont
  {{Lynch}}, \citenamefont {{McWhirter}}, \citenamefont {{Morgan}},
  \citenamefont {{Oberoi}}, \citenamefont {{Pathikulangara}}, \citenamefont
  {{Prabu}}, \citenamefont {{Remillard}}, \citenamefont {{Rogers}},
  \citenamefont {{Roshi}}, \citenamefont {{Salah}}, \citenamefont {{Sault}},
  \citenamefont {{Srivani}}, \citenamefont {{Stevens}}, \citenamefont
  {{Tingay}}, \citenamefont {{Waterson}}, \citenamefont {{Webster}},
  \citenamefont {{Whitney}}, \citenamefont {{Williams}},\ and\ \citenamefont
  {{Wyithe}}}]{2013ApJ...771..105B}%
  \BibitemOpen
  \bibfield  {author} {\bibinfo {author} {\bibfnamefont {G.}~\bibnamefont
  {{Bernardi}}}, \bibinfo {author} {\bibfnamefont {L.~J.}\ \bibnamefont
  {{Greenhill}}}, \bibinfo {author} {\bibfnamefont {D.~A.}\ \bibnamefont
  {{Mitchell}}}, \bibinfo {author} {\bibfnamefont {S.~M.}\ \bibnamefont
  {{Ord}}}, \bibinfo {author} {\bibfnamefont {B.~J.}\ \bibnamefont
  {{Hazelton}}}, \bibinfo {author} {\bibfnamefont {B.~M.}\ \bibnamefont
  {{Gaensler}}}, \bibinfo {author} {\bibfnamefont {A.}~\bibnamefont {{de
  Oliveira-Costa}}}, \bibinfo {author} {\bibfnamefont {M.~F.}\ \bibnamefont
  {{Morales}}}, \bibinfo {author} {\bibfnamefont {N.}~\bibnamefont {{Udaya
  Shankar}}}, \bibinfo {author} {\bibfnamefont {R.}~\bibnamefont
  {{Subrahmanyan}}}, \bibinfo {author} {\bibfnamefont {R.~B.}\ \bibnamefont
  {{Wayth}}}, \bibinfo {author} {\bibfnamefont {E.}~\bibnamefont {{Lenc}}},
  \bibinfo {author} {\bibfnamefont {C.~L.}\ \bibnamefont {{Williams}}},
  \bibinfo {author} {\bibfnamefont {W.}~\bibnamefont {{Arcus}}}, \bibinfo
  {author} {\bibfnamefont {B.~S.}\ \bibnamefont {{Arora}}}, \bibinfo {author}
  {\bibfnamefont {D.~G.}\ \bibnamefont {{Barnes}}}, \bibinfo {author}
  {\bibfnamefont {J.~D.}\ \bibnamefont {{Bowman}}}, \bibinfo {author}
  {\bibfnamefont {F.~H.}\ \bibnamefont {{Briggs}}}, \bibinfo {author}
  {\bibfnamefont {J.~D.}\ \bibnamefont {{Bunton}}}, \bibinfo {author}
  {\bibfnamefont {R.~J.}\ \bibnamefont {{Cappallo}}}, \bibinfo {author}
  {\bibfnamefont {B.~E.}\ \bibnamefont {{Corey}}}, \bibinfo {author}
  {\bibfnamefont {A.}~\bibnamefont {{Deshpande}}}, \bibinfo {author}
  {\bibfnamefont {L.}~\bibnamefont {{deSouza}}}, \bibinfo {author}
  {\bibfnamefont {D.}~\bibnamefont {{Emrich}}}, \bibinfo {author}
  {\bibfnamefont {R.}~\bibnamefont {{Goeke}}}, \bibinfo {author} {\bibfnamefont
  {D.}~\bibnamefont {{Herne}}}, \bibinfo {author} {\bibfnamefont {J.~N.}\
  \bibnamefont {{Hewitt}}}, \bibinfo {author} {\bibfnamefont {M.}~\bibnamefont
  {{Johnston-Hollitt}}}, \bibinfo {author} {\bibfnamefont {D.}~\bibnamefont
  {{Kaplan}}}, \bibinfo {author} {\bibfnamefont {J.~C.}\ \bibnamefont
  {{Kasper}}}, \bibinfo {author} {\bibfnamefont {B.~B.}\ \bibnamefont
  {{Kincaid}}}, \bibinfo {author} {\bibfnamefont {R.}~\bibnamefont {{Koenig}}},
  \bibinfo {author} {\bibfnamefont {E.}~\bibnamefont {{Kratzenberg}}}, \bibinfo
  {author} {\bibfnamefont {C.~J.}\ \bibnamefont {{Lonsdale}}}, \bibinfo
  {author} {\bibfnamefont {M.~J.}\ \bibnamefont {{Lynch}}}, \bibinfo {author}
  {\bibfnamefont {S.~R.}\ \bibnamefont {{McWhirter}}}, \bibinfo {author}
  {\bibfnamefont {E.}~\bibnamefont {{Morgan}}}, \bibinfo {author}
  {\bibfnamefont {D.}~\bibnamefont {{Oberoi}}}, \bibinfo {author}
  {\bibfnamefont {J.}~\bibnamefont {{Pathikulangara}}}, \bibinfo {author}
  {\bibfnamefont {T.}~\bibnamefont {{Prabu}}}, \bibinfo {author} {\bibfnamefont
  {R.~A.}\ \bibnamefont {{Remillard}}}, \bibinfo {author} {\bibfnamefont
  {A.~E.~E.}\ \bibnamefont {{Rogers}}}, \bibinfo {author} {\bibfnamefont
  {A.}~\bibnamefont {{Roshi}}}, \bibinfo {author} {\bibfnamefont {J.~E.}\
  \bibnamefont {{Salah}}}, \bibinfo {author} {\bibfnamefont {R.~J.}\
  \bibnamefont {{Sault}}}, \bibinfo {author} {\bibfnamefont {K.~S.}\
  \bibnamefont {{Srivani}}}, \bibinfo {author} {\bibfnamefont {J.}~\bibnamefont
  {{Stevens}}}, \bibinfo {author} {\bibfnamefont {S.~J.}\ \bibnamefont
  {{Tingay}}}, \bibinfo {author} {\bibfnamefont {M.}~\bibnamefont
  {{Waterson}}}, \bibinfo {author} {\bibfnamefont {R.~L.}\ \bibnamefont
  {{Webster}}}, \bibinfo {author} {\bibfnamefont {A.~R.}\ \bibnamefont
  {{Whitney}}}, \bibinfo {author} {\bibfnamefont {A.}~\bibnamefont
  {{Williams}}}, \ and\ \bibinfo {author} {\bibfnamefont {J.~S.~B.}\
  \bibnamefont {{Wyithe}}},\ }\href {\doibase 10.1088/0004-637X/771/2/105}
  {\bibfield  {journal} {\bibinfo  {journal} {\apj}\ }\textbf {\bibinfo
  {volume} {771}},\ \bibinfo {eid} {105} (\bibinfo {year} {2013})},\ \Eprint
  {http://arxiv.org/abs/1305.6047} {arXiv:1305.6047} \BibitemShut {NoStop}%
\bibitem [{\citenamefont {{Jeli{\'c}}}\ \emph {et~al.}(2014)\citenamefont
  {{Jeli{\'c}}}, \citenamefont {{de Bruyn}}, \citenamefont {{Mevius}},
  \citenamefont {{Abdalla}}, \citenamefont {{Asad}}, \citenamefont
  {{Bernardi}}, \citenamefont {{Brentjens}}, \citenamefont {{Bus}},
  \citenamefont {{Chapman}}, \citenamefont {{Ciardi}}, \citenamefont
  {{Daiboo}}, \citenamefont {{Fernandez}}, \citenamefont {{Ghosh}},
  \citenamefont {{Harker}}, \citenamefont {{Jensen}}, \citenamefont {{Kazemi}},
  \citenamefont {{Koopmans}}, \citenamefont {{Labropoulos}}, \citenamefont
  {{Martinez-Rubi}}, \citenamefont {{Mellema}}, \citenamefont {{Offringa}},
  \citenamefont {{Pandey}}, \citenamefont {{Patil}}, \citenamefont {{Thomas}},
  \citenamefont {{Vedantham}}, \citenamefont {{Veligatla}}, \citenamefont
  {{Yatawatta}}, \citenamefont {{Zaroubi}}, \citenamefont {{Alexov}},
  \citenamefont {{Anderson}}, \citenamefont {{Avruch}}, \citenamefont {{Beck}},
  \citenamefont {{Bell}}, \citenamefont {{Bentum}}, \citenamefont {{Best}},
  \citenamefont {{Bonafede}}, \citenamefont {{Bregman}}, \citenamefont
  {{Breitling}}, \citenamefont {{Broderick}}, \citenamefont {{Brouw}},
  \citenamefont {{Br{\"u}ggen}}, \citenamefont {{Butcher}}, \citenamefont
  {{Conway}}, \citenamefont {{de Gasperin}}, \citenamefont {{de Geus}},
  \citenamefont {{Deller}}, \citenamefont {{Dettmar}}, \citenamefont
  {{Duscha}}, \citenamefont {{Eisl{\"o}ffel}}, \citenamefont {{Engels}},
  \citenamefont {{Falcke}}, \citenamefont {{Fallows}}, \citenamefont
  {{Fender}}, \citenamefont {{Ferrari}}, \citenamefont {{Frieswijk}},
  \citenamefont {{Garrett}}, \citenamefont {{Grie{\ss}meier}}, \citenamefont
  {{Gunst}}, \citenamefont {{Hamaker}}, \citenamefont {{Hassall}},
  \citenamefont {{Haverkorn}}, \citenamefont {{Heald}}, \citenamefont
  {{Hessels}}, \citenamefont {{Hoeft}}, \citenamefont {{H{\"o}randel}},
  \citenamefont {{Horneffer}}, \citenamefont {{van der Horst}}, \citenamefont
  {{Iacobelli}}, \citenamefont {{Juette}}, \citenamefont {{Karastergiou}},
  \citenamefont {{Kondratiev}}, \citenamefont {{Kramer}}, \citenamefont
  {{Kuniyoshi}}, \citenamefont {{Kuper}}, \citenamefont {{van Leeuwen}},
  \citenamefont {{Maat}}, \citenamefont {{Mann}}, \citenamefont
  {{McKay-Bukowski}}, \citenamefont {{McKean}}, \citenamefont {{Munk}},
  \citenamefont {{Nelles}}, \citenamefont {{Norden}}, \citenamefont {{Paas}},
  \citenamefont {{Pandey-Pommier}}, \citenamefont {{Pietka}}, \citenamefont
  {{Pizzo}}, \citenamefont {{Polatidis}}, \citenamefont {{Reich}},
  \citenamefont {{R{\"o}ttgering}}, \citenamefont {{Rowlinson}}, \citenamefont
  {{Scaife}}, \citenamefont {{Schwarz}}, \citenamefont {{Serylak}},
  \citenamefont {{Smirnov}}, \citenamefont {{Steinmetz}}, \citenamefont
  {{Stewart}}, \citenamefont {{Tagger}}, \citenamefont {{Tang}}, \citenamefont
  {{Tasse}}, \citenamefont {{ter Veen}}, \citenamefont {{Thoudam}},
  \citenamefont {{Toribio}}, \citenamefont {{Vermeulen}}, \citenamefont
  {{Vocks}}, \citenamefont {{van Weeren}}, \citenamefont {{Wijers}},
  \citenamefont {{Wijnholds}}, \citenamefont {{Wucknitz}},\ and\ \citenamefont
  {{Zarka}}}]{2014A&A...568A.101J}%
  \BibitemOpen
  \bibfield  {author} {\bibinfo {author} {\bibfnamefont {V.}~\bibnamefont
  {{Jeli{\'c}}}}, \bibinfo {author} {\bibfnamefont {A.~G.}\ \bibnamefont {{de
  Bruyn}}}, \bibinfo {author} {\bibfnamefont {M.}~\bibnamefont {{Mevius}}},
  \bibinfo {author} {\bibfnamefont {F.~B.}\ \bibnamefont {{Abdalla}}}, \bibinfo
  {author} {\bibfnamefont {K.~M.~B.}\ \bibnamefont {{Asad}}}, \bibinfo {author}
  {\bibfnamefont {G.}~\bibnamefont {{Bernardi}}}, \bibinfo {author}
  {\bibfnamefont {M.~A.}\ \bibnamefont {{Brentjens}}}, \bibinfo {author}
  {\bibfnamefont {S.}~\bibnamefont {{Bus}}}, \bibinfo {author} {\bibfnamefont
  {E.}~\bibnamefont {{Chapman}}}, \bibinfo {author} {\bibfnamefont
  {B.}~\bibnamefont {{Ciardi}}}, \bibinfo {author} {\bibfnamefont
  {S.}~\bibnamefont {{Daiboo}}}, \bibinfo {author} {\bibfnamefont {E.~R.}\
  \bibnamefont {{Fernandez}}}, \bibinfo {author} {\bibfnamefont
  {A.}~\bibnamefont {{Ghosh}}}, \bibinfo {author} {\bibfnamefont
  {G.}~\bibnamefont {{Harker}}}, \bibinfo {author} {\bibfnamefont
  {H.}~\bibnamefont {{Jensen}}}, \bibinfo {author} {\bibfnamefont
  {S.}~\bibnamefont {{Kazemi}}}, \bibinfo {author} {\bibfnamefont {L.~V.~E.}\
  \bibnamefont {{Koopmans}}}, \bibinfo {author} {\bibfnamefont
  {P.}~\bibnamefont {{Labropoulos}}}, \bibinfo {author} {\bibfnamefont
  {O.}~\bibnamefont {{Martinez-Rubi}}}, \bibinfo {author} {\bibfnamefont
  {G.}~\bibnamefont {{Mellema}}}, \bibinfo {author} {\bibfnamefont {A.~R.}\
  \bibnamefont {{Offringa}}}, \bibinfo {author} {\bibfnamefont {V.~N.}\
  \bibnamefont {{Pandey}}}, \bibinfo {author} {\bibfnamefont {A.~H.}\
  \bibnamefont {{Patil}}}, \bibinfo {author} {\bibfnamefont {R.~M.}\
  \bibnamefont {{Thomas}}}, \bibinfo {author} {\bibfnamefont {H.~K.}\
  \bibnamefont {{Vedantham}}}, \bibinfo {author} {\bibfnamefont
  {V.}~\bibnamefont {{Veligatla}}}, \bibinfo {author} {\bibfnamefont
  {S.}~\bibnamefont {{Yatawatta}}}, \bibinfo {author} {\bibfnamefont
  {S.}~\bibnamefont {{Zaroubi}}}, \bibinfo {author} {\bibfnamefont
  {A.}~\bibnamefont {{Alexov}}}, \bibinfo {author} {\bibfnamefont
  {J.}~\bibnamefont {{Anderson}}}, \bibinfo {author} {\bibfnamefont {I.~M.}\
  \bibnamefont {{Avruch}}}, \bibinfo {author} {\bibfnamefont {R.}~\bibnamefont
  {{Beck}}}, \bibinfo {author} {\bibfnamefont {M.~E.}\ \bibnamefont {{Bell}}},
  \bibinfo {author} {\bibfnamefont {M.~J.}\ \bibnamefont {{Bentum}}}, \bibinfo
  {author} {\bibfnamefont {P.}~\bibnamefont {{Best}}}, \bibinfo {author}
  {\bibfnamefont {A.}~\bibnamefont {{Bonafede}}}, \bibinfo {author}
  {\bibfnamefont {J.}~\bibnamefont {{Bregman}}}, \bibinfo {author}
  {\bibfnamefont {F.}~\bibnamefont {{Breitling}}}, \bibinfo {author}
  {\bibfnamefont {J.}~\bibnamefont {{Broderick}}}, \bibinfo {author}
  {\bibfnamefont {W.~N.}\ \bibnamefont {{Brouw}}}, \bibinfo {author}
  {\bibfnamefont {M.}~\bibnamefont {{Br{\"u}ggen}}}, \bibinfo {author}
  {\bibfnamefont {H.~R.}\ \bibnamefont {{Butcher}}}, \bibinfo {author}
  {\bibfnamefont {J.~E.}\ \bibnamefont {{Conway}}}, \bibinfo {author}
  {\bibfnamefont {F.}~\bibnamefont {{de Gasperin}}}, \bibinfo {author}
  {\bibfnamefont {E.}~\bibnamefont {{de Geus}}}, \bibinfo {author}
  {\bibfnamefont {A.}~\bibnamefont {{Deller}}}, \bibinfo {author}
  {\bibfnamefont {R.-J.}\ \bibnamefont {{Dettmar}}}, \bibinfo {author}
  {\bibfnamefont {S.}~\bibnamefont {{Duscha}}}, \bibinfo {author}
  {\bibfnamefont {J.}~\bibnamefont {{Eisl{\"o}ffel}}}, \bibinfo {author}
  {\bibfnamefont {D.}~\bibnamefont {{Engels}}}, \bibinfo {author}
  {\bibfnamefont {H.}~\bibnamefont {{Falcke}}}, \bibinfo {author}
  {\bibfnamefont {R.~A.}\ \bibnamefont {{Fallows}}}, \bibinfo {author}
  {\bibfnamefont {R.}~\bibnamefont {{Fender}}}, \bibinfo {author}
  {\bibfnamefont {C.}~\bibnamefont {{Ferrari}}}, \bibinfo {author}
  {\bibfnamefont {W.}~\bibnamefont {{Frieswijk}}}, \bibinfo {author}
  {\bibfnamefont {M.~A.}\ \bibnamefont {{Garrett}}}, \bibinfo {author}
  {\bibfnamefont {J.}~\bibnamefont {{Grie{\ss}meier}}}, \bibinfo {author}
  {\bibfnamefont {A.~W.}\ \bibnamefont {{Gunst}}}, \bibinfo {author}
  {\bibfnamefont {J.~P.}\ \bibnamefont {{Hamaker}}}, \bibinfo {author}
  {\bibfnamefont {T.~E.}\ \bibnamefont {{Hassall}}}, \bibinfo {author}
  {\bibfnamefont {M.}~\bibnamefont {{Haverkorn}}}, \bibinfo {author}
  {\bibfnamefont {G.}~\bibnamefont {{Heald}}}, \bibinfo {author} {\bibfnamefont
  {J.~W.~T.}\ \bibnamefont {{Hessels}}}, \bibinfo {author} {\bibfnamefont
  {M.}~\bibnamefont {{Hoeft}}}, \bibinfo {author} {\bibfnamefont
  {J.}~\bibnamefont {{H{\"o}randel}}}, \bibinfo {author} {\bibfnamefont
  {A.}~\bibnamefont {{Horneffer}}}, \bibinfo {author} {\bibfnamefont
  {A.}~\bibnamefont {{van der Horst}}}, \bibinfo {author} {\bibfnamefont
  {M.}~\bibnamefont {{Iacobelli}}}, \bibinfo {author} {\bibfnamefont
  {E.}~\bibnamefont {{Juette}}}, \bibinfo {author} {\bibfnamefont
  {A.}~\bibnamefont {{Karastergiou}}}, \bibinfo {author} {\bibfnamefont
  {V.~I.}\ \bibnamefont {{Kondratiev}}}, \bibinfo {author} {\bibfnamefont
  {M.}~\bibnamefont {{Kramer}}}, \bibinfo {author} {\bibfnamefont
  {M.}~\bibnamefont {{Kuniyoshi}}}, \bibinfo {author} {\bibfnamefont
  {G.}~\bibnamefont {{Kuper}}}, \bibinfo {author} {\bibfnamefont
  {J.}~\bibnamefont {{van Leeuwen}}}, \bibinfo {author} {\bibfnamefont
  {P.}~\bibnamefont {{Maat}}}, \bibinfo {author} {\bibfnamefont
  {G.}~\bibnamefont {{Mann}}}, \bibinfo {author} {\bibfnamefont
  {D.}~\bibnamefont {{McKay-Bukowski}}}, \bibinfo {author} {\bibfnamefont
  {J.~P.}\ \bibnamefont {{McKean}}}, \bibinfo {author} {\bibfnamefont
  {H.}~\bibnamefont {{Munk}}}, \bibinfo {author} {\bibfnamefont
  {A.}~\bibnamefont {{Nelles}}}, \bibinfo {author} {\bibfnamefont {M.~J.}\
  \bibnamefont {{Norden}}}, \bibinfo {author} {\bibfnamefont {H.}~\bibnamefont
  {{Paas}}}, \bibinfo {author} {\bibfnamefont {M.}~\bibnamefont
  {{Pandey-Pommier}}}, \bibinfo {author} {\bibfnamefont {G.}~\bibnamefont
  {{Pietka}}}, \bibinfo {author} {\bibfnamefont {R.}~\bibnamefont {{Pizzo}}},
  \bibinfo {author} {\bibfnamefont {A.~G.}\ \bibnamefont {{Polatidis}}},
  \bibinfo {author} {\bibfnamefont {W.}~\bibnamefont {{Reich}}}, \bibinfo
  {author} {\bibfnamefont {H.}~\bibnamefont {{R{\"o}ttgering}}}, \bibinfo
  {author} {\bibfnamefont {A.}~\bibnamefont {{Rowlinson}}}, \bibinfo {author}
  {\bibfnamefont {A.~M.~M.}\ \bibnamefont {{Scaife}}}, \bibinfo {author}
  {\bibfnamefont {D.}~\bibnamefont {{Schwarz}}}, \bibinfo {author}
  {\bibfnamefont {M.}~\bibnamefont {{Serylak}}}, \bibinfo {author}
  {\bibfnamefont {O.}~\bibnamefont {{Smirnov}}}, \bibinfo {author}
  {\bibfnamefont {M.}~\bibnamefont {{Steinmetz}}}, \bibinfo {author}
  {\bibfnamefont {A.}~\bibnamefont {{Stewart}}}, \bibinfo {author}
  {\bibfnamefont {M.}~\bibnamefont {{Tagger}}}, \bibinfo {author}
  {\bibfnamefont {Y.}~\bibnamefont {{Tang}}}, \bibinfo {author} {\bibfnamefont
  {C.}~\bibnamefont {{Tasse}}}, \bibinfo {author} {\bibfnamefont
  {S.}~\bibnamefont {{ter Veen}}}, \bibinfo {author} {\bibfnamefont
  {S.}~\bibnamefont {{Thoudam}}}, \bibinfo {author} {\bibfnamefont
  {C.}~\bibnamefont {{Toribio}}}, \bibinfo {author} {\bibfnamefont
  {R.}~\bibnamefont {{Vermeulen}}}, \bibinfo {author} {\bibfnamefont
  {C.}~\bibnamefont {{Vocks}}}, \bibinfo {author} {\bibfnamefont {R.~J.}\
  \bibnamefont {{van Weeren}}}, \bibinfo {author} {\bibfnamefont {R.~A.~M.~J.}\
  \bibnamefont {{Wijers}}}, \bibinfo {author} {\bibfnamefont {S.~J.}\
  \bibnamefont {{Wijnholds}}}, \bibinfo {author} {\bibfnamefont
  {O.}~\bibnamefont {{Wucknitz}}}, \ and\ \bibinfo {author} {\bibfnamefont
  {P.}~\bibnamefont {{Zarka}}},\ }\href {\doibase 10.1051/0004-6361/201423998}
  {\bibfield  {journal} {\bibinfo  {journal} {\aap}\ }\textbf {\bibinfo
  {volume} {568}},\ \bibinfo {eid} {A101} (\bibinfo {year} {2014})},\ \Eprint
  {http://arxiv.org/abs/1407.2093} {arXiv:1407.2093} \BibitemShut {NoStop}%
\bibitem [{\citenamefont {{Moore}}\ \emph {et~al.}(2015)\citenamefont
  {{Moore}}, \citenamefont {{Aguirre}}, \citenamefont {{Kohn}}, \citenamefont
  {{Parsons}}, \citenamefont {{Ali}}, \citenamefont {{Bradley}}, \citenamefont
  {{Carilli}}, \citenamefont {{DeBoer}}, \citenamefont {{Dexter}},
  \citenamefont {{Gugliucci}}, \citenamefont {{Jacobs}}, \citenamefont
  {{Klima}}, \citenamefont {{Liu}}, \citenamefont {{MacMahon}}, \citenamefont
  {{Manley}}, \citenamefont {{Pober}}, \citenamefont {{Stefan}},\ and\
  \citenamefont {{Walbrugh}}}]{2015arXiv150205072M}%
  \BibitemOpen
  \bibfield  {author} {\bibinfo {author} {\bibfnamefont {D.}~\bibnamefont
  {{Moore}}}, \bibinfo {author} {\bibfnamefont {J.~E.}\ \bibnamefont
  {{Aguirre}}}, \bibinfo {author} {\bibfnamefont {S.}~\bibnamefont {{Kohn}}},
  \bibinfo {author} {\bibfnamefont {A.}~\bibnamefont {{Parsons}}}, \bibinfo
  {author} {\bibfnamefont {Z.}~\bibnamefont {{Ali}}}, \bibinfo {author}
  {\bibfnamefont {R.}~\bibnamefont {{Bradley}}}, \bibinfo {author}
  {\bibfnamefont {C.}~\bibnamefont {{Carilli}}}, \bibinfo {author}
  {\bibfnamefont {D.}~\bibnamefont {{DeBoer}}}, \bibinfo {author}
  {\bibfnamefont {M.}~\bibnamefont {{Dexter}}}, \bibinfo {author}
  {\bibfnamefont {N.}~\bibnamefont {{Gugliucci}}}, \bibinfo {author}
  {\bibfnamefont {D.}~\bibnamefont {{Jacobs}}}, \bibinfo {author}
  {\bibfnamefont {P.}~\bibnamefont {{Klima}}}, \bibinfo {author} {\bibfnamefont
  {A.}~\bibnamefont {{Liu}}}, \bibinfo {author} {\bibfnamefont
  {D.}~\bibnamefont {{MacMahon}}}, \bibinfo {author} {\bibfnamefont
  {J.}~\bibnamefont {{Manley}}}, \bibinfo {author} {\bibfnamefont
  {J.}~\bibnamefont {{Pober}}}, \bibinfo {author} {\bibfnamefont
  {I.}~\bibnamefont {{Stefan}}}, \ and\ \bibinfo {author} {\bibfnamefont
  {W.}~\bibnamefont {{Walbrugh}}},\ }\href@noop {} {\bibfield  {journal}
  {\bibinfo  {journal} {ArXiv e-prints}\ } (\bibinfo {year} {2015})},\ \Eprint
  {http://arxiv.org/abs/1502.05072} {arXiv:1502.05072} \BibitemShut {NoStop}%
\bibitem [{\citenamefont {{Jeli{\'c}}}\ \emph {et~al.}(2015)\citenamefont
  {{Jeli{\'c}}}, \citenamefont {{de Bruyn}}, \citenamefont {{Pandey}},
  \citenamefont {{Mevius}}, \citenamefont {{Haverkorn}}, \citenamefont
  {{Brentjens}}, \citenamefont {{Koopmans}}, \citenamefont {{Zaroubi}},
  \citenamefont {{Abdalla}}, \citenamefont {{Asad}}, \citenamefont {{Bus}},
  \citenamefont {{Chapman}}, \citenamefont {{Ciardi}}, \citenamefont
  {{Fernandez}}, \citenamefont {{Ghosh}}, \citenamefont {{Harker}},
  \citenamefont {{Iliev}}, \citenamefont {{Jensen}}, \citenamefont {{Kazemi}},
  \citenamefont {{Mellema}}, \citenamefont {{Offringa}}, \citenamefont
  {{Patil}}, \citenamefont {{Vedantham}},\ and\ \citenamefont
  {{Yatawatta}}}]{2015A&A...583A.137J}%
  \BibitemOpen
  \bibfield  {author} {\bibinfo {author} {\bibfnamefont {V.}~\bibnamefont
  {{Jeli{\'c}}}}, \bibinfo {author} {\bibfnamefont {A.~G.}\ \bibnamefont {{de
  Bruyn}}}, \bibinfo {author} {\bibfnamefont {V.~N.}\ \bibnamefont {{Pandey}}},
  \bibinfo {author} {\bibfnamefont {M.}~\bibnamefont {{Mevius}}}, \bibinfo
  {author} {\bibfnamefont {M.}~\bibnamefont {{Haverkorn}}}, \bibinfo {author}
  {\bibfnamefont {M.~A.}\ \bibnamefont {{Brentjens}}}, \bibinfo {author}
  {\bibfnamefont {L.~V.~E.}\ \bibnamefont {{Koopmans}}}, \bibinfo {author}
  {\bibfnamefont {S.}~\bibnamefont {{Zaroubi}}}, \bibinfo {author}
  {\bibfnamefont {F.~B.}\ \bibnamefont {{Abdalla}}}, \bibinfo {author}
  {\bibfnamefont {K.~M.~B.}\ \bibnamefont {{Asad}}}, \bibinfo {author}
  {\bibfnamefont {S.}~\bibnamefont {{Bus}}}, \bibinfo {author} {\bibfnamefont
  {E.}~\bibnamefont {{Chapman}}}, \bibinfo {author} {\bibfnamefont
  {B.}~\bibnamefont {{Ciardi}}}, \bibinfo {author} {\bibfnamefont {E.~R.}\
  \bibnamefont {{Fernandez}}}, \bibinfo {author} {\bibfnamefont
  {A.}~\bibnamefont {{Ghosh}}}, \bibinfo {author} {\bibfnamefont
  {G.}~\bibnamefont {{Harker}}}, \bibinfo {author} {\bibfnamefont {I.~T.}\
  \bibnamefont {{Iliev}}}, \bibinfo {author} {\bibfnamefont {H.}~\bibnamefont
  {{Jensen}}}, \bibinfo {author} {\bibfnamefont {S.}~\bibnamefont {{Kazemi}}},
  \bibinfo {author} {\bibfnamefont {G.}~\bibnamefont {{Mellema}}}, \bibinfo
  {author} {\bibfnamefont {A.~R.}\ \bibnamefont {{Offringa}}}, \bibinfo
  {author} {\bibfnamefont {A.~H.}\ \bibnamefont {{Patil}}}, \bibinfo {author}
  {\bibfnamefont {H.~K.}\ \bibnamefont {{Vedantham}}}, \ and\ \bibinfo {author}
  {\bibfnamefont {S.}~\bibnamefont {{Yatawatta}}},\ }\href {\doibase
  10.1051/0004-6361/201526638} {\bibfield  {journal} {\bibinfo  {journal}
  {\aap}\ }\textbf {\bibinfo {volume} {583}},\ \bibinfo {eid} {A137} (\bibinfo
  {year} {2015})},\ \Eprint {http://arxiv.org/abs/1508.06650}
  {arXiv:1508.06650} \BibitemShut {NoStop}%
\bibitem [{\citenamefont {{Roberts}}\ \emph {et~al.}(1975)\citenamefont
  {{Roberts}}, \citenamefont {{Cooke}}, \citenamefont {{Murray}}, \citenamefont
  {{Cooper}}, \citenamefont {{Roger}}, \citenamefont {{Ribes}},\ and\
  \citenamefont {{Biraud}}}]{1975AuJPh..28..325R}%
  \BibitemOpen
  \bibfield  {author} {\bibinfo {author} {\bibfnamefont {J.~A.}\ \bibnamefont
  {{Roberts}}}, \bibinfo {author} {\bibfnamefont {D.~J.}\ \bibnamefont
  {{Cooke}}}, \bibinfo {author} {\bibfnamefont {J.~D.}\ \bibnamefont
  {{Murray}}}, \bibinfo {author} {\bibfnamefont {B.~F.~C.}\ \bibnamefont
  {{Cooper}}}, \bibinfo {author} {\bibfnamefont {R.~S.}\ \bibnamefont
  {{Roger}}}, \bibinfo {author} {\bibfnamefont {J.-C.}\ \bibnamefont
  {{Ribes}}}, \ and\ \bibinfo {author} {\bibfnamefont {F.}~\bibnamefont
  {{Biraud}}},\ }\href {\doibase 10.1071/PH750325} {\bibfield  {journal}
  {\bibinfo  {journal} {Australian Journal of Physics}\ }\textbf {\bibinfo
  {volume} {28}},\ \bibinfo {pages} {325} (\bibinfo {year} {1975})}\BibitemShut
  {NoStop}%
\bibitem [{\citenamefont {{Wardle}}\ \emph {et~al.}(1998)\citenamefont
  {{Wardle}}, \citenamefont {{Homan}}, \citenamefont {{Ojha}},\ and\
  \citenamefont {{Roberts}}}]{1998Natur.395..457W}%
  \BibitemOpen
  \bibfield  {author} {\bibinfo {author} {\bibfnamefont {J.~F.~C.}\
  \bibnamefont {{Wardle}}}, \bibinfo {author} {\bibfnamefont {D.~C.}\
  \bibnamefont {{Homan}}}, \bibinfo {author} {\bibfnamefont {R.}~\bibnamefont
  {{Ojha}}}, \ and\ \bibinfo {author} {\bibfnamefont {D.~H.}\ \bibnamefont
  {{Roberts}}},\ }\href {\doibase 10.1038/26675} {\bibfield  {journal}
  {\bibinfo  {journal} {\nat}\ }\textbf {\bibinfo {volume} {395}},\ \bibinfo
  {pages} {457} (\bibinfo {year} {1998})}\BibitemShut {NoStop}%
\bibitem [{\citenamefont {{Homan}}\ and\ \citenamefont
  {{Wardle}}(1999)}]{1999AJ....118.1942H}%
  \BibitemOpen
  \bibfield  {author} {\bibinfo {author} {\bibfnamefont {D.~C.}\ \bibnamefont
  {{Homan}}}\ and\ \bibinfo {author} {\bibfnamefont {J.~F.~C.}\ \bibnamefont
  {{Wardle}}},\ }\href {\doibase 10.1086/301108} {\bibfield  {journal}
  {\bibinfo  {journal} {\aj}\ }\textbf {\bibinfo {volume} {118}},\ \bibinfo
  {pages} {1942} (\bibinfo {year} {1999})},\ \Eprint
  {http://arxiv.org/abs/astro-ph/0007396} {astro-ph/0007396} \BibitemShut
  {NoStop}%
\bibitem [{\citenamefont {{Sault}}\ and\ \citenamefont
  {{Macquart}}(1999)}]{1999ApJ...526L..85S}%
  \BibitemOpen
  \bibfield  {author} {\bibinfo {author} {\bibfnamefont {R.~J.}\ \bibnamefont
  {{Sault}}}\ and\ \bibinfo {author} {\bibfnamefont {J.-P.}\ \bibnamefont
  {{Macquart}}},\ }\href {\doibase 10.1086/312378} {\bibfield  {journal}
  {\bibinfo  {journal} {\apjl}\ }\textbf {\bibinfo {volume} {526}},\ \bibinfo
  {pages} {L85} (\bibinfo {year} {1999})},\ \Eprint
  {http://arxiv.org/abs/astro-ph/9910052} {astro-ph/9910052} \BibitemShut
  {NoStop}%
\bibitem [{\citenamefont {{Macquart}}\ and\ \citenamefont
  {{Melrose}}(2000{\natexlab{a}})}]{2000PhRvE..62.4177M}%
  \BibitemOpen
  \bibfield  {author} {\bibinfo {author} {\bibfnamefont {J.-P.}\ \bibnamefont
  {{Macquart}}}\ and\ \bibinfo {author} {\bibfnamefont {D.~B.}\ \bibnamefont
  {{Melrose}}},\ }\href {\doibase 10.1103/PhysRevE.62.4177} {\bibfield
  {journal} {\bibinfo  {journal} {\pre}\ }\textbf {\bibinfo {volume} {62}},\
  \bibinfo {pages} {4177} (\bibinfo {year} {2000}{\natexlab{a}})},\ \Eprint
  {http://arxiv.org/abs/astro-ph/0006437} {astro-ph/0006437} \BibitemShut
  {NoStop}%
\bibitem [{\citenamefont {{Macquart}}\ and\ \citenamefont
  {{Melrose}}(2000{\natexlab{b}})}]{2000ApJ...545..798M}%
  \BibitemOpen
  \bibfield  {author} {\bibinfo {author} {\bibfnamefont {J.-P.}\ \bibnamefont
  {{Macquart}}}\ and\ \bibinfo {author} {\bibfnamefont {D.~B.}\ \bibnamefont
  {{Melrose}}},\ }\href {\doibase 10.1086/317852} {\bibfield  {journal}
  {\bibinfo  {journal} {\apj}\ }\textbf {\bibinfo {volume} {545}},\ \bibinfo
  {pages} {798} (\bibinfo {year} {2000}{\natexlab{b}})},\ \Eprint
  {http://arxiv.org/abs/astro-ph/0007429} {astro-ph/0007429} \BibitemShut
  {NoStop}%
\bibitem [{\citenamefont {{Legg}}\ and\ \citenamefont
  {{Westfold}}(1968)}]{1968ApJ...154..499L}%
  \BibitemOpen
  \bibfield  {author} {\bibinfo {author} {\bibfnamefont {M.~P.~C.}\
  \bibnamefont {{Legg}}}\ and\ \bibinfo {author} {\bibfnamefont {K.~C.}\
  \bibnamefont {{Westfold}}},\ }\href {\doibase 10.1086/149777} {\bibfield
  {journal} {\bibinfo  {journal} {\apj}\ }\textbf {\bibinfo {volume} {154}},\
  \bibinfo {pages} {499} (\bibinfo {year} {1968})}\BibitemShut {NoStop}%
\bibitem [{\citenamefont {{de B{\'u}rca}}\ and\ \citenamefont
  {{Shearer}}(2015)}]{2015MNRAS.450..533D}%
  \BibitemOpen
  \bibfield  {author} {\bibinfo {author} {\bibfnamefont {D.}~\bibnamefont {{de
  B{\'u}rca}}}\ and\ \bibinfo {author} {\bibfnamefont {A.}~\bibnamefont
  {{Shearer}}},\ }\href {\doibase 10.1093/mnras/stv576} {\bibfield  {journal}
  {\bibinfo  {journal} {\mnras}\ }\textbf {\bibinfo {volume} {450}},\ \bibinfo
  {pages} {533} (\bibinfo {year} {2015})},\ \Eprint
  {http://arxiv.org/abs/1503.04722} {arXiv:1503.04722 [astro-ph.HE]}
  \BibitemShut {NoStop}%
\bibitem [{\citenamefont {{Haverkorn}}(2015)}]{2015ASSL..407..483H}%
  \BibitemOpen
  \bibfield  {author} {\bibinfo {author} {\bibfnamefont {M.}~\bibnamefont
  {{Haverkorn}}},\ }in\ \href {\doibase 10.1007/978-3-662-44625-6_17} {\emph
  {\bibinfo {booktitle} {Magnetic Fields in Diffuse Media}}},\ \bibinfo
  {series} {Astrophysics and Space Science Library}, Vol.\ \bibinfo {volume}
  {407},\ \bibinfo {editor} {edited by\ \bibinfo {editor} {\bibfnamefont
  {A.}~\bibnamefont {{Lazarian}}}, \bibinfo {editor} {\bibfnamefont {E.~M.}\
  \bibnamefont {{de Gouveia Dal Pino}}}, \ and\ \bibinfo {editor}
  {\bibfnamefont {C.}~\bibnamefont {{Melioli}}}}\ (\bibinfo {year} {2015})\ p.\
  \bibinfo {pages} {483},\ \Eprint {http://arxiv.org/abs/1406.0283}
  {arXiv:1406.0283} \BibitemShut {NoStop}%
\bibitem [{\citenamefont {{Sazonov}}(1969)}]{1969SvA....13..396S}%
  \BibitemOpen
  \bibfield  {author} {\bibinfo {author} {\bibfnamefont {V.~N.}\ \bibnamefont
  {{Sazonov}}},\ }\href@noop {} {\bibfield  {journal} {\bibinfo  {journal}
  {\sovast}\ }\textbf {\bibinfo {volume} {13}},\ \bibinfo {pages} {396}
  (\bibinfo {year} {1969})}\BibitemShut {NoStop}%
\bibitem [{\citenamefont {{Cordes}}\ and\ \citenamefont
  {{Lazio}}(2002)}]{2002astro.ph..7156C}%
  \BibitemOpen
  \bibfield  {author} {\bibinfo {author} {\bibfnamefont {J.~M.}\ \bibnamefont
  {{Cordes}}}\ and\ \bibinfo {author} {\bibfnamefont {T.~J.~W.}\ \bibnamefont
  {{Lazio}}},\ }\href@noop {} {\bibfield  {journal} {\bibinfo  {journal} {ArXiv
  Astrophysics e-prints}\ } (\bibinfo {year} {2002})},\ \Eprint
  {http://arxiv.org/abs/astro-ph/0207156} {astro-ph/0207156} \BibitemShut
  {NoStop}%
\bibitem [{\citenamefont {{Liu}}\ \emph {et~al.}(2009)\citenamefont {{Liu}},
  \citenamefont {{Tegmark}},\ and\ \citenamefont
  {{Zaldarriaga}}}]{2009MNRAS.394.1575L}%
  \BibitemOpen
  \bibfield  {author} {\bibinfo {author} {\bibfnamefont {A.}~\bibnamefont
  {{Liu}}}, \bibinfo {author} {\bibfnamefont {M.}~\bibnamefont {{Tegmark}}}, \
  and\ \bibinfo {author} {\bibfnamefont {M.}~\bibnamefont {{Zaldarriaga}}},\
  }\href {\doibase 10.1111/j.1365-2966.2009.14426.x} {\bibfield  {journal}
  {\bibinfo  {journal} {\mnras}\ }\textbf {\bibinfo {volume} {394}},\ \bibinfo
  {pages} {1575} (\bibinfo {year} {2009})},\ \Eprint
  {http://arxiv.org/abs/0807.3952} {arXiv:0807.3952} \BibitemShut {NoStop}%
\bibitem [{\citenamefont {Cohen}(2004)}]{LWA}%
  \BibitemOpen
  \bibfield  {author} {\bibinfo {author} {\bibfnamefont {A.}~\bibnamefont
  {Cohen}},\ }\href@noop {} {\emph {\bibinfo {title} {Estimates of the
  Classical Confusion Limit for the LWA}}} (\bibinfo {year} {2004}),\ \bibinfo
  {note}
  {\url{http://www.faculty.ece.vt.edu/swe/lwa/memo/lwa0017.pdf}}\BibitemShut
  {NoStop}%
\bibitem [{\citenamefont {{Ghosh}}\ \emph {et~al.}(2012)\citenamefont
  {{Ghosh}}, \citenamefont {{Prasad}}, \citenamefont {{Bharadwaj}},
  \citenamefont {{Ali}},\ and\ \citenamefont
  {{Chengalur}}}]{2012MNRAS.426.3295G}%
  \BibitemOpen
  \bibfield  {author} {\bibinfo {author} {\bibfnamefont {A.}~\bibnamefont
  {{Ghosh}}}, \bibinfo {author} {\bibfnamefont {J.}~\bibnamefont {{Prasad}}},
  \bibinfo {author} {\bibfnamefont {S.}~\bibnamefont {{Bharadwaj}}}, \bibinfo
  {author} {\bibfnamefont {S.~S.}\ \bibnamefont {{Ali}}}, \ and\ \bibinfo
  {author} {\bibfnamefont {J.~N.}\ \bibnamefont {{Chengalur}}},\ }\href
  {\doibase 10.1111/j.1365-2966.2012.21889.x} {\bibfield  {journal} {\bibinfo
  {journal} {\mnras}\ }\textbf {\bibinfo {volume} {426}},\ \bibinfo {pages}
  {3295} (\bibinfo {year} {2012})},\ \Eprint {http://arxiv.org/abs/1208.1617}
  {arXiv:1208.1617 [astro-ph.CO]} \BibitemShut {NoStop}%
\bibitem [{\citenamefont {{Rayner}}\ \emph {et~al.}(2000)\citenamefont
  {{Rayner}}, \citenamefont {{Norris}},\ and\ \citenamefont
  {{Sault}}}]{2000MNRAS.319..484R}%
  \BibitemOpen
  \bibfield  {author} {\bibinfo {author} {\bibfnamefont {D.~P.}\ \bibnamefont
  {{Rayner}}}, \bibinfo {author} {\bibfnamefont {R.~P.}\ \bibnamefont
  {{Norris}}}, \ and\ \bibinfo {author} {\bibfnamefont {R.~J.}\ \bibnamefont
  {{Sault}}},\ }\href {\doibase 10.1046/j.1365-8711.2000.03854.x} {\bibfield
  {journal} {\bibinfo  {journal} {\mnras}\ }\textbf {\bibinfo {volume} {319}},\
  \bibinfo {pages} {484} (\bibinfo {year} {2000})}\BibitemShut {NoStop}%
\bibitem [{\citenamefont {{Wang}}\ \emph {et~al.}(2006)\citenamefont {{Wang}},
  \citenamefont {{Tegmark}}, \citenamefont {{Santos}},\ and\ \citenamefont
  {{Knox}}}]{2006ApJ...650..529W}%
  \BibitemOpen
  \bibfield  {author} {\bibinfo {author} {\bibfnamefont {X.}~\bibnamefont
  {{Wang}}}, \bibinfo {author} {\bibfnamefont {M.}~\bibnamefont {{Tegmark}}},
  \bibinfo {author} {\bibfnamefont {M.~G.}\ \bibnamefont {{Santos}}}, \ and\
  \bibinfo {author} {\bibfnamefont {L.}~\bibnamefont {{Knox}}},\ }\href
  {\doibase 10.1086/506597} {\bibfield  {journal} {\bibinfo  {journal} {\apj}\
  }\textbf {\bibinfo {volume} {650}},\ \bibinfo {pages} {529} (\bibinfo {year}
  {2006})},\ \Eprint {http://arxiv.org/abs/astro-ph/0501081} {astro-ph/0501081}
  \BibitemShut {NoStop}%
\bibitem [{\citenamefont {{Vedantham}}\ and\ \citenamefont
  {{Koopmans}}(2016)}]{2016MNRAS.458.3099V}%
  \BibitemOpen
  \bibfield  {author} {\bibinfo {author} {\bibfnamefont {H.~K.}\ \bibnamefont
  {{Vedantham}}}\ and\ \bibinfo {author} {\bibfnamefont {L.~V.~E.}\
  \bibnamefont {{Koopmans}}},\ }\href {\doibase 10.1093/mnras/stw443}
  {\bibfield  {journal} {\bibinfo  {journal} {\mnras}\ }\textbf {\bibinfo
  {volume} {458}},\ \bibinfo {pages} {3099} (\bibinfo {year} {2016})},\ \Eprint
  {http://arxiv.org/abs/1512.00159} {arXiv:1512.00159 [astro-ph.IM]}
  \BibitemShut {NoStop}%
\bibitem [{\citenamefont {{Dodelson}}\ \emph {et~al.}(2003)\citenamefont
  {{Dodelson}}, \citenamefont {{Rozo}},\ and\ \citenamefont
  {{Stebbins}}}]{2003PhRvL..91b1301D}%
  \BibitemOpen
  \bibfield  {author} {\bibinfo {author} {\bibfnamefont {S.}~\bibnamefont
  {{Dodelson}}}, \bibinfo {author} {\bibfnamefont {E.}~\bibnamefont {{Rozo}}},
  \ and\ \bibinfo {author} {\bibfnamefont {A.}~\bibnamefont {{Stebbins}}},\
  }\href {\doibase 10.1103/PhysRevLett.91.021301} {\bibfield  {journal}
  {\bibinfo  {journal} {Physical Review Letters}\ }\textbf {\bibinfo {volume}
  {91}},\ \bibinfo {eid} {021301} (\bibinfo {year} {2003})},\ \Eprint
  {http://arxiv.org/abs/astro-ph/0301177} {astro-ph/0301177} \BibitemShut
  {NoStop}%
\bibitem [{\citenamefont {{Schmidt}}\ and\ \citenamefont
  {{Jeong}}(2012)}]{2012PhRvD..86h3513S}%
  \BibitemOpen
  \bibfield  {author} {\bibinfo {author} {\bibfnamefont {F.}~\bibnamefont
  {{Schmidt}}}\ and\ \bibinfo {author} {\bibfnamefont {D.}~\bibnamefont
  {{Jeong}}},\ }\href {\doibase 10.1103/PhysRevD.86.083513} {\bibfield
  {journal} {\bibinfo  {journal} {\prd}\ }\textbf {\bibinfo {volume} {86}},\
  \bibinfo {eid} {083513} (\bibinfo {year} {2012})},\ \Eprint
  {http://arxiv.org/abs/1205.1514} {arXiv:1205.1514 [astro-ph.CO]} \BibitemShut
  {NoStop}%
\bibitem [{\citenamefont {{Schmidt}}\ \emph {et~al.}(2014)\citenamefont
  {{Schmidt}}, \citenamefont {{Pajer}},\ and\ \citenamefont
  {{Zaldarriaga}}}]{2014PhRvD..89h3507S}%
  \BibitemOpen
  \bibfield  {author} {\bibinfo {author} {\bibfnamefont {F.}~\bibnamefont
  {{Schmidt}}}, \bibinfo {author} {\bibfnamefont {E.}~\bibnamefont {{Pajer}}},
  \ and\ \bibinfo {author} {\bibfnamefont {M.}~\bibnamefont {{Zaldarriaga}}},\
  }\href {\doibase 10.1103/PhysRevD.89.083507} {\bibfield  {journal} {\bibinfo
  {journal} {\prd}\ }\textbf {\bibinfo {volume} {89}},\ \bibinfo {eid} {083507}
  (\bibinfo {year} {2014})},\ \Eprint {http://arxiv.org/abs/1312.5616}
  {arXiv:1312.5616} \BibitemShut {NoStop}%
\bibitem [{\citenamefont {{Chisari}}\ \emph {et~al.}(2014)\citenamefont
  {{Chisari}}, \citenamefont {{Dvorkin}},\ and\ \citenamefont
  {{Schmidt}}}]{2014PhRvD..90d3527C}%
  \BibitemOpen
  \bibfield  {author} {\bibinfo {author} {\bibfnamefont {N.~E.}\ \bibnamefont
  {{Chisari}}}, \bibinfo {author} {\bibfnamefont {C.}~\bibnamefont
  {{Dvorkin}}}, \ and\ \bibinfo {author} {\bibfnamefont {F.}~\bibnamefont
  {{Schmidt}}},\ }\href {\doibase 10.1103/PhysRevD.90.043527} {\bibfield
  {journal} {\bibinfo  {journal} {\prd}\ }\textbf {\bibinfo {volume} {90}},\
  \bibinfo {eid} {043527} (\bibinfo {year} {2014})},\ \Eprint
  {http://arxiv.org/abs/1406.4871} {arXiv:1406.4871} \BibitemShut {NoStop}%
\bibitem [{\citenamefont {{Corbin}}\ and\ \citenamefont
  {{Cornish}}(2006)}]{2006CQGra..23.2435C}%
  \BibitemOpen
  \bibfield  {author} {\bibinfo {author} {\bibfnamefont {V.}~\bibnamefont
  {{Corbin}}}\ and\ \bibinfo {author} {\bibfnamefont {N.~J.}\ \bibnamefont
  {{Cornish}}},\ }\href {\doibase 10.1088/0264-9381/23/7/014} {\bibfield
  {journal} {\bibinfo  {journal} {Classical and Quantum Gravity}\ }\textbf
  {\bibinfo {volume} {23}},\ \bibinfo {pages} {2435} (\bibinfo {year}
  {2006})},\ \Eprint {http://arxiv.org/abs/gr-qc/0512039} {gr-qc/0512039}
  \BibitemShut {NoStop}%
\bibitem [{\citenamefont {{Kawamura}}\ \emph {et~al.}(2011)\citenamefont
  {{Kawamura}}, \citenamefont {{Ando}}, \citenamefont {{Seto}}, \citenamefont
  {{Sato}}, \citenamefont {{Nakamura}}, \citenamefont {{Tsubono}},
  \citenamefont {{Kanda}}, \citenamefont {{Tanaka}}, \citenamefont
  {{Yokoyama}}, \citenamefont {{Funaki}}, \citenamefont {{Numata}},
  \citenamefont {{Ioka}}, \citenamefont {{Takashima}}, \citenamefont
  {{Agatsuma}}, \citenamefont {{Akutsu}}, \citenamefont {{Aoyanagi}},
  \citenamefont {{Arai}}, \citenamefont {{Araya}}, \citenamefont {{Asada}},
  \citenamefont {{Aso}}, \citenamefont {{Chen}}, \citenamefont {{Chiba}},
  \citenamefont {{Ebisuzaki}}, \citenamefont {{Ejiri}}, \citenamefont
  {{Enoki}}, \citenamefont {{Eriguchi}}, \citenamefont {{Fujimoto}},
  \citenamefont {{Fujita}}, \citenamefont {{Fukushima}}, \citenamefont
  {{Futamase}}, \citenamefont {{Harada}}, \citenamefont {{Hashimoto}},
  \citenamefont {{Hayama}}, \citenamefont {{Hikida}}, \citenamefont
  {{Himemoto}}, \citenamefont {{Hirabayashi}}, \citenamefont {{Hiramatsu}},
  \citenamefont {{Hong}}, \citenamefont {{Horisawa}}, \citenamefont
  {{Hosokawa}}, \citenamefont {{Ichiki}}, \citenamefont {{Ikegami}},
  \citenamefont {{Inoue}}, \citenamefont {{Ishidoshiro}}, \citenamefont
  {{Ishihara}}, \citenamefont {{Ishikawa}}, \citenamefont {{Ishizaki}},
  \citenamefont {{Ito}}, \citenamefont {{Itoh}}, \citenamefont {{Izumi}},
  \citenamefont {{Kawano}}, \citenamefont {{Kawashima}}, \citenamefont
  {{Kawazoe}}, \citenamefont {{Kishimoto}}, \citenamefont {{Kiuchi}},
  \citenamefont {{Kobayashi}}, \citenamefont {{Kohri}}, \citenamefont
  {{Koizumi}}, \citenamefont {{Kojima}}, \citenamefont {{Kokeyama}},
  \citenamefont {{Kokuyama}}, \citenamefont {{Kotake}}, \citenamefont
  {{Kozai}}, \citenamefont {{Kunimori}}, \citenamefont {{Kuninaka}},
  \citenamefont {{Kuroda}}, \citenamefont {{Kuroyanagi}}, \citenamefont
  {{Maeda}}, \citenamefont {{Matsuhara}}, \citenamefont {{Matsumoto}},
  \citenamefont {{Michimura}}, \citenamefont {{Miyakawa}}, \citenamefont
  {{Miyamoto}}, \citenamefont {{Miyoki}}, \citenamefont {{Morimoto}},
  \citenamefont {{Morisawa}}, \citenamefont {{Moriwaki}}, \citenamefont
  {{Mukohyama}}, \citenamefont {{Musha}}, \citenamefont {{Nagano}},
  \citenamefont {{Naito}}, \citenamefont {{Nakamura}}, \citenamefont
  {{Nakano}}, \citenamefont {{Nakao}}, \citenamefont {{Nakasuka}},
  \citenamefont {{Nakayama}}, \citenamefont {{Nakazawa}}, \citenamefont
  {{Nishida}}, \citenamefont {{Nishiyama}}, \citenamefont {{Nishizawa}},
  \citenamefont {{Niwa}}, \citenamefont {{Noumi}}, \citenamefont {{Obuchi}},
  \citenamefont {{Ohashi}}, \citenamefont {{Ohishi}}, \citenamefont {{Ohkawa}},
  \citenamefont {{Okada}}, \citenamefont {{Okada}}, \citenamefont {{Oohara}},
  \citenamefont {{Sago}}, \citenamefont {{Saijo}}, \citenamefont {{Saito}},
  \citenamefont {{Sakagami}}, \citenamefont {{Sakai}}, \citenamefont
  {{Sakata}}, \citenamefont {{Sasaki}}, \citenamefont {{Sato}}, \citenamefont
  {{Shibata}}, \citenamefont {{Shinkai}}, \citenamefont {{Shoda}},
  \citenamefont {{Somiya}}, \citenamefont {{Sotani}}, \citenamefont
  {{Sugiyama}}, \citenamefont {{Suwa}}, \citenamefont {{Suzuki}}, \citenamefont
  {{Tagoshi}}, \citenamefont {{Takahashi}}, \citenamefont {{Takahashi}},
  \citenamefont {{Takahashi}}, \citenamefont {{Takahashi}}, \citenamefont
  {{Takahashi}}, \citenamefont {{Takahashi}}, \citenamefont {{Takahashi}},
  \citenamefont {{Akiteru}}, \citenamefont {{Takano}}, \citenamefont
  {{Tanaka}}, \citenamefont {{Taniguchi}}, \citenamefont {{Taruya}},
  \citenamefont {{Tashiro}}, \citenamefont {{Torii}}, \citenamefont
  {{Toyoshima}}, \citenamefont {{Tsujikawa}}, \citenamefont {{Tsunesada}},
  \citenamefont {{Ueda}}, \citenamefont {{Ueda}}, \citenamefont {{Utashima}},
  \citenamefont {{Wakabayashi}}, \citenamefont {{Yagi}}, \citenamefont
  {{Yamakawa}}, \citenamefont {{Yamamoto}}, \citenamefont {{Yamazaki}},
  \citenamefont {{Yoo}}, \citenamefont {{Yoshida}}, \citenamefont {{Yoshino}},\
  and\ \citenamefont {{Sun}}}]{2011CQGra..28i4011K}%
  \BibitemOpen
  \bibfield  {author} {\bibinfo {author} {\bibfnamefont {S.}~\bibnamefont
  {{Kawamura}}}, \bibinfo {author} {\bibfnamefont {M.}~\bibnamefont {{Ando}}},
  \bibinfo {author} {\bibfnamefont {N.}~\bibnamefont {{Seto}}}, \bibinfo
  {author} {\bibfnamefont {S.}~\bibnamefont {{Sato}}}, \bibinfo {author}
  {\bibfnamefont {T.}~\bibnamefont {{Nakamura}}}, \bibinfo {author}
  {\bibfnamefont {K.}~\bibnamefont {{Tsubono}}}, \bibinfo {author}
  {\bibfnamefont {N.}~\bibnamefont {{Kanda}}}, \bibinfo {author} {\bibfnamefont
  {T.}~\bibnamefont {{Tanaka}}}, \bibinfo {author} {\bibfnamefont
  {J.}~\bibnamefont {{Yokoyama}}}, \bibinfo {author} {\bibfnamefont
  {I.}~\bibnamefont {{Funaki}}}, \bibinfo {author} {\bibfnamefont
  {K.}~\bibnamefont {{Numata}}}, \bibinfo {author} {\bibfnamefont
  {K.}~\bibnamefont {{Ioka}}}, \bibinfo {author} {\bibfnamefont
  {T.}~\bibnamefont {{Takashima}}}, \bibinfo {author} {\bibfnamefont
  {K.}~\bibnamefont {{Agatsuma}}}, \bibinfo {author} {\bibfnamefont
  {T.}~\bibnamefont {{Akutsu}}}, \bibinfo {author} {\bibfnamefont {K.-s.}\
  \bibnamefont {{Aoyanagi}}}, \bibinfo {author} {\bibfnamefont
  {K.}~\bibnamefont {{Arai}}}, \bibinfo {author} {\bibfnamefont
  {A.}~\bibnamefont {{Araya}}}, \bibinfo {author} {\bibfnamefont
  {H.}~\bibnamefont {{Asada}}}, \bibinfo {author} {\bibfnamefont
  {Y.}~\bibnamefont {{Aso}}}, \bibinfo {author} {\bibfnamefont
  {D.}~\bibnamefont {{Chen}}}, \bibinfo {author} {\bibfnamefont
  {T.}~\bibnamefont {{Chiba}}}, \bibinfo {author} {\bibfnamefont
  {T.}~\bibnamefont {{Ebisuzaki}}}, \bibinfo {author} {\bibfnamefont
  {Y.}~\bibnamefont {{Ejiri}}}, \bibinfo {author} {\bibfnamefont
  {M.}~\bibnamefont {{Enoki}}}, \bibinfo {author} {\bibfnamefont
  {Y.}~\bibnamefont {{Eriguchi}}}, \bibinfo {author} {\bibfnamefont {M.-K.}\
  \bibnamefont {{Fujimoto}}}, \bibinfo {author} {\bibfnamefont
  {R.}~\bibnamefont {{Fujita}}}, \bibinfo {author} {\bibfnamefont
  {M.}~\bibnamefont {{Fukushima}}}, \bibinfo {author} {\bibfnamefont
  {T.}~\bibnamefont {{Futamase}}}, \bibinfo {author} {\bibfnamefont
  {T.}~\bibnamefont {{Harada}}}, \bibinfo {author} {\bibfnamefont
  {T.}~\bibnamefont {{Hashimoto}}}, \bibinfo {author} {\bibfnamefont
  {K.}~\bibnamefont {{Hayama}}}, \bibinfo {author} {\bibfnamefont
  {W.}~\bibnamefont {{Hikida}}}, \bibinfo {author} {\bibfnamefont
  {Y.}~\bibnamefont {{Himemoto}}}, \bibinfo {author} {\bibfnamefont
  {H.}~\bibnamefont {{Hirabayashi}}}, \bibinfo {author} {\bibfnamefont
  {T.}~\bibnamefont {{Hiramatsu}}}, \bibinfo {author} {\bibfnamefont {F.-L.}\
  \bibnamefont {{Hong}}}, \bibinfo {author} {\bibfnamefont {H.}~\bibnamefont
  {{Horisawa}}}, \bibinfo {author} {\bibfnamefont {M.}~\bibnamefont
  {{Hosokawa}}}, \bibinfo {author} {\bibfnamefont {K.}~\bibnamefont
  {{Ichiki}}}, \bibinfo {author} {\bibfnamefont {T.}~\bibnamefont {{Ikegami}}},
  \bibinfo {author} {\bibfnamefont {K.~T.}\ \bibnamefont {{Inoue}}}, \bibinfo
  {author} {\bibfnamefont {K.}~\bibnamefont {{Ishidoshiro}}}, \bibinfo {author}
  {\bibfnamefont {H.}~\bibnamefont {{Ishihara}}}, \bibinfo {author}
  {\bibfnamefont {T.}~\bibnamefont {{Ishikawa}}}, \bibinfo {author}
  {\bibfnamefont {H.}~\bibnamefont {{Ishizaki}}}, \bibinfo {author}
  {\bibfnamefont {H.}~\bibnamefont {{Ito}}}, \bibinfo {author} {\bibfnamefont
  {Y.}~\bibnamefont {{Itoh}}}, \bibinfo {author} {\bibfnamefont
  {K.}~\bibnamefont {{Izumi}}}, \bibinfo {author} {\bibfnamefont
  {I.}~\bibnamefont {{Kawano}}}, \bibinfo {author} {\bibfnamefont
  {N.}~\bibnamefont {{Kawashima}}}, \bibinfo {author} {\bibfnamefont
  {F.}~\bibnamefont {{Kawazoe}}}, \bibinfo {author} {\bibfnamefont
  {N.}~\bibnamefont {{Kishimoto}}}, \bibinfo {author} {\bibfnamefont
  {K.}~\bibnamefont {{Kiuchi}}}, \bibinfo {author} {\bibfnamefont
  {S.}~\bibnamefont {{Kobayashi}}}, \bibinfo {author} {\bibfnamefont
  {K.}~\bibnamefont {{Kohri}}}, \bibinfo {author} {\bibfnamefont
  {H.}~\bibnamefont {{Koizumi}}}, \bibinfo {author} {\bibfnamefont
  {Y.}~\bibnamefont {{Kojima}}}, \bibinfo {author} {\bibfnamefont
  {K.}~\bibnamefont {{Kokeyama}}}, \bibinfo {author} {\bibfnamefont
  {W.}~\bibnamefont {{Kokuyama}}}, \bibinfo {author} {\bibfnamefont
  {K.}~\bibnamefont {{Kotake}}}, \bibinfo {author} {\bibfnamefont
  {Y.}~\bibnamefont {{Kozai}}}, \bibinfo {author} {\bibfnamefont
  {H.}~\bibnamefont {{Kunimori}}}, \bibinfo {author} {\bibfnamefont
  {H.}~\bibnamefont {{Kuninaka}}}, \bibinfo {author} {\bibfnamefont
  {K.}~\bibnamefont {{Kuroda}}}, \bibinfo {author} {\bibfnamefont
  {S.}~\bibnamefont {{Kuroyanagi}}}, \bibinfo {author} {\bibfnamefont {K.-i.}\
  \bibnamefont {{Maeda}}}, \bibinfo {author} {\bibfnamefont {H.}~\bibnamefont
  {{Matsuhara}}}, \bibinfo {author} {\bibfnamefont {N.}~\bibnamefont
  {{Matsumoto}}}, \bibinfo {author} {\bibfnamefont {Y.}~\bibnamefont
  {{Michimura}}}, \bibinfo {author} {\bibfnamefont {O.}~\bibnamefont
  {{Miyakawa}}}, \bibinfo {author} {\bibfnamefont {U.}~\bibnamefont
  {{Miyamoto}}}, \bibinfo {author} {\bibfnamefont {S.}~\bibnamefont
  {{Miyoki}}}, \bibinfo {author} {\bibfnamefont {M.~Y.}\ \bibnamefont
  {{Morimoto}}}, \bibinfo {author} {\bibfnamefont {T.}~\bibnamefont
  {{Morisawa}}}, \bibinfo {author} {\bibfnamefont {S.}~\bibnamefont
  {{Moriwaki}}}, \bibinfo {author} {\bibfnamefont {S.}~\bibnamefont
  {{Mukohyama}}}, \bibinfo {author} {\bibfnamefont {M.}~\bibnamefont
  {{Musha}}}, \bibinfo {author} {\bibfnamefont {S.}~\bibnamefont {{Nagano}}},
  \bibinfo {author} {\bibfnamefont {I.}~\bibnamefont {{Naito}}}, \bibinfo
  {author} {\bibfnamefont {K.}~\bibnamefont {{Nakamura}}}, \bibinfo {author}
  {\bibfnamefont {H.}~\bibnamefont {{Nakano}}}, \bibinfo {author}
  {\bibfnamefont {K.}~\bibnamefont {{Nakao}}}, \bibinfo {author} {\bibfnamefont
  {S.}~\bibnamefont {{Nakasuka}}}, \bibinfo {author} {\bibfnamefont
  {Y.}~\bibnamefont {{Nakayama}}}, \bibinfo {author} {\bibfnamefont
  {K.}~\bibnamefont {{Nakazawa}}}, \bibinfo {author} {\bibfnamefont
  {E.}~\bibnamefont {{Nishida}}}, \bibinfo {author} {\bibfnamefont
  {K.}~\bibnamefont {{Nishiyama}}}, \bibinfo {author} {\bibfnamefont
  {A.}~\bibnamefont {{Nishizawa}}}, \bibinfo {author} {\bibfnamefont
  {Y.}~\bibnamefont {{Niwa}}}, \bibinfo {author} {\bibfnamefont
  {T.}~\bibnamefont {{Noumi}}}, \bibinfo {author} {\bibfnamefont
  {Y.}~\bibnamefont {{Obuchi}}}, \bibinfo {author} {\bibfnamefont
  {M.}~\bibnamefont {{Ohashi}}}, \bibinfo {author} {\bibfnamefont
  {N.}~\bibnamefont {{Ohishi}}}, \bibinfo {author} {\bibfnamefont
  {M.}~\bibnamefont {{Ohkawa}}}, \bibinfo {author} {\bibfnamefont
  {K.}~\bibnamefont {{Okada}}}, \bibinfo {author} {\bibfnamefont
  {N.}~\bibnamefont {{Okada}}}, \bibinfo {author} {\bibfnamefont
  {K.}~\bibnamefont {{Oohara}}}, \bibinfo {author} {\bibfnamefont
  {N.}~\bibnamefont {{Sago}}}, \bibinfo {author} {\bibfnamefont
  {M.}~\bibnamefont {{Saijo}}}, \bibinfo {author} {\bibfnamefont
  {R.}~\bibnamefont {{Saito}}}, \bibinfo {author} {\bibfnamefont
  {M.}~\bibnamefont {{Sakagami}}}, \bibinfo {author} {\bibfnamefont {S.-i.}\
  \bibnamefont {{Sakai}}}, \bibinfo {author} {\bibfnamefont {S.}~\bibnamefont
  {{Sakata}}}, \bibinfo {author} {\bibfnamefont {M.}~\bibnamefont {{Sasaki}}},
  \bibinfo {author} {\bibfnamefont {T.}~\bibnamefont {{Sato}}}, \bibinfo
  {author} {\bibfnamefont {M.}~\bibnamefont {{Shibata}}}, \bibinfo {author}
  {\bibfnamefont {H.}~\bibnamefont {{Shinkai}}}, \bibinfo {author}
  {\bibfnamefont {A.}~\bibnamefont {{Shoda}}}, \bibinfo {author} {\bibfnamefont
  {K.}~\bibnamefont {{Somiya}}}, \bibinfo {author} {\bibfnamefont
  {H.}~\bibnamefont {{Sotani}}}, \bibinfo {author} {\bibfnamefont
  {N.}~\bibnamefont {{Sugiyama}}}, \bibinfo {author} {\bibfnamefont
  {Y.}~\bibnamefont {{Suwa}}}, \bibinfo {author} {\bibfnamefont
  {R.}~\bibnamefont {{Suzuki}}}, \bibinfo {author} {\bibfnamefont
  {H.}~\bibnamefont {{Tagoshi}}}, \bibinfo {author} {\bibfnamefont
  {F.}~\bibnamefont {{Takahashi}}}, \bibinfo {author} {\bibfnamefont
  {K.}~\bibnamefont {{Takahashi}}}, \bibinfo {author} {\bibfnamefont
  {K.}~\bibnamefont {{Takahashi}}}, \bibinfo {author} {\bibfnamefont
  {R.}~\bibnamefont {{Takahashi}}}, \bibinfo {author} {\bibfnamefont
  {R.}~\bibnamefont {{Takahashi}}}, \bibinfo {author} {\bibfnamefont
  {T.}~\bibnamefont {{Takahashi}}}, \bibinfo {author} {\bibfnamefont
  {H.}~\bibnamefont {{Takahashi}}}, \bibinfo {author} {\bibfnamefont
  {T.}~\bibnamefont {{Akiteru}}}, \bibinfo {author} {\bibfnamefont
  {T.}~\bibnamefont {{Takano}}}, \bibinfo {author} {\bibfnamefont
  {N.}~\bibnamefont {{Tanaka}}}, \bibinfo {author} {\bibfnamefont
  {K.}~\bibnamefont {{Taniguchi}}}, \bibinfo {author} {\bibfnamefont
  {A.}~\bibnamefont {{Taruya}}}, \bibinfo {author} {\bibfnamefont
  {H.}~\bibnamefont {{Tashiro}}}, \bibinfo {author} {\bibfnamefont
  {Y.}~\bibnamefont {{Torii}}}, \bibinfo {author} {\bibfnamefont
  {M.}~\bibnamefont {{Toyoshima}}}, \bibinfo {author} {\bibfnamefont
  {S.}~\bibnamefont {{Tsujikawa}}}, \bibinfo {author} {\bibfnamefont
  {Y.}~\bibnamefont {{Tsunesada}}}, \bibinfo {author} {\bibfnamefont
  {A.}~\bibnamefont {{Ueda}}}, \bibinfo {author} {\bibfnamefont {K.-i.}\
  \bibnamefont {{Ueda}}}, \bibinfo {author} {\bibfnamefont {M.}~\bibnamefont
  {{Utashima}}}, \bibinfo {author} {\bibfnamefont {Y.}~\bibnamefont
  {{Wakabayashi}}}, \bibinfo {author} {\bibfnamefont {K.}~\bibnamefont
  {{Yagi}}}, \bibinfo {author} {\bibfnamefont {H.}~\bibnamefont {{Yamakawa}}},
  \bibinfo {author} {\bibfnamefont {K.}~\bibnamefont {{Yamamoto}}}, \bibinfo
  {author} {\bibfnamefont {T.}~\bibnamefont {{Yamazaki}}}, \bibinfo {author}
  {\bibfnamefont {C.-M.}\ \bibnamefont {{Yoo}}}, \bibinfo {author}
  {\bibfnamefont {S.}~\bibnamefont {{Yoshida}}}, \bibinfo {author}
  {\bibfnamefont {T.}~\bibnamefont {{Yoshino}}}, \ and\ \bibinfo {author}
  {\bibfnamefont {K.-X.}\ \bibnamefont {{Sun}}},\ }\href {\doibase
  10.1088/0264-9381/28/9/094011} {\bibfield  {journal} {\bibinfo  {journal}
  {Classical and Quantum Gravity}\ }\textbf {\bibinfo {volume} {28}},\ \bibinfo
  {eid} {094011} (\bibinfo {year} {2011})}\BibitemShut {NoStop}%
\bibitem [{\citenamefont {{Masui}}\ and\ \citenamefont
  {{Pen}}(2010)}]{2010PhRvL.105p1302M}%
  \BibitemOpen
  \bibfield  {author} {\bibinfo {author} {\bibfnamefont {K.~W.}\ \bibnamefont
  {{Masui}}}\ and\ \bibinfo {author} {\bibfnamefont {U.-L.}\ \bibnamefont
  {{Pen}}},\ }\href {\doibase 10.1103/PhysRevLett.105.161302} {\bibfield
  {journal} {\bibinfo  {journal} {Physical Review Letters}\ }\textbf {\bibinfo
  {volume} {105}},\ \bibinfo {eid} {161302} (\bibinfo {year} {2010})},\ \Eprint
  {http://arxiv.org/abs/1006.4181} {arXiv:1006.4181 [astro-ph.CO]} \BibitemShut
  {NoStop}%
\bibitem [{\citenamefont {{Book}}\ \emph {et~al.}(2012)\citenamefont {{Book}},
  \citenamefont {{Kamionkowski}},\ and\ \citenamefont
  {{Schmidt}}}]{2012PhRvL.108u1301B}%
  \BibitemOpen
  \bibfield  {author} {\bibinfo {author} {\bibfnamefont {L.}~\bibnamefont
  {{Book}}}, \bibinfo {author} {\bibfnamefont {M.}~\bibnamefont
  {{Kamionkowski}}}, \ and\ \bibinfo {author} {\bibfnamefont {F.}~\bibnamefont
  {{Schmidt}}},\ }\href {\doibase 10.1103/PhysRevLett.108.211301} {\bibfield
  {journal} {\bibinfo  {journal} {Physical Review Letters}\ }\textbf {\bibinfo
  {volume} {108}},\ \bibinfo {eid} {211301} (\bibinfo {year} {2012})},\ \Eprint
  {http://arxiv.org/abs/1112.0567} {arXiv:1112.0567 [astro-ph.CO]} \BibitemShut
  {NoStop}%
\end{thebibliography}%

\end{document}